**An integrated *P-T*-H$_2$O-lattice strain model to quantify the role of clinopyroxene fractionation on REE+Y/HFSE patterns of mafic alkaline magmas: Application to eruptions at Mt. Etna**

Silvio Mollo[1,2], Jonathan Blundy[3], Piergiorgio Scarlato[2], Serena Pia De Cristofaro[4], Vanni Tecchiato[1], Flavio Di Stefano[1], Francesco Vetere[5], Francois Holtz[6], Olivier Bachmann[7]

[1]Dipartimento di Scienze della Terra, Sapienza-Università di Roma, P.le Aldo Moro 5, 00185 Roma, Italy

[2]Istituto Nazionale di Geofisica e Vulcanologia, Via di Vigna Murata 605, 00143 Rome, Italy

[3]School of Earth Sciences, University of Bristol, Wills Memorial Building, Bristol BS8 1RJ, UK

[4]Dipartimento di Scienze della Terra, Università degli Studi di Torino, Via Valperga Caluso 35, 10125 Torino, Italy

[5]Dipartimento di Fisica e Geologia, Università di Perugia, Piazza Università 1, 06100 Perugia, Italy

[6]Institute of Mineralogy, Leibniz University of Hannover, Callinstrasse 3, 30167 Hannover, Germany

[7]Institute of Geochemistry and Petrology, ETH Zürich, Clausiusstrasse 25, 8092 Zürich, Switzerland

**Abstract**

A correct description and quantification of the geochemical behaviour of REE+Y (rare earth elements and Y) and HFSE (high field strength elements) is a key requirement for modelling petrological and volcanological aspects of magma dynamics. In this context, mafic alkaline magmas (MAM) are characterized by the ubiquitous stability of clinopyroxene from mantle depths to shallow crustal levels. On one hand, clinopyroxene incorporates REE+Y/HFSE at concentration levels that are much higher than those measured for olivine, plagioclase, and magnetite. On the other hand, the composition of clinopyroxene is highly sensitive to variations in pressure, temperature, and melt-water content, according to exchange-equilibria between jadeite and melt, and between jadeite/Ca-Tschermak and diopside-hedenbergite. As a consequence, the dependence of the partition coefficient on the physicochemical state of the system results in a variety of $D_{REE+Y}/D_{HFSE}$ values that are sensitive to the magmatic conditions at which clinopyroxenes nucleate and grow.

In order to better explore magma dynamics using clinopyroxene, a new $P$-$T$-$H_2O$-lattice strain model specific to MAM compositions has been developed. The model combines a set of refined clinopyroxene-based barometric, thermometric and hygrometric equations with thermodynamically-derived expressions for the three lattice strain parameters, i.e., the strain-free partition coefficient ($D_0$), the site radius ($r_0$), and the effective elastic modulus ($E$). The accuracy of the model has been tested against experimental and thermodynamic data, whereas its applicability to natural environments has been verified using clinopyroxene-melt pairs from a great number of volcanic eruptions. Results from these calculations show that the entry of REE+Y/HFSE into M2/M1 octahedral sites of clinopyroxene is determined by a variety of effects that may or may not change simultaneously during magma differentiation. In accordance with thermodynamic principles, $D_{REE+Y}/D_{HFSE}$ are positively correlated with $^TAl$ in clinopyroxene and negatively correlated with $T$ and $H_2O$ dissolved in the melt. Despite the increase in partition coefficients increase with increasing pressure due to the positive volume of fusion of silicate minerals, changes in $D_{REE+Y}/D_{HFSE}$ with $P$ are prevalently attributed to the covarying temperature and bulk composition. Due to the electrostatic work done by placing a cation in a charge-balanced/imbalanced crystal site, crystal electrostatic effects also have a great influence on REE+Y/HFSE incorporation. Additionally, melt depolymerization increases the number of large structural sites critically important to accommodate trace element cations in the melt phase rather than in the crystal lattice. When $D_{REE+Y}/D_{HFSE}$ values recovered by the $P$-$T$-$H_2O$-lattice strain model are used as input data to quantify fractional crystallization processes in natural MAM compositions from Mt. Etna volcano (Sicily, Italy), it is found that the concentration of REE+Y/HFSE in the magma is primary controlled by the geochemical evolution of clinopyroxene in terms of major cation exchange-equilibria and trace cation lattice strain properties.

# 1. Introduction

Mafic alkaline magmas (MAM) are typical of intraplate settings such as oceanic islands, intra-continental volcanoes, and continental rift zones (Pilet, 2015). Generally, natural rocks comprise basalts, trachybasalts, and basaltic trachyandesites, as members of the same differentiation series with either K- or Na-affinities. Although the abundance of MAM is relatively small at the Earth's surface, alkaline volcanism is an important of magma generation by mantle processes and further differentiation at crustal levels, due to the different mineralogical, petrological, and geochemical aspects of the erupted products (e.g., Eggler and Holloway, 1977; Dasgupta and Hirschmann, 2007).

The mineral association of olivine + clinopyroxene + plagioclase + magnetite is recognized as one of the most common parageneses controlling the geochemical evolution of MAM at both mantle and crustal depths (e.g., Armienti et al., 2013). Among these mineral phases, the crystallization of clinopyroxene impacts significantly the composition of magma in terms of REE+Y/HFSE due to the preferential incorporation of these trace elements into clinopyroxene. As documented by a number of studies (D'Orazio et al., 1998; Wood and Blundy, 1997, 2001, 2002; Bedard, 2005, 2014; Sun and Liang, 2012, 2014; Dohmen and Blundy, 2014; Laubier et al., 2014; Sun et al., 2017), values of clinopyroxene $D_{REE+Y}/D_{HFSE}$ [$D_i = {}^{xls}(I)/{}^{melt}(I)$ on a weight basis where $i$ refers to the cation of interest] are considerably higher those for olivine, plagioclase, and magnetite. This is better evidenced when the hypothetical weight fractions of olivine (${}^{ol}X = 0.05\text{-}0.15$), clinopyroxene (${}^{cpx}X = 0.05\text{-}0.25$), plagioclase (${}^{pl}X = 0.05\text{-}0.25$), and magnetite (${}^{mt}X = 0.05$) that may contribute to MAM evolution are multiplied by the corresponding partition coefficients measured for REE (i.e., ${}^{ol}D_{La} = 0.01$, ${}^{cpx}D_{La} = 0.2$, ${}^{pl}D_{La} = 0.01$, and ${}^{mt}D_{La} = 0.1$) and HFSE (i.e., ${}^{ol}D_{Zr} = 0.04$, ${}^{cpx}D_{Zr} = 0.4$, ${}^{pl}D_{Zr} = 0.03$, and ${}^{mt}D_{Zr} = 0.3$; partitioning data from D'Orazio et al., 1998). Results from this simple calculation show that, as a function of the different mineral proportions, the effect of clinopyroxene fractionation on the final concentration of La and Zr in the magma can be one, two or even three orders of magnitude greater than that of olivine, plagioclase or magnetite (Fig. 1).

Considering that REE+Y are weakly mobile in fluid-rich magmatic environments and that HFSE are practically fluid-immobile (Tsay et al., 2014), it is unsurprising that the partitioning of trace elements between clinopyroxene and melt is frequently used in petrological and geochemical investigations dealing with MAM differentiation processes, such as fractional or equilibrium crystallization, assimilation, and partial melting (e.g., Di Rocco et al., 2012; Del Bello et al., 2014; Vetere et al., 2015). However, during magma modeling, the partition coefficient cannot be assumed constant and independent of pressure, temperature, melt-water content, and bulk composition (i.e., crystal/melt chemistry). To do this would represent a crude oversimplification that may potentially lead to miscalculations and erroneous conclusions (cf. Wood and Trigila, 2001; Blundy and Wood 2003). Clinopyroxene is highly sensitive to variations in crystallization histories of magmas and, accordingly, several solution models (i.e., exchange-equilibria between jadeite and melt, and between jadeite/Ca-Tschermak and diopside-hedenbergite) have been proposed over time to derive more accurate barometric, thermometric, and hygrometric equations (Putirka et al., 1996, 2003; Putirka, 2008; Masotta et al., 2013; Mollo et al., 2013a; Neave and Putirka, 2017; Perinelli et al., 2016). A number of efforts have been made to parameterize the partitioning of REE+Y/HFSE between clinopyroxene and melt as a function of the physicochemical state of the system (Wood and Blundy, 1997, 2001, 2002; Hill et al., 2011; Sun and Liang, 2012; Bedard, 2014; Mollo et al., 2016). As a general rule, partition coefficients decrease with increasing temperature due to the positive entropy of fusion of silicate minerals and increase with increasing pressure due to the positive volume of fusion (Hill et al., 2011 and references therein). However, the solution of jadeite/Ca-Tschermak components into diopside-hedenbergite may greatly influence the number of charge-balanced configurations able to accommodate the REE+Y/HFSE into clinopyroxene crystal lattice (Hill et al., 2000; Wood and Trigila, 2001; Bennett et al., 2004; Sun and Liang, 2012; Mollo et al., 2013b, 2017; Scarlato et al., 2014). Despite this compositional effect, there is no obvious correlation between jadeite/Ca-Tschermak contents and $D_{REE+Y}/D_{HFSE}$, in view of more complex cation incorporation mechanisms due to the electrostatic work done in placing trace element cations into charge-balanced/imbalanced structural sites (Wood and Blundy, 2001; Hill et al., 2011) and the melt structural influence on trace element partitioning (Gaetani et al., 2003; Gaetani, 2004; Huang et al., 2006; Qian et al., 2015). Thus, modeling the evolutionary behaviour of $D_{REE+Y}/D_{HFSE}$ along the $P$-$T$-$H_2O$ differentiation path of magma is not a trivial task, especially because major cation exchange-equilibria and thermodynamic descriptions for the partitioning of REE+Y/HFSE have been systematically treated as separate topics.

The present work provides an updated review of previous clinopyroxene-based barometers, thermometers, and hygrometers, in order to derive a set of integrated predictive equations specific to MAM compositions. These equations are then combined with different thermodynamically-derived REE+Y/HFSE lattice strain expressions, in order to quantify the change of $D_{REE+Y}/D_{HFSE}$ as a function of the physicochemical state of the system. The validity of the $P$-$T$-$H_2O$-lattice strain model is tested through the use of an impressively broad array of natural compositions available for eruptions at Mt. Etna (Sicily, Italy), the largest volcano in Europe and one of the most monitored volcanoes of the world. Through this approach, it is documented as the incorporation of REE+Y/HFSE into clinopyroxene crystallographic sites is strictly controlled by the covarying effects of temperature, pressure, and melt-water content, as well as the effects of tetrahedrally-coordinated aluminum, crystal electrostatic work, and melt structure. Evidently, Mt. Etna volcano is an excellent case study due to the ubiquitous crystallization of clinopyroxene from mantle depths to shallow crustal levels, with the convenience of a great number of chemical analyses for the historic and recent eruptions. However, the $P$-$T$-$H_2O$-lattice strain model is able to track the geochemical variation of REE+Y/HFSE in MAM from different and complex intraplate settings where clinopyroxene fractionation contributes significantly to compositional evolution.

## 2. Mt. Etna volcano

Mt. Etna is a ~3,340-m-high composite stratovolcano (Fig. 2), covering ~1,250 km$^2$ on the eastern coast of Sicily (Italy) at the intersection of two major fault zones trending NNW–SSE and NNE–SSW. This area represents a suture zone between the African and European plates, where the geodynamic setting is dominated by the compressional front of the Appennine–Maghrebian Chain, the transform zone in NE Sicily, and the subduction of the Ionian slab (Cristofolini et al., 1985). Volcanic activity of Mt. Etna began with fissural emission of sub-marine and sub-aerial tholeiitic lavas (~500 ka; Branca et al., 2004). Subsequently, the erupted products gradually shifted towards transitional and Na-alkaline compositions (~220 ka; Tanguy et al., 1997).

The eruptive centers can be schematically grouped into main volcano-stratigraphic units (Cristofolini and Romano 1982). The Ancient Alkaline Centers (180–100 ka) unit corresponds to the transition from subalkaline to alkaline products, marking a change in style from fissure to central eruptions. Thereafter, three volcanic centers (i.e., Monte Po, Calanna, and Trifoglietto I) developed in the Val Calanna and Valle del Bove areas. The Trifoglietto (80–60 ka) unit marks the formation of the composite stratovolcano by superimposition of some small volcanic centers located on the south-western sector of the Valle del Bove (Serra Giannicola Piccola, Trifoglietto II, Zoccolaro, Vavalaci, and Belvedere). The Ancient Mongibello (35–15 ka) unit formed by the volcanic activity of two distinct eruptive centers (Ellittico and Leone). The Recent Mongibello (15 ka to Present) unit, responsible for the construction of the summit cones, was characterized by a wide range of eruptive styles, from effusive and mildly Strombolian to sub-Plinian. Products of Ancient and Recent Mongibello cover ~85% of the present volcano surface (Catalano et al., 2004). During this time span, magmatic intrusions occurred mainly in the NE and S rift zone of Mt. Etna edifice (Branca and Del Carlo, 2005). From the second half of the 17$^{th}$ century, volcanic activity was characterized by both periods of explosive (summit craters) and effusive (flank eruptions) eruptions. After 1971, a significant increase in eruption frequency and eruption rate was observed, with large volumes of magma produced with the 1983 eruption on the S flank and the 1991–1993 lateral eruptions in Valle del Bove (Branca and Del Carlo, 2005).

The structure of the present feeding system is characterized by an open-conduit system that is persistently filled with magma feeding eruptions from the summit craters. This fresh magma undergoes continuous degassing and rises from the deeper portion of the plumbing system (6–15 km) into more shallow reservoirs (3–5 km) (Patanè et al., 2003). Explosive Strombolian eruptions and lava fountains characterize the two summit craters of Bocca Nuova (BN, including two pit craters BN1 and BN2) and Chasm or Voragine (VOR), and the two parasitic cones of NE Crater (NEC) and SE Crater (SEC) (Fig. 2). In contrast, effusive flank eruptions occur as both lateral eruptions draining magma from the central conduit (Corsaro et al., 2009) and as eccentric (peripheral) eruptions bypassing the central conduit and draining magma from the deeper feeding system (Andronico et al., 2005).

A peculiar characteristic of Etnean volcanic rocks is that both explosive and effusive products share a common mineral assemblage of plagioclase + clinopyroxene + olivine + titanomagnetite. Only rarely are a few amphibole and orthopyroxene crystals are also observed. The historic (pre-1971) products show a distinctive Na-affinity, being mostly hawaiites with subordinate mugearites. Conversely, the recent (post-1971) products gradually shift towards a K-affinity and are prevalently trachybasalts that sporadically evolve towards basaltic trachyandesites (Clocchiatti et al., 1988). As a whole, the volcanic rocks at Mt. Etna exhibit a geochemical signature that is transitional between ocean island basalts (OIB) and island arc basalts (Cristofolini et al., 1987; Schiano et al., 2001). However, recent eruptions display a progressive enrichment in K, Rb, Cs, and radiogeneic Sr isotopes coupled with depletion in Th and radiogenic Nd–Pb–Hf isotopes (Tanguy et al., 1997). This geochemical variation is highly debated in literature due to the complex geodynamic setting of the area. Indeed, the magmatic activity of Mt. Etna is interpreted as the effect of rollback-induced upper mantle upwelling around the Ionian slab edge (Schellart, 2010), or the progressive migration of the subducting Ionian slab (Schiano et al., 2001), or the opening of a slab window (Doglioni et al., 2001). The progressive enrichment in LILE with respect to HFSE is tentatively addressed to a great number of crustal-, fluid-, and mantle-related phenomena, such as the contamination of magma by crustal materials of the sub-volcanic sedimentary basement (e.g., Clocchiatti et al., 1988; Michaud, 1994; Condomines et al., 1995), the interaction of magma with supercritical fluids carrying chlorine and alkalis (Ferlito and Lanzafame, 2010; Ferlito et al., 2014), the interaction between an OIB-type mantle source and a slab-derived component (Schiano et al., 2001), the metasomatic action in the mantle of slab-derived fluids enriched in B, $\delta^{11}B$, and fluid-mobile elements (Armienti et al., 2004; Tonarini et al., 2001), the partial melting of a heterogeneous mantle source containing variable amounts of hydrous phases (i.e., phlogopite/amphibole; Viccaro and Cristofolini, 2008) and, finally, the partial melting of a $^{87}$Sr–Cl-rich clinopyroxenite-veined mantle (i.e., a Hyblean peridotite-type source) metasomatized by slab-derived fluids selectively enriched in LILE (Corsaro and Metrich, 2016).

## 3. Methods

*3.1 Starting materials*

In order to effectively track the whole differentiation path of MAM, the starting materials used in this study represent some of the most common basalts, trachybasalts, and basaltic trachyandesites erupted at Mt. Etna volcano (Table 1S).

The basalt (BA; MgO ≈ 8 wt.%) belongs to the Monte Maletto lava dated 7000 years BP (Armienti et al., 1988). Monte Maletto has been widely adopted as the primitive composition of historic and recent Etnean eruptions, as indicated by (i) mass balance and trace element calculations, (ii) MELTS simulations, (iii) partial melting modeling of a phlogopite/amphibole-bearing mantle source, and (iv) phase equilibrium experiments (Tanguy et al., 1997; Armienti et al., 1988, 1997, 2007, 2004, 2013; Metrich and Rutherford, 1998; Tonarini et al., 2001; Ferlito and Lanzafame, 2010; Ferlito et al., 2014). It has been also argued that the isotopic signature ($^{87}$Sr/$^{86}$Sr and $\delta^{11}$B) of Monte Maletto is consistent with that of pre-1971 eruptions. However, due to the metasomatic action of fluids in the mantle source, post-1971 products are notably enriched in K and Rb, departing from the LILE signature of pre-1971 magmas (Corsaro and Metrich, 2016). Indeed, post-1971 products may result from the differentiation of a primitive magma similar to that of sample 041102A (MgO ≈ 7.5 wt.%) in Corsaro et al. (2009) and erupted from the southern fissures opened during 2002-2003 volcanic activity (R. A. Corsaro, personal communication). In terms of REE and HFSE ratios, this fissure eruption (La/Yb = 23 and Zr/Hf = 44) is very similar to Monte Maletto (La/Yb = 20 and Zr/Hf = 43), suggesting that mantle source modifications are limited to the selective transfer of fluid-mobile elements (cf. Corsaro and Metrich, 2016).

The trachybasalt (TB; MgO ≈ 5.5 wt.%) consists of scoria clasts erupted during 2011-2013 lava fountains (Mollo et al., 2015a). Major (CaO/Al$_2$O$_3$ = 0.68) and trace element (La/Yb = 25 and Zr/Hf = 45) concentrations identify TB as one of the less differentiated trachybasalts erupted from the summit craters since post-1971 volcanic activity (cf. Corsaro et al., 2013).

The basaltic trachyandesite (BT; MgO ≈ 4 wt.%) is representative of Etnean mugearites younger than the 16$^{th}$ century and erupted during the Recent Mongibello volcanic activity (Corsaro and Cristofolini, 1996). With respect to the more differentiated products, the selected basaltic trachyandesite exhibits relatively low REE (La/Yb = 41) and HFSE (Zr/Hf = 50) ratios, indicative of low degrees of crystal fractionation before eruption to the surface.

*3.2 Experimental strategy*

Crystallization experiments were carried out in an internally heated pressure vessel (IHPV) using Ar as pressure medium at the Institute for Mineralogy, Leibniz University of Hannover (Germany). The three starting materials were powdered and loaded in $Au_{80}Pd_{20}$ capsules containing 2 and 3 wt.% $H_2O$ added as deionized water (Table 1S). The volatile content is consistent with both $H_2O$ measurements in melt inclusions from Etnean olivines (Spilliaert et al., 2006; Collins et al., 2009) and $H_2O$ estimates based on clinopyroxene/plagioclase-melt hygrometry (Armienti et al., 2013; Mollo et al., 2015b; Perinelli et al., 2016). The charges were heated directly to the superliquidus temperature of 1,200 °C (Vetere et al., 2015) that was maintained constant for 1 h. The temperature was then decreased to the final target temperature of 1,050 °C. This value corresponds to direct measurements of inner lava flow temperatures (Tanguy and Clocchiatti, 1984) and, thus, to the closure temperature of the system (i.e., the overall crystallization path of magma at the time of eruption). The experiments consisted of four different experimental sets (Run#1, Run#2, Run#3, and Run#4) designed to reproduce isothermal/isobaric and isothermal/decompression conditions at Mt. Etna volcano (Table 1S). In Run#1, the charges were kept at 400 MPa for 24 h to reproduce magma equilibrium crystallization at depth. In Run#2, the charges were kept at 50 MPa for 24 h to reproduce magma equilibrium crystallization in a shallow reservoir. In Run#3, the charges were kept at 400 MPa for 24 h, decompressed to 50 MPa at a rate of 0.06 MPa/s, and then immediately quenched to reproduce relatively rapid ascent conditions controlled by the kinetic effects associated with magma undercooling. In Run#4, the charges were kept at 400 MPa for 24 h, decompressed to 50 MPa at a rate of 0.06 MPa/s, kept at 50 MPa for 24 h, and then quenched to reproduce the ascent of a deep-seated magma and its subsequent storage in a shallow reservoir where equilibrium crystallization occurred. All experiments were quenched using a rapid-quench device leading to a cooling rate of approximately 150 °C/s (Berndt et al., 2002).

According to previous data from phase equilibrium experiments, thermodynamic and solubility modeling, and *P-T*-$H_2O$-*f*$O_2$ estimates (Mollo et al., 2015b), the physicochemical state of the Etnean system changes from $H_2O$-undersaturated (i.e., fluid-absent experiments at 400 MPa) to $H_2O$-saturated (i.e., fluid-present experiments at 50 MPa) regimes driving the differentiation of magmas. Note that, at 1,050 °C, the solubility of $H_2O$ in MAM compositions is ≤9 and ≤3 wt.% at 400 and 50 MPa, respectively (data from the thermodynamically-derived formalism of Duan, 2014). Moreover, the decompression rate of 0.06 MPa/s translates to magma ascent velocity of 0.12 m/s, in agreement with that measured for eruptions at Mt. Etna (Aloisi et al., 2006; Armienti et al., 2014; Mollo et al., 2015a). The use of Ar-$H_2$ mixture as a pressure medium provided the possibility to adjust $fH_2$ in the vessel (cf. Berndt et al., 2002). The experiments were conducted at an oxygen fugacity NNO +1 log unit, corresponding to the redox state of the natural products (Armienti et al., 1994, 2004, 2013; Giacomoni et al., 2014; Mollo et al., 2011, 2013c, 2015b). Within the sample capsule, the hydrogen fugacity was affected by permeation of hydrogen through the capsule walls driven by the $fH_2$ gradient between the pressure medium and the capsule interior. This in turn, controls the fugacity of oxygen inside the capsule through the equilibrium reaction of water formation ($H_2 + 1/2\ O_2 \leftrightarrow H_2O$). At the beginning of the experiment the $fH_2$ inside the capsule is in disequilibrium with the pressure medium, as well as dynamic redox conditions that occur during decompression and volatile exsolution. Thus, the redox state of the charges were estimated at the end of the experiments through the oxygen barometers of Lundgaard and Tegner (2004) and Ishibashi (2013) based on plagioclase-melt (error ±0.6) and spinel-melt (error ±0.8) equilibria, respectively. These models provided estimates from NNO+0.5 to NNO+1.5 that, within the calibration error, are comparable to the redox state of Etnean compositions (Table 1S).

To verify the attainment of equilibrium conditions and the supply of equilibrium cation proportions to the growing crystals, the Fe-Mg exchange reactions of $^{ol-melt}Kd_{Fe-Mg} = 0.30\pm0.03$ (Roeder and Emslie, 1970) and $^{cpx-melt}Kd_{Fe-Mg} = 0.27\pm0.03$ (Putirka et al., 2003) were used for olivine and clinopyroxene, respectively. Results from experiments ($^{ol-melt}Kd_{Fe-Mg} = 0.27$-$0.33$ and $^{cpx-melt}Kd_{Fe-Mg} = 0.24$-$0.28$) conform to those expected for equilibrium crystal-melt pairs (Table 1S). Conversely, the equilibrium crystallization of plagioclase was tested through the empirical (*T*-independent) and thermodynamic (*T*-dependent) equations of Namur et al. (2012) that predict the anorthite (An) content in plagioclase from the activity of An in the melt. The difference (Δ) between measured and predicted An contents is relatively low (ΔAn = 0.00-0.09), suggesting equilibrium to near-equilibrium crystallization conditions (Table 1S)

Modal phase proportions in the experimental charges (Table 1S) were derived by mass balance calculations (Stormer & Nicholls, 1978), yielding fairly good residual sum of squares ($\Sigma r^2$) below 0.25.

*3.3 Chemical analyses*

Major element concentrations (Table 2S) were obtained at the HP-HT Laboratory of Experimental Volcanology and Geophysics of the Istituto Nazionale di Geofisica e Vulcanologia in Roma (Italy), using a Jeol-JXA8200 equipped with five spectrometers. For glasses, a slightly defocused electron beam with a size of 3 μm was used with a counting time of 5 s on background and 15 s on peak. For crystals, the beam size was 1 μm with a counting time of 20 and 10 s on peaks and background respectively. The following standards have been adopted: jadeite (Si and Na), corundum (Al), forsterite (Mg), andradite (Fe), rutile (Ti), orthoclase (K), barite (Ba), apatite (P), spessartine (Mn) and chromite (Cr). Sodium and potassium were analyzed first to prevent alkali migration effects. The precision of the microprobe was measured through the analysis of well-characterized synthetic standards. Data quality was ensured by analyzing these standards as unknowns. Based on counting statistics, analytical uncertainties relative to their reported concentrations indicate that precision and accuracy were better than 5% for all cations (cf. Iezzi et al., 2008, 2011).

Trace element analyses (Table 3S) were conducted at the Institute of Geochemistry and Petrology of the ETH Zürich (Switzerland) with a 193 nm excimer laser coupled with a second generation two-volume constant geometry ablation cell (Resonetics:S-155LR) and a high-sensitivity, sector-field inductively-coupled plasma mass spectrometer (ICP-MS; Thermo:Element XR). Points with a spot size of 15 μm were set on chemically homogeneous portions of the material previously analyzed by EPMA, and ablated with a pulse rate of 10 Hz and an energy density of 3.5 J/cm$^3$ for 40 sec. The isotopes were analyzed relative to an internal standard of known composition (i.e., NIST612). A second standard (i.e., GSD-1G) was used as an unknown to check data quality during each analytical run. $^{43}$Ca or $^{29}$Si were used as internal standards for clinopyroxene and glasses, respectively, in order to recover the concentrations of REE and HFSE. The precision of individual analyses varied depending upon a number of factors, e.g., the element and isotope analyzed as well as the homogeneity of the ablated material. The 1 sigma errors calculated from variations in replicate analyses were invariably several times larger than the fully integrated 1 sigma errors determined from counting statistics alone (cf. Mollo et al., 2016).

*3.4 Calculation of clinopyroxene and melt components*

Clinopyroxene components, on the basis of six oxygen atoms, were determined using the procedure reported in Putirka et al. (1996) and subsequently reappraised in Putirka (1999). The charge balance equation of Lindsley (1983) was applied to all the analyses for determining Fe$^{3+}$. Jadeite (Jd, NaAlSi$_2$O$_6$) corresponds to the amount of Na or octahedral Al ($^{M1}$Al = $^{TOT}$Al - $^{T}$Al; $^{T}$Al = 2-Si). CrCa-Tschermak (CrCaTs, CaCr$_2$SiO$_6$ = Cr/2) is calculated at first, then Ca-Tschermak (CaTs, $^{M1}$Ca$^{T}$Al$^{M1}$AlSiO$_6$) is equal to any remaining $^{M1}$Al (CaTs = $^{M1}$Al - Jd). $^{M1}$Al in excess is used to form CaTi-Tschermak (CaTiTs, CaTiAl$_2$O$_6$ = ($^{T}$Al – CaTs)/2) and CaFe-Tschermak (CaFeTs, CaFeSiAlO$_6$). All Ca remaining after forming Ts, i.e. the sum of CaTs, CaFeTs, CrCaTs and CaTiTs, gives diopside (Di, CaMgSi$_2$O$_6$) and hedenbergite (Hd, CaFeSi$_2$O$_6$) components, i.e., DiHd

= Ca – Ts. Only Mg and $Fe^{2+}$ are used for calculation of the enstatite (En, $Mg_2Si_2O_6$) and ferrosilite (Fs, $Fe_2Si_2O_6$). The enstatite-ferrosilite (EnFs) component is equal to one-half the FeO + MgO (Fm) component remaining after forming DiHd (EnFs = (Fm - DiHd)/2). At equilibrium it is expected that the sum of clinopyroxene components calculated following this scheme is very close to unity (see also Putirka et al., 1996; Putirka, 1999).

Melt components were calculated as cation fractions following the scheme reported in Putirka (1999). However, in the case of natural Etnean products, the occurrence of residual melts is extremely rare due to abundant crystallization of matrix minerals at both syn- and post-eruptive conditions (Corsaro et al., 2013). Therefore, it is difficult to recognize the chemical correspondence between a zoned clinopyroxene and the melt from which it crystallized. In general, the bulk rock analysis is assumed as the equilibrium magma composition feeding the early growth of clinopyroxene cores (Putirka et al., 2008). However, this strategy cannot be safely applied when minerals are not the liquidus phase and/or magma mixing takes place. In relation to magma dynamics at Mt. Etna volcano, the clinopyroxene crystallizes cotectically with olivine (Armienti et al., 2013) and the plumbing system is characterized by transfer of magma through several compositionally different sub-volcanic environments (Kahl et al., 2015). This means that thermometers, barometers, hygrometers, and lattice-strain models based on clinopyroxene-melt equilibrium can be affected by great uncertainties. To obviate this issue, the compositions of the melts coexisting with clinopyroxene phenocrysts have been recalculated by the excel spreadsheet (https://academic.oup.com/petrology/article/54/4/795/1549443/A-New-Model-to-Estimate-Deep-level-Magma-Ascent#22552129) developed by Armienti et al. (2013). This algorithm corrects the bulk-rock composition by reconstructing the differentiation path of Etnean magmas evolving along the olivine-clinopyroxene-plagioclase cotectic. The recalculation method is based on mineral-melt distribution coefficients and mineral stoichiometry, in conformity with the formalism of Pearce (1978). The algorithm has been tested with chemical data from either phase equilibria experiments or natural historic eruptions from Mt. Etna, providing reliable melt corrections for degrees of crystallization lower than 15% (Armienti et al., 2013). In the present study, the melt recalculation was restricted to a maximum degree of crystallization of ~6% (Table 5).

*3.5 Selection of the natural compositions*

Clinopyroxene compositions and bulk rock analyses of eruptions at Mt. Etna volcano (Table 5) were recovered from the GEOROC database (http://georoc.mpch-mainz.gwdg.de/georoc). Inspection of the analytical dataset evidences that 1) a moderate number of clinopyroxene compositions are generally provided by the authors but, in some cases, only a few data are found, 2) bulk rock analyses are frequently incomplete for REE+Y/HFSE (particularly for Pr, Gd, Dy, Ho, Er, Tm, Y, Nb, Ta, and Hf), 3) a complete bulk rock dataset does not necessarily match with an adequate number of clinopyroxene compositions, and 4) some analytical biases are found due to the use of different laboratories, techniques, and standards. While discrepant bulk rock analyses and non-stoichiometric clinopyroxene compositions can be detected and excluded, the lack of a good correspondence between clinopyroxene and bulk rock data represents the most critical restriction. As a consequence, this study deals only with Etnean products showing an appropriate number of clinopyroxene and bulk rock analyses for the modeling of $P$, $T$, $H_2O$, and lattice strain parameters, provided that the achievement of equilibrium crystallization conditions is tested for each clinopyroxene-melt pair used as input data for the calculation (see below).

The activity of Mt. Etna is characterized by a highly dynamic regime resulting from continuous recharge of the volcanic conduit due to rapid input of new magma batches from depth. Under such dynamic conditions, the crystal growth of Etnean phenocrysts is strongly influenced by volatile loss (i.e., degassing) and heat dissipation (i.e., cooling) due to variable degrees of undercooling experienced by magma (Lanzafame et al., 2013; Mollo et al., 2013a, 2013c, 2015b; Vetere et al., 2015; Giuffrida et al., 2017). In particular, the occurrence of zoned phenocrysts testifies to cation redistribution reactions at the crystal-melt interface that are controlled by kinetic pathways in which the crystal compositions attempt to equilibrate with an ever-changing $P$-$T$-$H_2O$ path bounded between the mineral saturation temperature and the closure temperature for crystal growth (Mollo et al., 2013a, 2015a). For the case of clinopyroxene, it has been demonstrated that values of $D_{Fe}$ and $D_{Mg}$ tend towards an almost constant $D_{Fe}/D_{Mg}$ ratio when the crystal growth rate is governed by undercooling (Mollo and Hammer, 2017). This means that the exchange of Fe and Mg between clinopyroxene and melt is not a reliable indicator of equilibrium crystallization for rapidly growing crystals (Baker and Grove, 1985; Putirka, 2008; Mollo et al., 2010, 2012; Mollo and Hammer, 2017). The most striking consequence of the lack of equilibrium between zoned clinopyroxene phenocrysts and undercooled magmas is that barometers, thermometers, and hygrometers fail with errors up to ±550 MPa, ±95 °C, and ±3 wt.% for $H_2O$, respectively (Mollo et al., 2010, 2012; Hammer et al., 2016). For this reason, Mollo et al. (2013b) developed a new clinopyroxene-based model that tests for equilibrium through the early formalism of Putirka (1999): Eqn. DiHd with $R^2$ = 0.92 and SEE = ±0.06 (where R is the correlation coefficient and SEE is the standard error of estimate):

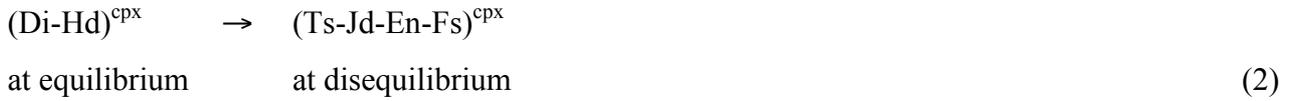

$$\ln(X_{DiHd}^{cpx}) = -2.18 - 3.16 X_{TiO_2}^{melt} - 0.365 \ln(X_{AlO_{1.5}}^{melt}) + 0.05 \ln(X_{MgO}^{melt})$$
$$- 3858.2 \frac{(X_{EnFs}^{cpx})^2}{T} + \frac{2107.4}{T} - 17.64 \frac{P}{T} \quad (1)$$

According to the original calibration, $P$ and $T$ are expressed in kbar and K, respectively. The validity of Eqn. DiHd has been tested with clinopyroxene-melt pairs either from experiments conducted on Etnean compositions or from naturally undercooled Etnean magmas during eruption to the surface (cf. Mollo et al., 2013a, 2013b). The equilibrium model is based on the difference between Di and Hd (ΔDiHd) components predicted for clinopyroxene via regression analysis of clinopyroxene-melt pairs in equilibrium conditions, and those measured in the analyzed natural crystals. From a theoretical point of view, the equilibrium condition is achieved when the value of ΔDiHd goes to zero, minimizing the error of estimate of clinopyroxene-based models (Mollo and Masotta, 2014). The value of ΔDiHd provides a more robust test for equilibrium than $^{cpx-melt}Kd_{Fe-Mg}$ due to the fact that deviation from equilibrium occurs in response to the following kinetically-controlled exchange:

(Di-Hd)$^{cpx}$ → (Ts-Jd-En-Fs)$^{cpx}$

at equilibrium    at disequilibrium    (2)

Concentration-dependent partitioning produces clinopyroxene growth layers that respond to chemical gradients in the melt, producing compositions that are progressively enriched in $^TAl$, Ti and Na and depleted in $Fe^{2+}$, Ca and Mg with increasing the degree of undercooling (Mollo et al., 2010, 2011, 2012). Therefore, Di and Hd components decrease, whereas the opposite occurs for Ts, Jd, En, and Fs, according to Eqn. (2). Clinopyroxene chemistry is controlled by the exchange of $^{M2}(Mg, Fe^{2+})$ with $^{M1}(Al, Fe^{3+})$ coupled with the substitution of Si with Al in the tetrahedral site to form the Tschermak components. Highly charged cations, such as titanium and trivalent iron, are also accommodated in the M1 site of clinopyroxene to balance the charge deficiency caused by the increasing concentration of aluminium (Scarlato et al., 2014). Since deviations from equilibrium account for significant aluminium enrichments, Eqn. DiHd allows to effectively identify clinopyroxene compositions approaching equilibrium partitioning (Mollo et al., 2013b, 2016, 2017). As documented by a number of authors (Lindstrom, 1976; Ray et al., 1983; Hart and Dunn, 1993; Forsythe et al., 1994; Lundstrom et al., 1994, 1998; Skulski et al., 1994; Blundy et al., 1998; Hill et al., 2000; Bedard, 2014), $D_{REE+Y}/D_{HFSE}$ increases systematically with increasing $^TAl$ due to charge imbalanced configurations caused by cation substitutions in both tetrahedral and octahedral sites of clinopyroxene. REE+Y replacing Ca in M2 are charge-balanced by Al replacing Si in T or Na replacing Ca in M2 (Gaetani and Grove, 1995; Blundy et al., 1998; Schosnig and Hoffer, 1998; Bennett et al., 2004; Marks et al., 2004; Francis and Minarik, 2008; Sun and Liang, 2012). This reflects an increased ease of locally balancing the excess charge at M2 as the number of surrounding $^T$Al atoms increases (Hill et al., 2000; Mollo et al., 2016, 2017). Conversely, the entry of HFSE onto M1 is more favored because the average charge on M1 increases with increasing Ts due to replacement of $Mg^{2+}$ by $Fe^{3+}$ and $Al^{3+}$ (Wood and Trigila, 2001; Marks et al., 2004). Since $D_{REE+Y}/D_{HFSE}$ are related to charge balance mechanisms primarily controlled by $^T$Al, the use of disequilibrium ($\Delta DiHd \gg 0$) clinopyroxenes as input data for the lattice strain model may be the source of important miscalculations (Blundy and Wood, 1994, 2003; Wood and Blundy, 1997, 2002; Hill et al., 2011). In the present study, clinopyroxene phenocrysts from Mt. Etna eruptions were selected in order to ensure values of $\Delta DiHd \leq 0.1$ (see Table 5), attesting to slow crystal growth conditions and (near-)equilibrium major and trace element partitioning (cf. Mollo et al., 2013b, 2017). For these selected natural compositions (i.e., 505 clinopyroxene-melt pairs from Table 5), the regression analysis of $lnD_{Ti}$ vs. $[(Di+Hd)-(Ts+Jd+En+Fs)]^{cpx}$ yields fairly acceptable regression statistics ($R^2 = 0.47$ and SEE = 0.15) that confirm the ability of Eqn. DiHd to prevent partition coefficient miscalculations. The effect of tetrahedrally-coordinated aluminium cations on the partitioning behavior of Ti (as representative of the HFSE group) is also demonstrated by the regression fit of $lnD_{Ti}$ vs. $^T$Al, providing regression statistics ($R^2 = 0.53$ and SEE = 0.14) comparable to those ($R^2 = 0.67$ and SEE = 0.33) reported in the review of Bedard (2014), wherein $D_{Ti}$ was parameterized using an experimental dataset of clinopyroxenes in equilibrium with primitive melts ($SiO_2$ = 42.5-44.99 wt.%).

## 4. Towards an integrated *P*-*T*-H$_2$O-lattice strain model

*4.1 Experimental constraints*

Experimental runs produced the typical phase assemblage of Etnean products; olivine, clinopyroxene, plagioclase, and titanomagnetite. The phase proportions and compositions for each run are listed in Tables 2S, 3S, and 4S. The most important experimental results can be summarized as follows:
- clinopyroxene and titanomagnetite are ubiquitous in all the runs;
- the primitive BA magma more favorably crystallizes olivine whose stability field expands at the expense of clinopyroxene and plagioclase with increasing H$_2$O and decreasing pressure (cf. Sisson and Grove, 1993; Metrich and Rutherford, 1998);

- the more evolved BT magma does not crystallize olivine and is preferentially saturated with plagioclase, especially at lower $H_2O$ contents and pressures (cf. Mollo et al., 2015b);
- the crystal content of decompression experiments (Run#3 and Run#4) is always lower than that of isobaric runs (Run#1 and Run#2) due to a delay in crystal nucleation, as typically observed for rapidly undercooled MAM compositions (Cashman et al., 1988, 1999; Simakin et al., 1999; 2003; Simakin and Salova, 2004; Baker et al., 2008; Del Gaudio et al., 2010; Mollo and Hammer, 2017);
- for a given run, the degree of crystallization decreases with increasing the amount of $H_2O$ added to the experimental charge;
- the overall compositions of olivine ($Fo_{76-82}$), clinopyroxene ($Di_{41-76}$), plagioclase ($An_{44-66}$), and titanomagnetite ($Usp_{12-47}$) match either those from previous crystallization experiments conducted on Etnean compositions (Métrich and Rutherford, 1998; Mollo et al., 2015b; Vetere et al., 2015; Perinelli et al., 2016), or those analyzed in naturally solidified historic and recent magmas (Armienti et al., 1988, 1994; Ferlito et al., 2012; Corsaro et al., 2013; Kahl et al., 2015);
- in terms of major and trace element concentrations, the experimental melt compositions are strictly controlled by the type and amount of crystals formed in each run.

The TAS (total alkali vs. silica) diagram (Fig. 3a) shows that, under both isobaric and decompression conditions, differentiation of the primitive BA magma reproduces the trachybasaltic compositions of post-1971 products (cf. Armienti et al., 2004; Corsaro et al., 2013; Mollo et al., 2015b), including also the TB magma erupted during 2001-2003. Similarly, the crystallization of the more evolved BT magma reconstructs most of the historic basaltic trachyandesites of the Recent Mongibello unit (Fig. 3a). Conversely, the evolutionary path derived for the TB magma shows compositions that, in some cases, deviate towards alkali-rich melts with tephritic affinity never observed at Mt. Etna volcano (Fig. 3a).

The MgO vs. CaO diagram (Fig. 3b) shows that, at 400 MPa and $H_2O$-undersaturated conditions (Run#1), the low degree (< 5%) of olivine crystallization shifts the BA magma towards post-1971 trachybasaltic compositions (MgO ≈ 5 wt.%; cf. Armienti et al., 2013; Corsaro et al., 2013; Mollo et al., 2015b). In contrast, enlargement of the olivine stability field at 50 MPa and $H_2O$-saturated conditions (Run#2) produces residual melts that do not match fully the chemistry of natural products (Fig. 3b). This occurs also when the BA magma is decompressed from 400 MPa and kept at 50 MPa for 24 h (Run#4). The effect of delayed nucleation during decompression is overprinted by further isobaric/isothermal crystallization over time (cf. Mollo et al., 2012) and, consequently, the formation of supplementary crystals drives the starting magma towards alkali-rich,

MgO-poor compositions (Fig. 3b). In contrast, when the BA magma is decompressed (Run#3) and then immediately quenched, the degree of crystallization is low, showing that dynamic crystallization conditions more closely reproduce the compositions of Etnean trachybasaltic magmas ascending towards the surface (Fig. 3b).

La vs. Yb (Fig. 4a) and Zr vs. Hf (Fig. 4b) diagrams reveal that BA, TB, and BT magmas equilibrated at 50 MPa for 24 h (Run#2 and Run#4) become dramatically enriched in trace elements with respect to natural concentration levels. This conforms to the observation that olivine and plagioclase contents in Run#2 and Run#4 are higher than those measured in Run#1 and Run#3 (Table 2S). Specifically, the ratio of [(olivine + plagioclase) / (clinopyroxene + titanomagnetite)] in BA, TB, and BT runs, increases from 0.21 to 0.48, from 0.46 to 1.58, and from 0.89 to 1.18, respectively. With respect to clinopyroxene and titanomagnetite, REE+Y/HFSE are weakly incorporated into olivine and plagioclase (Fig. 1), leading to striking La, Yb, Zr and Hf enrichments during magma crystallization (Fig. 4).

On one hand, mass balance, fractional crystallization, and thermodynamic calculations from previous studies (Metrich and Rutherford, 1998; Armienti et al., 2004; Corsaro et al., 2013; Vetere et al., 2013; Mollo et al., 2015b) demonstrate that post-1971 trachybasaltic magmas stored in shallow reservoirs at Mt. Etna volcano cannot be derived from primitive basaltic liquids when the degree of fractional crystallization is higher than 20-25%. On the other hand, thermometric, barometric, and hygrometric calculations (Armienti et al., 2013; Mollo et al., 2015a, 2015b; Perinelli et al., 2016) attest that most of the crystallization is driven by a dynamic process due to volatile exsolution at very shallow pressures ($P < 100$ MPa) during magma ascent in the conduit and emplacement to the surface. The system undergoes little crystal fractionation at depth, until most of the $H_2O$ rapidly exsolves during decompression (cf. Vetere et al., 2015). This causes strong magma undercooling, and the crystal growth rate increases by 3-4 orders of magnitude (Applegarth et al., 2013), facilitating the entrapment of abundant melt inclusions (Spilliaert et al., 2006; Giacomoni et al., 2014). Under such circumstances, the rapidly growing crystals are not fractionated from the host liquid but rather are transported towards the surface by the ascending magma. Coherently, the compositions of clinopyroxene phenocrysts from one single eruption at Mt. Etna do not record a constant $H_2O$-content in equilibrium at isothermal/isobaric conditions, but rather they describe a continuous $P$-$T$-$H_2O$ array (Armienti et al., 2013; Mollo et al., 2015a, 2015b; Perinelli et al., 2016), accounting for variable degrees of magma cooling, decompression, and degassing (see below).

*4.2 Thermometry, barometry and hygrometry*

Clinopyroxene is a common phenocryst phase in volcanic rocks and its compositional changes along the $P$-$T$-$H_2O$ differentiation path of magma provides a method for determining depths, temperatures, and volatile contents from mantle conditions to shallow crustal levels (Putirka et al., 2008). Generally, thermobarometric equations from literature are based on thermodynamic principles describing the relationship between the equilibrium constant of a specific clinopyroxene-melt exchange reaction and the intensive variables of the system. The most common exchange equilibria are: 1) Jd-melt where the formation of Jd is accompanied by a great change in molar volume, resulting in a $P$-dependent reaction (e.g. Blundy et al, 1995), and 2) Jd-DiHd and CaTs-DiHd where the volume changes for the solution of Jd and CaTs into DiHd are small, leading to $T$-dependent reactions (see Putirka et al., 1996 for further details). Thermodynamic description of these equations require quantification of the change in enthalpy and entropy as the $P$-$T$ conditions change in a thermodynamically reversible reaction. If these quantities are unknown for the investigated clinopyroxene and melt components, the thermobarometric equation is derived from regression analysis of equilibrium compositions obtained in laboratory at the $P$-$T$ conditions of interest (Putirka et al., 2008). In a recent review Perinelli et al. (2016) observed that ~50% of natural clinopyroxene phenocrysts from MAM exhibit Hd contents higher than 0.14, and such compositions are weakly represented in the calibration dataset used to derive clinopyroxene-based barometers, thermometers, and hygrometers (Putirka et al., 2008; Armienti et al., 2013). Notably, the precision of these regression models mostly depends on the number of experimental data included into the calibration dataset, and the compositional range of this dataset relative to the natural compositions studied (Putirka 2008; Masotta et al., 2013). Given the use of restricted compositional bounds, the refined predictive equations may potentially show much lower errors than the original models (cf. Masotta et al., 2013). Therefore, Perinelli et al. (2016) compiled a new calibration dataset specific to MAM (http://www.minsocam.org/msa/AmMin/TOC/2016/Dec2016_data/AM-16-125916.zip), comprising experimental clinopyroxene-melt pairs that more closely reproduce the crystallization conditions and compositions of natural products (i.e., $P$ = 27–800 MPa, $T$ = 1000–1175 °C, $H_2O$ = 1-5 wt.%, and $fO_2$ = QFM-NNO+2). Through regression analyses of these compositions, it has been derived the following $P$-$T$-dependent hygrometer:

- Eqn. $H_{MAM}$ of Perinelli et al. (2016) with $R^2$ = 0.79 and SEE= ±0.45 wt.%,

$$H_2O = 39.6 DiHd + 29.48 EnFs + 41.76 CaTs + 39.58 Jd + 0.44 CaTi + 0.14 \ln P - 0.01 T - 27.3 \tag{3}$$

Where *P*, *T*, and H$_2$O are expressed in MPa, °C and, wt.%, respectively, as reported in Perinelli et al. (2016). Following this approach, previous barometric and thermometric equations from literature (Putirka et al., 1996, 2003; Putirka, 2008; Neave and Putirka, 2017) have been tested and refined by extending the calibration dataset for MAM by Perinelli et al. (2016) to the clinopyroxene-melt pairs from this study (Table 3S). To prevent potential discrepancies, all the literature equations are here treated using their original notations (i.e., mole fractions, cation fractions, and clinopyroxene components), units (i.e, kbar and K for *P* and *T*, respectively), and equation labels (i.e., Eqn. P1, T1, etc.).

The literature barometric equations and their original regression statistics are:

- Eqn. P1 of Putirka et al. (1996) with $R^2 = 0.98$ and SEE = ±1.36 kbar,

$$P = -54.3 + 299\frac{T}{10^4} + 36.4\frac{T}{10^4}\ln\left[\frac{X^{cpx}_{NaAlSi_2O_6}}{X^{melt}_{NaO_{0.5}} X^{melt}_{AlO_{1.5}} \left(X^{melt}_{SiO_2}\right)^2}\right] \quad (4)$$
$$+ 367\ln(X^{melt}_{NaO_{0.5}} X^{melt}_{AlO_{1.5}})$$

- Eqn. A of Putirka et al. (2003) with $R^2 = 0.99$ and SEE = ±1.3 kbar,

$$P = -88.3 + 28.2\frac{T}{10^4}\ln\left[\frac{X^{cpx}_{NaAlSi_2O_6}}{X^{melt}_{NaO_{0.5}} X^{melt}_{AlO_{1.5}} \left(X^{melt}_{SiO_2}\right)^2}\right] \quad (5)$$
$$+ 219\frac{T}{10^4} - 25.1\ln(X^{melt}_{CaO} X^{melt}_{SiO_2}) + 7.03(Mg\#^{melt}) + 12.4\ln(X^{melt}_{CaO})$$

- Eqn. 30 of Putirka (2008) with $R^2 = 0.97$ and SEE = ±1.6 kbar,

$$P = -48.7 + 271\frac{T}{10^4} + 32\frac{T}{10^4}\ln\left[\frac{X^{cpx}_{NaAlSi_2O_6}}{X^{melt}_{NaO_{0.5}} X^{melt}_{AlO_{1.5}} \left(X^{melt}_{SiO_2}\right)^2}\right]$$
$$- 8.2\ln(X^{melt}_{FeO}) + 4.6\ln(X^{melt}_{MgO}) - 0.96\ln(X^{melt}_{KO_{0.5}}) \quad (6)$$
$$- 2.2\ln(X^{cpx}_{DiHd}) - 31(Mg\#^{melt}) + 56(X^{melt}_{NaO_{0.5}} + X^{melt}_{KO_{0.5}}) + 0.76(H_2O^{melt})$$

- Eqn. 31 of Putirka (2008) with $R^2 = 0.89$ and SEE = ±2.9 kbar,

$$P = -40.73 + 358\frac{T}{10^4} + 21.69\frac{T}{10^4}\ln\left[\frac{X^{cpx}_{NaAlSi_2O_6}}{X^{melt}_{NaO_{0.5}} X^{melt}_{AlO_{1.5}} \left(X^{melt}_{SiO_2}\right)^2}\right]$$
$$- 105.7(X^{melt}_{CaO}) - 165.5(X^{melt}_{NaO_{0.5}} + X^{melt}_{KO_{0.5}})^2 - 50.15(X^{melt}_{SiO_2})(X^{melt}_{FeO} + X^{melt}_{MgO}) \quad (7)$$
$$- 3.178\ln(X^{cpx}_{DiHd}) - 2.205\ln(X^{cpx}_{EnFs}) + 0.864\ln(X^{cpx}_{TOT\,Al}) + 0.3962(H_2O^{melt})$$

- Eqn. 32a of Putirka (2008) with $R^2 = 0.92$ and SEE = ±2.2 kbar,

$$P = 3205 + 0.384T - 518\ln T - 5.62(X_{Mg}^{cpx}) + 83.2(X_{Na}^{cpx}) \\ + 68.2(X_{DiHd}^{cpx}) + 2.52\ln(X_{M1\_Al}^{cpx}) - 51.1(X_{DiHd}^{cpx})^2 + 34.8(X_{EnFs}^{cpx})^2 \qquad (8)$$

- Eqn. 32b of Putirka (2008) with $R^2 = 0.91$ and SEE = ±2 kbar,

$$P = 1458 + 0.197T - 241\ln T + 0.453(H_2O^{melt}) + 55.5(X_{M1\_Al}^{cpx}) + 8.05(X_{Fe}^{cpx}) \\ - 277(X_K^{cpx}) + 18(X_{Jd}^{cpx}) + 44.1(X_{DiHd}^{cpx}) + 2.2\ln(X_{Jd}^{cpx}) - 17.7(X_{Al}^{cpx})^2 \\ + 97.3(X_{M2\_Fe}^{cpx})^2 + 30.7(X_{M2\_Mg}^{cpx})^2 - 27.6(X_{DiHd}^{cpx})^2 \qquad (9)$$

- Eqn. 1 of Neave and Putirka (2017) with $R^2 = 0.94$ and SEE = ±1.4 kbar,

$$P = -26.27 + 39.16\frac{T}{10^4}\ln\left[\frac{X_{NaAlSi_2O_6}^{cpx}}{X_{NaO_{0.5}}^{melt} X_{AlO_{1.5}}^{melt} (X_{SiO_2}^{melt})^2}\right] \\ - 4.22\ln(X_{DiHd}^{cpx}) + 78.43\ln(X_{AlO_{1.5}}^{melt}) + 393.81(X_{NaO_{0.5}}^{melt} X_{KO_{0.5}}^{melt})^2 \qquad (10)$$

When the dataset based on MAM compositions is used to test these equations, regression analysis of the data indicates lower accuracy than those originally derived (i.e., SEE increases up to ±4.06 kbar; Fig. 5), showing that mafic alkaline products are not adequately represented in the calibration dataset (cf. Perinelli et al., 2016). Among all the tested equations, the best predictive barometer is the $T$-dependent, H$_2$O-independent Eqn. A of Putirka et al. (2003), yielding $R^2 = 0.57$ and SEE = ±2.1 kbar (Fig. 5). This model has been therefore recalibrated to derive a new equation specific to MAM compositions:

- Eqn. A$_{MAM}$ with $R^2 = 0.66$ and SEE = ±1.5 kbar,

$$P = -88.08 + 3.25\frac{T}{10^4}\ln\left[\frac{X_{NaAlSi_2O_6}^{cpx}}{X_{NaO_{0.5}}^{melt} X_{AlO_{1.5}}^{melt} (X_{SiO_2}^{melt})^2}\right] \\ + 0.04T - 17.17\ln(X_{CaO}^{melt} X_{SiO_2}^{melt}) + 0.46(Mg\#^{melt}) + 7.75\ln(X_{CaO}^{melt}) \qquad (11)$$

The accuracy of Eqn. A$_{MAM}$ has been verified following the procedure reported in Putirka (2008): 1) ~20% of clinopyroxene–melt pairs were randomly subtracted from the Etnean dataset before the regression analysis, 2) ~80% of the residual clinopyroxene–melt pairs were regressed to refine the model ($R^2 = 0.61$ and SEE = ±1.71 kbar; Fig. 5), and 3) the subtracted clinopyroxene–melt pairs were used as test data for the model (Fig. 5), providing regression statistics ($R^2 = 0.73$ and SEE = ±1.16 kbar) comparable to those obtained through regression fit of the whole dataset ($R^2 = 0.66$ and SEE = ±1.5 kbar).

The literature thermometric equations and their original regression statistics are:

- Eqn. T1 of Putirka et al. (1996) with $R^2 = 0.98$ and SEE = ±27 °C,

$$\frac{10^4}{T} = 6.73 - 0.26\ln\left(\frac{X_{Jd}^{cpx} X_{CaO}^{melt} X_{Fm}^{melt}}{X_{DiHd}^{cpx} X_{Na}^{melt} X_{Al}^{melt}}\right) - 0.86\ln(Mg\#^{melt}) + 0.52\ln(X_{CaO}^{melt}) \tag{12}$$

- Eqn. B of Putirka et al. (2003) with $R^2 = 0.97$ and SEE = ±34 °C,

$$\frac{10^4}{T} = 4.6 - 0.437\ln\left(\frac{X_{Jd}^{cpx} X_{CaO}^{melt} X_{Fm}^{melt}}{X_{DiHd}^{cpx} X_{Na}^{melt} X_{Al}^{melt}}\right) - 0.654\ln(Mg\#^{melt}) - 0.326\ln(X_{NaO_{0.5}}^{melt})$$
$$- 63.2\frac{P}{10^4} - 0.92\ln(X_{SiO_2}^{melt}) + 27.4\ln(X_{Jd}^{cpx}) \tag{13}$$

- Eqn. 32d of Putirka (2008) (based on the clinopyroxene-only thermometer of Nimis and Taylor, 2000) with $R^2 = 0.82$ and SEE = ±58 °C,

$$T = \frac{93100 + 544P}{61.1 + 36.6\left(X_{Ti}^{cpx}\right) + 10.9\left(X_{Fe}^{cpx}\right) - 0.95\left(X_{Al}^{cpx} + X_{Cr}^{cpx} - X_{Na}^{cpx} - X_{K}^{cpx}\right) + 0.395\left[\ln\left(a_{En}^{cpx}\right)\right]^2} \tag{14}$$

- Eqn. 33 of Putirka (2008) with $R^2 = 0.81$ and SEE = ±32 °C,

$$\frac{10^4}{T} = 7.53 - 0.14\ln\left(\frac{X_{Jd}^{cpx} X_{CaO}^{melt} X_{Fm}^{melt}}{X_{DiHd}^{cpx} X_{Na}^{melt} X_{Al}^{melt}}\right) + 0.07\left(H_2O^{melt}\right) - 14.9\left(X_{CaO}^{melt} X_{SiO_2}^{melt}\right)$$
$$- 0.08\ln\left(X_{TiO_2}^{melt}\right) - 3.62\left(X_{NaO_{0.5}}^{melt} + X_{KO_{0.5}}^{melt}\right) - 1.1\left(Mg\#^{melt}\right) - 0.18\ln\left(X_{EnFs}^{cpx}\right) - 0.027P \tag{15}$$

- Eqn. 34 of Putirka (2008) with $R^2 = 0.81$ and SEE = ±40 °C,

$$\frac{10^4}{T} = 6.39 + 0.076\left(H_2O^{melt}\right) - 5.55\left(X_{CaO}^{melt} X_{SiO_2}^{melt}\right) - 0.386\ln\left(X_{MgO}^{melt}\right)$$
$$- 0.046P + 2.2\frac{P^2}{10^4} \tag{16}$$

As for the case of barometers, MAM compositions are poorly reproduced by the thermometric equations, with SEE values up to ±86 °C (Fig. 6). The best predictive models are the $P$-$H_2O$-dependent Eqn. 33 and Eqn. 34, with SEE = ±39 °C and SEE = ±31 °C respectively (Fig. 6).

However, half of the regression terms of Eqn. 34 includes $P$ and $H_2O$, making the regression fit highly dependent on these parameters. Additionally, Eqn. 34 is independent of clinopyroxene chemistry, meaning that it predicts the temperature at which the melt becomes saturated with clinopyroxene, rather than the thermal path recorded by the growing phenocryst (cf. Putirka, 2008). In contrast, Eqn. 33 incorporates $P$, $H_2O$, and both clinopyroxene and melt compositions. Regression analysis of the data indicates that the omission of $P$ and $H_2O$ as regression terms decreases the predictive ability of Eqn. 33 by only ±2 °C. This effect is counterbalanced by the addition of $X_{TiO2}^{cpx}$ as a new predictor, with improvement for the regression fit of ±2 °C. Therefore, the following new equations can be derived:

- Eqn. 33$_{MAM}$ with $R^2$ = 0.55 and SEE = ±28 °C,

$$\frac{10^4}{T} = 5.63 - 0.15\ln\left(\frac{X_{Jd}^{cpx} X_{CaO}^{melt} X_{Fm}^{melt}}{X_{DiHd}^{cpx} X_{Na}^{melt} X_{Al}^{melt}}\right) + 4.78\left(X_{CaO}^{melt} X_{SiO_2}^{melt}\right) - 0.21\ln\left(X_{TiO_2}^{melt}\right)$$
$$- 0.002\left(X_{NaO_{0.5}}^{melt} + X_{KO_{0.5}}^{melt}\right) - 0.52\left(Mg\#^{melt}\right) + 0.07\ln\left(X_{EnFs}^{cpx}\right) - 0.16\ln\left(X_{TiO_2}^{cpx}\right)$$

(17)

- Eqn. 34$_{MAM}$ with $R^2$ = 0.78 and SEE = ±20 °C,

$$\frac{10^4}{T} = 7.4 + 0.08\left(H_2O^{melt}\right) - 7.9\left(X_{CaO}^{melt} X_{SiO_2}^{melt}\right) - 0.09\ln\left(X_{MgO}^{melt}\right)$$
$$- 0.007P - 65.08\frac{P^2}{10^4}$$

(18)

The accuracy of the $P$-$H_2O$-independent Eqn. 33$_{MAM}$ and $P$-$H_2O$-dependent Eqn. 34$_{MAM}$ has been verified following the procedure reported in Putirka (2008). Regression statistics derived for the calibration dataset are in good agreement with those retrieved for the test dataset (Fig. 6), ensuring that no systematic overestimates or underestimates result from miscalibration (c.f. Masotta et al., 2013).

An immediate outcome from these new regressions is that the $T$-dependent Eqn. A$_{MAM}$ (error ±1.5 kbar) can be combined with the $P$-$H_2O$-independent Eqn. 33$_{MAM}$ (error ±28 °C) to determine the pressure and temperature of the system. Once the $P$-$T$ conditions are known, the amount of $H_2O$ dissolved in the melt during clinopyroxene formation can be retrieved through the $P$-$T$-dependent Eqn. H$_{MAM}$ (error ±0.45 wt.%). At the same time, the achievement of (near-) equilibrium crystallization conditions between natural clinopyroxene phenocrysts and host magmas can be assessed by Eqn. DiHd (error ±0.06). All these models are reported in the Excel spreadsheet submitted online as supplementary material.

*4.3. Lattice strain theory*

It has long been known that plots of partition coefficient versus ionic radius for suites of isovalent cations exhibit a parabolic trend in response to the lattice strain in the crystal structure (Nagasawa, 1966; Onuma et al., 1968). Later, Blundy and Wood (1994) provided a quantitative expression relating the variation in D with ionic radius to the physical properties of the host crystal through the model of Brice (1975) based on the strain energy of cation substitution. It has been found that the partitioning of a substituent cation of radius $r_i$ onto a structural site is related to the optimum site radius $r_0$, the elastic response of that site $E$ (GPa) as measured by its Young's Modulus, and the strain-free partition coefficient $D_0$ for a fictive cation with radius $r_0$:

$$D_i = D_0 \exp\left(\frac{-4\pi E N_A \left(\frac{r_0}{2}(r_i - r_0)^2 + \frac{1}{3}(r_i - r_0)^3\right)}{RT}\right) \quad (19)$$

where $N_A$ is the Avogadro's number (6.022 ×10$^{23}$ mol$^{-1}$), $R$ is the universal gas constant (0.0083145 kJ mol$^{-1}$), and $T$ (K) is temperature.

In order to develop a predictive model for the partitioning of REE+Y between clinopyroxene and silicate melt, Wood and Blundy (1997) used thermodynamic descriptions to estimate the free energy of fusion (ΔG$_f$) of a hypothetical (REE+Y)MgAlSiO$_6$ component. Propagating ΔG$_f$ into the model of Brice (1975), they derived an expression for $D_0^{+3}$ in terms of the atomic fraction of Mg on the clinopyroxene M1 site ($X_{Mg}^{M1}$), the *Mg#* of the melt [$Mg\#^{melt} = X_{Mg}^{melt}/(X_{Fe}^{melt} + X_{Mg}^{melt})$], $P$ (GPa), and $T$ (K):

$$RT \ln\left(\frac{D_0^{+3} X_{Mg}^{M1}}{Mg\#^{melt}}\right) - 7{,}050P + 770P^2 = 88{,}750 - 65.644T \quad (20)$$

Where $X_{Mg}^{M1}$ and $Mg\#^{melt}$ are expressions of the activity-composition relations of clinopyroxene ($a_{REE+YMgAlSiO6}^{cpx} = X_{REE+Y}^{M2} X_{Mg}^{M1}$) and melt ($a_{REE+YMgAlSiO6}^{melt} = X_{REE+Y}^{melt} Mg\#^{melt}$) respectively. $D_0^{+3}$ is the ratio of $X_{REE+Y}^{M2}$ to $X_{REE+Y}^{melt}$ assuming that crystal chemistry influences both the dimensions of the M2 site, into which REE+Y partition, and the molar fraction of the hypothetical REE+Y end-member, such as (REE+Y)MgAlSiO$_6$. By fitting experimental data to the temperature and pressure derivatives of the bulk and shear moduli of diopside, Wood and Blundy (1997) derived an equation for calculating the Young's Modulus for 3+ trace cations in the M2 site of clinopyroxene:

$$E_{M2}^{+3} = 318.6 + 6.9P - 0.036T \tag{21}$$

The change of $r_{0,M2}^{+3}$ was derived through linear regression of optimum M2 site radii to show the site radius is related to the Al content of M1, and the Ca content M2:

$$r_{0,M2}^{+3} = 0.974 + 0.067 X_{Ca}^{M2} - 0.051 X_{Al}^{M1} \tag{22}$$

Once $D_0^{+3}$, $E_{M2}^{+3}$, and $r_{0,M2}^{+3}$ are integrated into a magma crystallization model, it is possible to track the evolutionary path of $D_{REE+Y}$ during differentiation, provided that the crystallization pressures and temperatures are known for each clinopyroxene phenocryst in equilibrium with the host melt. However, due to assumed complete short-range order between REE+Y on M2 site and Al in T site in the (REE+Y)MgAlSiO$_6$ molecule, $D_0^{+3}$ and $D_{REE+Y}$ are taken to be independent of the $^T$Al content in clinopyroxene (Wood and Blundy, 1997). This is an apparent oversimplification of the model that does not consider charge-balance mechanisms more complex than simple short-range order (Wood and Blundy, 2001). $D_0^{+3}$ increases due to the increasing probability of cations entering a locally charge-balanced site as $^T$Al increases, causing that the height of the partitioning parabola to shift to higher $D_{REE+Y}$ values (Hill et al., 2000; Wood and Blundy, 2001; Wood and Trigila, 2001; Tuff and Gibson, 2007, Sun and Liang, 2012; Mollo et al., 2013b, 2017; Scarlato et al., 2014). Thus, the electrostatic work done on transferring one mole of ions from silicate melt to the clinopyroxene M2 site (i.e., $\Delta G_{elec}^{cpx}$) must also be considered. Through this approach, the predictive model for $D_{REE+Y}$ accounts for REE+Y substitution onto any M2 site by including the electrostatic work involved in dissipating any local charge imbalance. The concentrations of REE+Y are computed for trace elements that substitute into sites of the "correct" -3 charge (cf. Wood and Blundy, 2001), assuming local charge balance between REE+Y in the M2 site and adjacent $^T$Al within clinopyroxene (i.e., *[REE+Y]$^{cpx}$/[X]$^{3+}$*, where *[X]$^{3+}$* is the proportion of M2 sites charge-balanced by a cation of +3 charge). Trace cations that substitute into M2 sites that do not have the correct charge incur electrostatic work in proportion to the square of the charge mismatch and the fraction of sites with 0, -1, -2, and -4 charges (i.e., *[X]$^{0+}$, [X]$^{+1}$, [X]$^{+2}$, [X]$^{+4}$*). In practice, the original expression of Wood and Blundy (1997) for a hypothetical partition coefficient of REE+Y between clinopyroxene and melt can be reformulated considering the crystal electrostatic effects on partitioning as follows:

$$D_{REE+Y} = \left(\frac{D_0^{+3} X_{Mg}^{M1}}{Mg\#^{melt}}\right)\left\{\begin{array}{l}[X]^{3+} + ([X]^{2+} + [X]^{4+})e^{\left(\frac{-\Delta G_{elec}^{cpx}}{RT}\right)} \\ + [X]^{+}e^{\left(\frac{-4\Delta G_{elec}^{cpx}}{RT}\right)} + [X]^{0+}e^{\left(\frac{-9\Delta G_{elec}^{cpx}}{RT}\right)}\end{array}\right\} \quad (23)$$

As derived by Wood and Blundy (2001), the electrostatic energy of substitution is $\Delta G_{elec}^{cpx}$ = 28 kJ. In Eqn. (23) the total concentration of REE+Y in the clinopyroxene involves a term for each different structural site multiplied by the relative probability for a cation to occupy that specific site. The probability that a charge-balanced/imbalanced configuration occurs in structural sites with charges 0, -1, -2, -3 and -4 is expressed as:

$$[X]^{0+} = X_{Ti}^{M1}(X_{Si}^{T})^2, \quad (24)$$

$$[X]^{+} = X_{Al}^{M1}(X_{Si}^{T})^2 + 2X_{Ti}^{M1}(X_{Al}^{T})(X_{Si}^{T}), \quad (25)$$

$$[X]^{2+} = (1 - X_{Al}^{M1} - X_{Cr}^{M1} - X_{Fe^{3+}}^{M1} - X_{Ti}^{M1})(X_{Si}^{T})(X_{Si}^{T}) \\ + 2(X_{Al}^{M1} + X_{Cr}^{M1} + X_{Fe^{3+}}^{M1})(X_{Al}^{T})(X_{Si}^{T}) + X_{Ti}^{M1}(X_{Al}^{T})^2, \quad (26)$$

$$[X]^{3+} = (1 - X_{Al}^{M1} - X_{Cr}^{M1} - X_{Fe^{3+}}^{M1} - X_{Ti}^{M1})(X_{Al}^{T})(X_{Si}^{T}) \\ + (X_{Al}^{M1} + X_{Cr}^{M1} + X_{Fe^{3+}}^{M1})(X_{Al}^{T})^2, \text{ and} \quad (27)$$

$$[X]^{4+} = (1 - X_{Al}^{M1} - X_{Cr}^{M1} - X_{Fe^{3+}}^{M1})(X_{Al}^{T})^2. \quad (28)$$

For simplicity, Wood & Blundy (1997) assumed that the activity of REE+Y in the melt is equal to its (molar) concentration. However, the magnitude of the partition coefficient may correlate with the degree of melt polymerization due to a melt structural influence on trace element partitioning (Blundy et al., 1996; Gaetani et al., 2003; Gaetani, 2004; Schmidt et al., 2006; Qian et al., 2015). Although $D_0^{+3}$ tends to increase with decreasing the value of the number of non-bridging oxygens per tetrahedral cations (i.e., *NBO/T*), it is also true that the number of large structural sites critically important to accommodate REE+Y increases with increasing the concentration of CaO in the melt as a network-modifier cation (Huang et al., 2006). Due to similarity in ionic radius and charge, REE+Y preferentially substitute for $Ca^{2+}$ in the melt, as described by following exchange reaction:

$$[REE+Y_{2/3}MgSi_2O_6]^{cpx} + [Ca-NBO]^{melt} \leftrightarrow [CaMgS_2O_6]^{cpx} + [REE+Y_{2/3}-NBO]^{melt} \qquad (29)$$

In consideration of this, $D_0^{+3}$ is better correlated with the ratio of molar $[Ca^{2+}/(M^+ + M^{2+})]^{melt}$, where $M^+$ and $M^{2+}$ are, respectively, $Na^+$ and $K^+$, and $Fe^{2+}$, $Ca^{2+}$ and $Mg^{2+}$ of the melt given as percentages (Huang et al., 2006). This melt structure parameter is incorporated into the lattice strain equation of Wood and Blundy (1997), so that the simple concentration of REE+Y in the melt (i.e., $X_{REE+Y}^{melt}$) is now described as $[REE+Y]^{melt}/[Ca^{2+}/(M^+ + M^{2+})]^{melt}$. In alkaline systems, when the effects of $T$ and $^TAl$ in clinopyroxene are constrained in a narrow range, it has been found that $[Ca^{2+}/(M^+ + M^{2+})]^{melt}$ is effectively correlated with $D_{REE+Y}$ (Mollo et al., 2016).

It is well known that the liquidus temperature of the melt is not a linear function of $H_2O$ concentration (Burnham, 1975). Intuitively, the value of $D_0^{+3}$ decreases with decreasing the liquidus temperature as the amount of $H_2O$ dissolved in the melt increases. However, if $H_2O$ is added at fixed temperature, its principal effect is to lower the activity coefficient ($\gamma$) of REE+Y clinopyroxene component (REE+YMgAlSiO_6) in the melt, thus decreasing the value of $D_0^{+3}$. If the effect of $H_2O$ on $\gamma_{REE+YMgAlSiO_6}^{melt}$ component is ignored, the prediction of $D_{REE+Y}$ can be overestimated up to one order of magnitude (e.g., Sun and Liang, 2012). In this framework, Wood and Blundy (2002) developed a thermodynamic method for separating the effects of $H_2O$ from those of $T$, assuming that the influence exercised by $H_2O$ on the minor (REE+Y)MgAlSiO_6 component is virtually identical that exercised on the major $CaMgSi_2O_6$ component in the melt. Through the melting temperatures of diopside on the join $CaMgSi_2O_6$-$H_2O$ (cf. Eggler and Rosenhauer, 1978), Wood & Blundy (2002) found that:

$$\gamma_{CaMgSi_2O_6}^{melt} = -0.3208 + \frac{0.0563}{1 - X_{H_2O}^{melt}} + 1.8452(1 - X_{H_2O}^{melt}) - 0.5807(1 - X_{H_2O}^{melt})^2 \qquad (30)$$

If $\gamma^{melt}_{REE+YMgAlSiO_6} = \gamma^{melt}_{CaMgSi_2O_6}$, the values of $D_{REE+Y}$ for clinopyroxenes coexisting with hydrous melts can be reappraised through the following correction equation (Wood & Blundy, 2002):

$$D^{+3}_{0,hydrous} = D^{+3}_0 \gamma^{melt}_{REE+YMgAlSiO_6} (1 - X^{melt}_{H_2O}) \frac{100}{(100 - H_2O^{melt})} \qquad (31)$$

$X^{melt}_{H_2O}$ and $H_2O^{melt}$ are the mole fraction and the weight percent of H$_2$O dissolved in the melt, respectively. $D^{+3}_{0,hydrous}$ refers to the strain-free partition coefficient when the melt is hydrous, whereas $D^{+3}_0$ is the value predicted through the model Wood and Blundy (1997) that, in the present study, includes also the electrostatic work done in substituting the REE+Y cations into each configuration and the melt structure parameter.

Modeling the partitioning behavior of HFSE is also important for a better understanding of magma dynamics, especially given fact that Ti, Zr, and Hf are relatively immobile in hydrous fluids, and the incompatible character of Zr and Hf (as elements with similar ionic radii) provides indications for magma differentiation processes and source compositions. In this regard, Hill et al. (2011) parameterized the value of Young Modulus for tetravalent cations in the small M1 site of clinopyroxene to obtain the following best-fit equation:

$$E^{+4}_{M1} = 10{,}473 - 5.09T - 201.54P + 14{,}633 X^T_{Al} \qquad (32)$$

Reformulating the clinopyroxene composition in terms of Di-Jd-CaTs solid solutions, Hill et al. (2011) provided also a regression function for the radius of the tetravalent lattice site:

$$r^{+4}_{0,M1} = 0.664 - 0.008P + 0.071 X^{M1}_{Al} \qquad (33)$$

Accounting for the combination of $P$, $T$, the electrostatic work done by placing a tetravalent cation in a charge-balanced/imbalanced crystal site, and the electrostatic effect produced by insertion of a HFSE cation in the melt, Hill et al. (2011) derived the following predictive model for $D_{Ti}$:

$$-RT \ln \frac{D_{Ti}}{\left\{ \begin{array}{l} [X]^{4+} + [X]^{3+} e^{\left(\frac{-\Delta G_{elec}}{RT}\right)} \\ + [X]^{2+} e^{\left(\frac{-4\Delta G_{elec}}{RT}\right)} \end{array} \right\}} = -37{,}800 + 5{,}375P + 660P^2 - 22{,}800 \Delta Z^2_{melt} \qquad (34)$$

$\Delta G_{elec}^{cpx}$ = 16.5 kJ takes in to account the square of the charge difference between the substituent ion and the ideal charge of the site. The effective ionic charge of the melt is $\Delta Z_{melt}^2 = (Z_{Ti} - Z_0)^2$ where $Z_{Ti}$ is the charge of titanium cations placed in the melt with average charge $Z_0$. By excluding the contributions of $Al^{3+}$ and $Si^{4+}$, $Z_0$ is described as [$\Sigma( X_{melt}^{cation}$ × charge) / $\Sigma$charges]. $X_{1+}^{M2}$ refers to the atomic fraction of cations with +1 charges entering the M2 site and the proportions of M1 sites charge-balanced/imbalanced by tetravalent cations are expressed as:

$$[X]^{2+} = (1 - X_{1+}^{M2})(X_{Si}^T)^2, \tag{35}$$

$$[X]^{3+} = \left(X_{1+}^{M2}(X_{Si}^T)^2 + 2(1 - X_{1+}^{M2})X_{Al}^{M1}(X_{Si}^T)(X_{Al}^T)\right), \text{ and} \tag{36}$$

$$[X]^{4+} = (1 - X_{1+}^{M2})(X_{Al}^T)^2 + 2X_{1+}^{M2}(X_{Al}^T)(X_{Si}^T). \tag{37}$$

Since the early work of Forsythe et al. (1994), it is known that the overall behavior of $D_{HFSE}$ can be approximated to that of $D_{Ti}$ over a wide range of $T$, $P$, and bulk composition. This has been recently confirmed by Bedard (2014) through global regression analyses of a broad $D_{HFSE}$ dataset from literature that comprises a wide range of natural and experimental crystallization conditions for clinopyroxene. Similarly, Blundy and Wood (2003) argued that $D_{Hf}$ and $D_{Zr}$ are well correlated with $D_{Ti}$, and that partition coefficients of ''near-neighbor'' cations show good intercorrelations. On this basis, $D_{Ti}$ can serve as a proxy for $D_{Hf}$ and $D_{Zr}$ through the simple application of lattice strain theory:

$$\frac{D_{Ti}}{D_{Hf/Zr}} = \exp\left(\frac{-4\pi E_{M1}^{4+} N_A \left(\frac{r_{0,M1}^{+4}}{2}(r_{Hf/Zr}^2 - r_{Ti}^2) + \frac{1}{3}(r_{Hf/Zr}^2 - r_{Ti}^2)^3\right)}{RT}\right) \tag{38}$$

where $r_{Ti}$, $r_{Hf}$, and $r_{Zr}$ are 0.605 Å, 0.71 Å, and 0.72 Å, respectively, as reported in Shannon (1976).

Due to the paucity of data from literature, the lattice strain model for HFSE does not consider the role played by $H_2O$ on trace element partitioning between clinopyroxene and melt. To investigate this effect, some values of $D_{HFSE}$ (see Table 6S; data from Skulsi et al., 1994; Hauri et al., 1994; Salters and Longhi, 1999; McDade et al., 2003a, 2003b) have been selected for basaltic melts (MgO = 7.74-12.1) coexisting with clinopyroxenes (MgO = 16.86-18.86) equilibrated at variable $P$ (1-1.7 GPa), $T$ (1245-1502 °C), $H_2O$ (0-4.8 wt.%), and $fO_2$ (QFM-2 - NNO+1). Regression of the data (Fig. 7) shows that $H_2O$ does not correlate with $D_{Ti}$ ($R^2_{Ti}$ = 0.09), whereas moderate correlations are found with $D_{Hf}$ ($R^2_{Hf}$ = 0.72) and $D_{Zr}$ ($R^2_{Zr}$ = 0.60). $D_{Hf}$ and $D_{Zr}$ decrease by 68% and 67%, respectively, when the $H_2O$ content increases from 0 to ~5 wt.%. However, this effect is statistically significant only for $H_2O$ > 4 wt.%. Therefore, below this water threshold, it can is assumed that the lattice strain model for HFSE predicts reliable partition coefficients.

In accordance with thermodynamic principles, 3+ and 4+ trace cation partition coefficients are positively correlated with $^TAl$ in clinopyroxene and negatively correlated with $T$ and $H_2O$ dissolved in the melt (e.g., Sun and Liang, 2012). In this context, the validity of the lattice strain models for REE+Y/HFSE has been tested using as input data MELTS (Ghiorso and Sack, 1995) thermodynamic simulations conducted on TB magma at 400 MPa, 970-1,120 °C, 0-3 wt.% $H_2O$, and NNO buffer (Table 7). The aim of these simulations was to recover equilibrium clinopyroxene-melt pairs that minimize and/or isolate the combined effects of $^TAl$, $T$, and $H_2O$ on trace element partitioning. For example, at constant $T$ = 1,120 °C, the height of the partitioning parabola for REE+Y decreases when the system shifts from anhydrous to hydrous conditions, but the change of $^TAl$ is restricted to only ~2% (Fig. 8a). This suggests that the lattice strain model accounts for the concentration of $H_2O$ dissolved in the melt and its effect on the activity of (REE+Y)MgAlSiO$_6$ (Wood and Blundy, 2002). Under anhydrous conditions, $D_{REE+Y}$ decreases with increasing $T$ and decreasing $^TAl$ (Fig. 8b), in agreement with the influence of these parameters on $D_0^{+3}$ (Wood and Blundy, 1997; Hill et al., 2010). For the HFSE, at constant $T$ = 1,090 °C, $^TAl$ decreases by ~28% causing the partitioning parabola to move downwards (Fig. 8c). The entry of HFSE into the M1 octahedral site is facilitated as the average charge on this site increases with CaTs and CaFeTs substitutions (Wood and Trigila, 2001). The strong control of $^TAl$ on HFSE incorporation overwhelms the effect of $H_2O$ dissolved in the melt (Fig. 8c). This finding corroborates the experimental trends of Fig. 7, showing that $H_2O$ contents below 4 wt.% do not influence the partitioning behaviour of HFSE. According to anhydrous data from Lundstrom et al. (1994) and Blundy et al. (1995), it is also confirmed that $D_{HFSE}$ decreases with increasing $T$ and decreasing $^TAl$ (Fig. 8d).

The accuracy of the lattice strain model for REE+Y/HFSE has been verified for MAM compositions using values of $D_{REE+Y}/D_{HFSE}$ measured for BA, TB, and BT from Run#1 (Table 8S). At 400 MPa, the system is $H_2O$-undersaturated, so that the $H_2O$ content in equilibrium with the bubble-free melt can be adequately determined by mass balance calculations during formation of nominally anhydrous minerals (see for further details the studies of Masotta et al., 2013, Ushioda et al., 2014, Mollo et al., 2015c, and Perinelli et al., 2016). $D_{REE+Y}/D_{HFSE}$ values for Run#1 have been compared with those predicted by the lattice strain model. The regression fit of the data for $D_{La}$, $D_{Yb}$, $D_{Hf}$ and $D_{Zr}$ (Fig. 9) yields relatively high correlation coefficients ($R^2_{La}$ = 0.98, $R^2_{Yb}$ = 0.98, $R^2_{Hf}$ = 0.95, and $R^2_{Zr}$ = 0.94) and low standard error of estimates ($SEE_{La}$ = ±0.005, $SEE_{Yb}$ = ±0.04, $SEE_{Hf}$ = ±0.12, and $SEE_{Zr}$ = ±0.04). The mean absolute percentage errors from the lattice strain models are also good, i.e., 6-9% ($D_{REE+Y}$) and 9-12% ($D_{HFSE}$). As the starting melt composition shifts from basalt (BA) to trachybasalt (TB) to basaltic trachyandesite (BT), $D_{REE+Y}/D_{HFSE}$ progressively increase (Fig. 9), reflecting the control of melt composition and structure on element partitioning (Gaetani, 2004; Huang et al., 2006; Qian et al., 2015). $[Ca^{2+}/(M^{+}+M^{2+})]^{melt}$ is negatively correlated with $D_{REE+Y}$ ($R^2_{REE+Y}$ = 0.53-57) and $D_{HFSE}$ ($R^2_{HFSE}$ = 0.52-63), consistent with an increase in the number of large structural sites in the melt rather than in the crystal (Huang et al., 2006). The moderate values of the measured correlation coefficients conform to thermodynamic data from Schmidt et al. (2006), showing that a regular solution model for the effect of melt structure on trace element partition coefficient can be formulated effectively only when the melt composition changes drastically from gabbro to granite. In contrast, the crystal electrostatic effects are positively correlated with $D_{REE+Y}$ ($R^2_{REE+Y}$ = 0.85-91) and $D_{HFSE}$ ($R^2_{HFSE}$ = 0.93-94), providing evidence that both electrostatic and lattice strain energies have a major influence on partitioning behavior. The high correlation between $D_{REE+Y}/D_{HFSE}$ and the crystal electrostatic effects denotes that the predictive ability of the lattice strain models is greatly enhanced once cation incorporation at charge-imbalanced sites is accounted for (Wood and Blundy, 2001; Hill et al., 2011).

## 5. Application of the *P-T*-H$_2$O-lattice strain model to Mt. Etna

*5.1. Overview of the eruptions selected as test data*

*5.1.1. Historic eruptions*

The historic eruptions object of this study include undated (5-3 ka BP) and dated (from 1329 to 1886) lavas sampled on the SE flank of Mt. Etna volcano and emplaced during the Recent Mongibello activity (Corsaro and Cristofolini, 1996). These products were responsible for the construction of the summit cones by the emplacement of primitive to more evolved (MgO = 3-7 wt.%) lava flows with Na-affinity. The studied volcanic rocks share the common mineral association of Etnean magmas. Products earlier than the 16$^{th}$ century are hawaiites and mugearites, whereas the younger lavas are hawaiites (Corsaro and Cristofolini, 1996). Bulk rock major and trace element concentrations, as well as Sr isotopes reveal that simple crystal fractionation is not responsible for the differentiation of magma (Corsaro and Cristofolini, 1996). Instead, magma mixing phenomena have been recognized, accounting for successive influxes of mafic melts with slightly different geochemical and isotopic features (Armienti et al., 1984; Corsaro and Cristofolini, 1996). K, Rb, and Cs contents tend to increase progressively over time, but more pronounced changes are measured for post-1971 eruptions, in concert with a net shift of $K_2O/Na_2O$ ratio towards a remarkable K-affinity (Armienti et al., 1994). The temporal correlation of $\delta^{11}B$ with Sr-Nd isotopes has been attributed to addition of a fluid component to an asthenospheric mantle source (Armienti et al., 1994). In this view, the temporal enrichment in fluid-mobile elements is related to the occurrence of dehydration reactions in the subducting Ionian oceanic slab (Tonarini et al., 2001).

*5.1.2. 1991-1993 eruptions*

On December 1991, the opening and propagation of a fissure near SEC caused a short episode of lava fountains and a small lava flow (Barberi et al. 1992). Afterwards, eruptive activity proceeded in the lower portion of the fracture (on the western wall of Valle del Bove) with continuous lava flows accompanied by weak Strombolian activity. This eruptive period ended on March 1993, with the noteworthy production of $\sim300\times10^6$ $m^3$ of material (Barberi et al. 1992). Lavas are mildly alkaline porphyritic hawaiites (MgO < 6 wt.%) with a uniform modal and chemical composition (Armienti et al., 2013). The paragenesis is dominated by the crystallization of plagioclase with subordinated clinopyroxene, olivine, and titanomagnetite. Since September 1992, the erupted products showed an increase in the degree of evolution and a contrasting Sr isotope signature, due to arrival of a distinct batch of magma (Armienti et al., 1994). Olivine phenocrysts exhibit complex zoning patterns (i.e., crystal portions with both normally and reversely zoned compositions), recording variable degrees of interaction between at least three distinct magmas (Kahl et al., 2011). It has been argued that the initial recharge event started in 1990 with the arrival of a primitive magma from depth. Subsequently, from September 1991, the uppermost part of the plumbing system was characterized by continuous recharge (Kahl et al., 2011). This is also confirmed by Sr isotope disequilibrium between bulk rocks and separated clinopyroxene phenocrysts, as the expression of incorporation of minerals crystallized in isotopically distinct magma batches rising from a heterogeneous mantle source (Armienti et al., 2004, 2007).

*5.1.3. 2001 eruptions*

In July 2001, eruptive activity of Mt. Etna started at the SEC with lava fountaining followed by a complex system of eruptive fissures on the NE and S flanks. The NE flank was characterized by effusive eruptions from upper vents (UV) that produced porphyritic trachybasalts with a mineral assemblage dominated by plagioclase and subordinate clinopyroxene, olivine and titanomagnetite, typical of summit activity in 1995-2000 (Corsaro et al., 2007). Conversely, lava flows and explosive eruptions occurred at the S flank from lower vents (LV) producing more primitive trachybasalts dominated by large clinopyroxene and olivine phenocrysts, and the presence of uncommon minerals (amphibole, apatite, orthopyroxene), siliceous xenoliths, and cognate xenoliths heralding an intrusive gabbroic body at ~190-400 MPa (~5-13 km depth; Corsaro et al., 2014). These petrographically distinct products were interpreted as the result of two separate magma storage zones: a shallower reservoir providing a more evolved and degassed UV magma, and a deeper plumbing system providing a more primitive and volatile-rich LV magma (Corsaro et al., 2007). The former is representative of small and shallow reservoirs inside the volcanic piles that are directly related to the central feeding system. Conversely, the latter is responsible for eccentric activity that occurs as rare, explosive flank eruptions fed by basaltic to trachybasaltic dikes rising from depth and bypassing the central volcano conduits (Rittmann, 1965; Tanguy, 1980; Armienti et al., 1988). While the stability of plagioclase in UV magma is enhanced at $P < 100$ MPa and $H_2O < 1$ wt.%, clinopyroxene and olivine are more favored in LV magma at $P > 100$ MPa and $H_2O > 1$ wt.% (Corsaro et al., 2007). The occurrence of two distinct $H_2O$-poor and $H_2O$-rich magmas is also supported by melt inclusion data from olivines found in NE and S flank products, respectively (Metrich et al., 2004). For the LV eruptions, melt inclusions suggest the ascent of a primitive magma from ~12 to ~6.5 km depth, intruding and mixing with a slightly more evolved magma at ~5 km depth. Coherently, geophysical data evidence the intrusion of a large magma body beneath the volcano at 6-15 km depth and the subsequent injections of magmas into a storage zone at 3-5 km depth (Patanè et al., 2003). In terms of bulk rock geochemistry, UV (MgO < 6 wt.%) and LV (MgO > 6 wt.%) products are characterized by distinct major element compositions, but their trace element concentrations are virtually identical. These magmas exhibit enrichments in fluid-mobile elements (e.g., K, Rb, and Pb) and their low Nb/Rb ratios are interpreted as progressive modification of the mantle source region by subduction derived fluids (Armienti et al., 1989; Beccaluva et al., 1982; Cristofolini et al., 1987; Schiano et al., 2001; Tonarini et al., 2001; Clocchiatti et al., 2004). Conversely, ratios of highly incompatible elements (e.g., Ce/Yb and Th/Ta) do not change significantly, being comparable with those measured in post-1971 products. Therefore, according to some authors (Viccaro et al., 2006; Corsaro et al., 2007), 2001 eruptions can derive from the same primary magma that underwent variable degrees of fractionation crystallization.

*5.1.4. 2002-2003 eruptions*

In 2002-2003, fourteen eruptive vents formed along the NE rift system (NERS) of Mt. Etna volcano (Ferlito et al., 2009). Until 5 November 2002, lava flows and near continuous explosive fire fountaining occurred on both the NE and S flanks. Afterwards, between 13 November 2002 and 28 January 2003, the eruptions continued exclusively on the S flank (Andronico et al., 2005). Flank eruptions from southern fissures (SF) show petrographic and geochemical characteristics similar to those of primitive magmas feeding 2001 LV and eccentric eruptions. SF products are sub-aphyric (5–10% phenocrysts) alkali basalts to trachybasalts characterized by primitive compositions (MgO > 6 wt.%) and prevalent mafic mineralogy (including amphibole and orthopyroxene) with rare plagioclase (Clocchiatti et al., 2004; Metrich et al., 2004; Spilliaert et al., 2006; Corsaro et al., 2007). It is argued that these SF eruptions were fed by the 2001 LV magma that ascended from the deeper portion of the plumbing system (i.e., ~10 km depth; Spilliaert et al., 2006; Corsaro et al., 2009), bypassing the shallow region of magma storage (Corsaro et al., 2007). In contrast, flank eruptions from the northern fissures (NF) share similarities with magmas feeding 2001 UV eruptions and the recent explosive activity at the summit craters (i.e., BN and NEC). NF products are more differentiated (MgO < 6 wt.%) and degassed trachybasalts to trachyandesites, showing porphyritic (plagioclase-phyric) textures that resemble those of trachybasalts erupted during the past decades at the summit craters (Clocchiatti et al., 2004; Ferlito et al., 2008; Corsaro et al., 2009; Coulson et al., 2011). In terms of incompatible trace elements, LILE, Sr-Nd isotopes, and radioactive disequilibria, both the 2001 UV and 2002-2003 NF products conform to the post-1971 volcanic activity (Clocchiatti et al., 2004). Conversely, 2001 LV and 2002-2003 SF products are the most femic but also the most enriched in K, Rb, Ra, and $^{87}$Sr (Clocchiatti et al., 2004), perhaps due to melting of a metasomatized mantle source containing hydrous phases (phlogopite/amphibole) or direct transfer to the melting region of fluid-mobile elements produced by dehydration of the subducting Ionian plate (Tonarini et al., 2001). It has been also suggested that 2001 LV and 2002-2003 SF magmas tapped a heterogeneous, clinopyroxenite-veined mantle source which experienced several stages of metasomatism by slab-derived fluids selectively enriched in Rb and K (Corsaro and Metrich, 2016). However, melt inclusions in olivines from all the 2001 and 2002-2003 flank eruptions do not show obvious differences in the $^3$He/$^4$He ratios (Coulson et al., 2011), being also identical to the helium isotopes of products erupted during 2001–2005 (Nuccio et al., 2008) and in the last 0.5 Ma (Marty et al., 1994). This points out that the magmatic source might not have changed over the past 0.5 Ma and/or that subduction-related metasomatic fluids have minimal effect on helium isotope signature (Nuccio et al., 2008; Coulson et al., 2011). In other words, the petrological dissimilarity of 2001 and 2002-2003 flank eruptions does not necessarily correspond to significant variability in the mantle source, but rather may reflect the different evolutionary history of cogenetic magmas during crystal fractionation and degassing phenomena at different levels within the crust (Coulson et al., 2011). Importantly, a peculiar feature of 2002–2003 flank eruptions was the occurrence of low-potassium oligophyric (LKO) magmas erupted only in the uppermost part of NERS. While SF and NF products result from variable degrees of fractionation from a single parental magma (Ferlito et al., 2008; Corsaro et al., 2009; Coulson et al., 2011), LKO trachybasalts exhibit lower K, Rb, and Nb contents that exclude a cogenetic origin with SF and NF magmas (Ferlito et al., 2008). Indeed, LKO compositions are more similar to historic (pre-1971) hawaiitic to mugearitic lavas from Mt. Etna (Tanguy et al., 1997; Schiano et al., 2001; Armienti et al., 2004). In this view, the NERS is part of more complex subvolcanic structure in which magmas with different parental affinities may rise and differentiate independently from the main open conduit system (Ferlito et al., 2008).

*5.1.5. 2004-2005 eruptions*

On 7 September 2004, a new fracture field opened on the flank of SEC and extended towards the rim of Valle del Bove, resuming effusive activity of Mt. Etna for ~6 months, until 8 March 2005 (Corsaro and Miraglia, 2005). Products erupted from different vents show an identical mineral assemblage dominated by plagioclase phenocrysts, suggesting magma storage in the shallow plumbing system accompanied by strong degassing (Corsaro and Miraglia, 2005). The 2004-2005 trachybasalts (MgO = 5-6 wt.%) show compositions comparable to those of 2001 and 2002-2003 products. However, the incompatible trace element ratios of Ta/Tb (i.e., elements that do not change during fractional crystallization) and Rb/Nb (i.e., elements modified by the contribution of aqueous fluids) in the 2004–2005 eruptive products can be categorized into three distinctive temporal phases of magmatic activity (Corsaro et al., 2009): Phase 1 with the highest Rb/Nb and lowest Ta/Tb ratios, Phase 2 with decreasing Rb/Nb ratios and increasing Ta/Tb ratios, and Phase 3 with constantly low Rb/Nb ratios and high Ta/Tb ratios. These distinct phases of activity are also evidenced by the temporal variation of Sr-Nd isotopes, demonstrating that Phase 2 magmas result from mixing between Phase 1 and Phase 3 magmas. In a more general view, Phase 1 products are the most evolved and radiogenic, resembling geochemical data from 2001 LV and 2002-2003 SF eruptions and denoting the presence of a small reservoir (at ~300–400 m below the summit craters) in which a residual magma batch was stored (Corsaro et al., 2009). Indeed, modeling of major and trace elements indicates that Phase 1 products are reproduced by fractionation of plagioclase +

clinopyroxene + olivine + titanomagnetite from the most primitive magma erupted during 2002-2003 SF activity (Corsaro et al., 2009). In contrast, Phase 3 products show affinity with the 2001 UV and 2002-2003 NF eruptions. This means that, during Phase 2, a new magma rising in the central conduits mixed with the residual and more differentiated magma batch left in the shallow plumbing system after the 2002-2003 SF eruptions (Corsaro et al., 2009). With respect to the intensely explosive stage of 2001 and 2002–2003 eruptions, it has been also argued that 2004-2005 effusive products represent the late stage of a recharging process following arrival of a geochemically distinct magma into the uppermost segment of the open-conduit system (Ferlito et al., 2012). This volatile-rich magma gradually moved and degassed from the eccentric system towards the uppermost part of the main open-conduit, also modifying the sub-volcanic fracture system to form a new vent (SEC 2) at the eastern base of the SEC. At first, the volatile-rich magma interacted with the eccentric reservoirs located in the shallow part of the plumbing system (<10 km), giving rise to the 2001 and 2002–2003 explosive eruptions. Subsequently, the magma reached the open-conduit system and partially lost its volatile content through the summit craters, feeding the 2004–2005 effusive eruptions (Ferlito et al., 2012).

*5.1.6. 2006 eruptions*

On 14 July 2006, the volcanic activity at Mt. Etna restarted with the opening of an eruptive fissure on the eastern flank of SEC (Ferlito et al., 2010; Sciotto et al., 2011). In contrast to 2001, 2002-2003, and 2004-2005 events, the volcanic phenomena displayed a variety of styles, from purely effusive to highly explosive, until 15 December. The early lava effusions were followed by Strombolian eruptions and sustained lava fountains. Then, a pulsating regime took place with eruptions of different frequency and intensity, as well as with the opening of new vents in Valle del Bove and on the southern flanks of BN and SEC. The predominant mineral assemblage of erupted products comprises phenocrysts of plagioclase followed by clinopyroxene. With respect to previous 2001, 2002-2003, and 2004-2005 events, the 2006 volcanic rocks are the most compositionally evolved, showing the lowest MgO (4-5 wt.%) and CaO (10-11 wt.%) contents, especially at the onset of the volcanic activity. These early-erupted products derive by fractional crystallization from a residual magma that was stored in the open-conduit at the end of 2004–2005 eruptions (Nicotra and Viccaro, 2012). Moreover, olivine-hosted melt inclusions in 2006 products denote $H_2O$ concentrations (<2 wt.%) significantly lower than those (~3.5 wt.% H2O) measured for 2001 and 2002–2003 eruptions (Métrich et al., 2004; Spilliaert et al., 2006; Collins et al., 2009). This is ascribed to magma dehydration events due to fluxing by $CO_2$ and/or the arrival of $CO_2$-rich magma batches with distinctive geochemical and isotopic signatures. $CO_2$ fluxing is consistent with a number of positive peaks in the $CO_2/SO_2$ molar ratios of the gas plume that, in turn, are temporally correlated with the pulsating regime of the eruptions (Aiuppa et al., 2007). On the other hand, towards the middle of the 2006 volcanic activity, a new recharging event was identified with ingress into the plumbing system of a more basic, $CO_2$-rich magma showing a distinctive isotopic signature (Nicotra and Viccaro, 2012). Indeed, bulk rock data evidence a temporal shift in composition towards trachybasaltic magmas characterized by negative trends of LREE, HFSE (e.g., Nb, Zr, and Y), and Pb isotopes, coupled with positive trends of incompatible element ratios (e.g., Rb/La and Zr/Nb) and Sr isotopes. These geochemical characteristics closely match with those outlined for the volatile-rich magma batch that, firstly, ascended into the shallow eccentric reservoirs triggering the 2001 and 2002–03 explosive events and, secondly, fed the 2004–2005 effusive eruptions after substantial degassing into the main open-conduit system (Ferlito et al., 2012).

*5.1.7. 2007-2008 and 2008-2009 eruptions*

Summit eruptions of lava fountains and lavas that flowed towards the depression of Valle del Bove took place at SEC from March 2007 to May 2008. On September 2007, the most powerful paroxysm was also accompanied by a plume up to ~2 km high that produced fall-out deposits at distances of several tens of kilometers from SEC (Andronico et al., 2008). Subsequent 2008–2009 flank activity ran from 13 May 2008 to 6 July 2009. This eruptive phase involved a fissure system developed on the flank of the volcanic edifice (Bonaccorso et al., 2011). From the lower portion of the eruptive fissures, lava flows propagated and expanded towards Valle del Bove, to form a large lava delta. From a compositional point of view, trachybasalts (MgO = 5-6 wt.%) erupted during 2007-2008 and 2008-2009 events exhibit the typical mineral assemblage of historic summit and flank eruptions fed by magma ascending through the central conduit system (Corsaro and Pompilio, 2004). However, the chemistry of olivine and plagioclase reveals magma mixing phenomena, with mineral cores in disequilibrium with the host rock but in equilibrium with a more primitive magma (Corsaro and Miraglia, 2014). Major and trace element contents of 2007–2008 SEC episodic eruptions reveal that the original magma becomes progressively more evolved over time, even though this compositional change cannot be modeled by simple fractional crystallization (Corsaro and Miraglia, 2014). Conversely, the more complex compositional variation of 2008–2009 flank eruptions reveals the occurrence of three distinct temporal phases: Phase 1 fed the early eruption of a magma more evolved than that of the last 2007–2008 paroxysm, Phase 2 was due to the gradual shift towards the composition of a more primitive magma, and Phase 3 was characterized by progressive differentiation of this latter primitive magma (Corsaro and Miraglia, 2014). Results from trace element modeling demonstrate that the wide spectrum of compositions erupted during both 2007–2008 and 2008–2009 events can be reproduced by two distinct magma batches that mixed together in variable proportions over time (Corsaro and Miraglia, 2014). In particular, the early 2007 paroxysm at SEC was triggered by a single magma recharge due to the rapid injection of a primitive, volatile-rich ascending from depth. This magma passed through the shallow reservoir, and mixed with the more evolved magma already stored in the reservoir, probably since the last volcanic activity occurred at the end of 2006 (Corsaro and Miraglia, 2014).

*5.1.8. 2011-2013 eruptions*

In the period from January 2011 to April 2013, several sequences of paroxysmal eruptions with intermittent character occurred mostly at the newly-formed summit crater named New South East Crater (NSEC). The early activity was characterized by pulsating degassing and episodes of Strombolian explosions accompanied by lava flows directed towards the Valle del Bove (Viccaro et al., 2014). During paroxysmal activity at NSEC, intense degassing and ash emissions also occurred at NEC, BN, and VOR. Trachybasalts (MgO = 3.9-5.5) erupted during 2011-2013 are porphyritic rocks dominated by plagioclase crystals. In terms of both major and trace elements, 2011-2013 magmas display compositions similar to those of 2007-2008 and 2008-2009 eruptions. However, the variable degree of evolution of 2011-2013 products and the disequilibrium features of minerals indicate that fractional crystallization was periodically interrupted by mixing phenomena due to replenishment events in the Etnean feeding system (Giacomoni et al., 2017; Giuffrida and Viccaro, 2017). Magma recharge was controlled by the input of a primitive and volatile-rich melt that is similar in composition to the most basic products emplaced during 2007 and 2012 paroxysmal eruptions (Giuffrida et al., 2017). After a phase of major magma recharge into the deep portions of the plumbing system (>6 km), the pulsating magma transferred from the intermediate (2–6 km depth; Patané et al., 2006; Bonforte et al., 2008; Bonaccorso et al., 2011) to the shallow (1–2 km within the volcanic edifice; Palano et al., 2008; Patanè et al., 2013) portions of the plumbing system, feeding the paroxysmal activity at NSEC.

*5.2. Modeling the control of clinopyroxene on REE+Y/HFSE concentrations in magmas*

The lattice strain expressions for $D_{REE+Y}/D_{HFSE}$ have been integrated with the clinopyroxene-based barometric (Eqn. $A_{MAM}$), thermometric (Eqn. $33_{MAM}$), hygrometric (Eqn. H), and equilibrium (Eqn. DiHd) equations, to derive an overall $P$-$T$-$H_2O$-lattice strain model reported in the Excel spreadsheet submitted online as supplementary material. The natural compositions of clinopyroxene phenocrysts and melts from eruptions at Mt. Etna volcano have been used as input data (Table 5S) and results from calculations are plotted in $P$ vs. $T$ and $P$ vs. $H_2O$ diagrams (Fig. 10), showing that the geochemical evolution of clinopyroxene reconstructs the crystallization path of each single eruption. The linear regression fit of the data evidences a high correlation between $P$ and $T$ ($R^2$ = 85-99), and a moderate to high correlation between $P$ and $H_2O$ ($R^2$ = 0.47-0.98). However, it must be noted that the correlation coefficient and the length/slope of each $P$-$T$-$H_2O$ segment are mostly determined by the quality and statistical number of clinopyroxene analyses. Considering the internal calibration error of the predictive models, the Etnean magmas record variable crystallization conditions from 53 to 634 MPa, 1,019 to 1,152 °C, and 0.7 to 3.7 wt.% $H_2O$ (Fig. 10). These estimates are consistent with the general observation that magma crystallizes throughout the entire length of a vertically-developed plumbing system characterized by multiple storage zones, wherein magmas undergo fractional crystallization, degassing, and mixing processes through the effect of different $P$-$T$-$H_2O$ paths (e.g., Armienti et al., 2004; Mollo et al., 2015b; Corsaro et al., 2013; Kahl et al., 2016; Ferlito, 2017). For the deeper portions of the plumbing system, Armienti et al. (2013) modeled the ascending path of fluid-undersaturated magmas feeding some key eruptions, showing that the clinopyroxene liquidus is constrained to lie between 500–900 MPa, 1,100–1,180 °C, 3–4 wt.% $H_2O$. Conversely, at shallower crustal levels, Mollo et al. (2015b) and Perinelli et al. (2016) showed that the saturation temperature of clinopyroxene progressively decreases down to ~1,050 °C along a continuous decompression path where most of the $H_2O$ release (~2.5 wt.%) occurs at $P$ < 100 MPa. The transport of mafic phenocrysts and their carrier melts towards the surface is dictated by a continuous recharging mechanism due to frequent inputs from mantle depths of primitive, volatile-rich magmas into shallower crustal reservoirs (Armienti et al., 2004; Ferlito et al., 2008; Corsaro et al., 2013; Kahl et al., 2015; Ferlito, 2017). Through independent evidence based on crystal size distribution and isotope analyses of products erupted during 1852–2004 volcanic activity, Armienti et al. (2007) identified distinct clinopyroxene populations hosted in the same magmas. This testifies to a highly dynamic regime, wherein the entrapment of pre-existing phenocrysts in new magma batches is determined by high input rates of primitive magmas from depth, in the absence of a large magma chamber (Armienti et al., 2007). The persistent supply of fresh magma in the uppermost section of the plumbing system effectively buffers the erupted products to trachybasalt compositions (Armienti et al., 2013). Experiments from this study (Figs. 3 and 4) and previous differentiation models based on natural data (Armienti et al., 2004; Corsaro et al., 2013; Mollo et al., 2015b) indicate that the segregation of ~20-25% of mafic phenocrysts from primitive basaltic magmas is the maximum admissible value to explain the trachybasaltic compositions of historic and recent eruptions. Moreover, clinopyroxene phenocrysts formed at $P$ > 400 MPa and $T$ > 1,150 °C (Fig. 10), evidence that most of the $H_2O$ (> 2.5 wt.%) is retained in the deep-seated magmas, corroborating solubility data modeled by Armienti et al. (2013) for a non-ideal $H_2O$-$CO_2$ mixture. In contrast, substantial degassing ($H_2O$ < 1 wt.%) may take place only in the uppermost portion ($P$ < 100 MPa) of the volcanic conduit (Metrich and Rutherford, 1998). Noteworthy, melt inclusions in olivines (Métrich et al., 2004; Spilliaert et al., 2006; Collins et al., 2009) show that volatile-rich magmas travelling towards the surface become extensively depleted in $CO_2$ (~0.05 wt.%) but maintain a relatively high $H_2O$ content (2-2.5 wt.%). This implies that degassing-driven crystallization occurs in the uppermost part of the conduit system when most of the $H_2O$ dissolved in the melt is rapidly exsolved, controlling the final crystal growth of phenocryst rims, microphenocrysts, and microlites (Lanzafame et al., 2013, 2017; Mollo et al. 2015b; Moretti et al., 2017; Ferlito, 2017). Under such circumstances, the near-equilibrium crystal growth in the magmatic reservoir shifts to rapid crystal growth conditions from the conduit to the surface. This is particularly evident for the crystallization of plagioclase and titanomagnetite during magma undercooling and decompression (Applegarth et al., 2013; Mollo et al., 2015a, 2015b; Lanzafame et al., 2017). On eruption, the phenocryst and microphenocryst content of magma is up to 50% but a great number of plagioclase and titanomagnetite nucleate and grow by degassing- and cooling-driven crystallization mechanisms (Applegarth et al., 2013). While plagioclase textures and compositions reflect dynamic conditions in the final segment of the volcanic conduit (Giacomoni et al., 2014), clinopyroxene phenocrysts start to equilibrate at depth and so are able to record the entire $P$-$T$-$H_2O$ paths of magmas erupted at Mt. Etna (Fig. 10).

According to the compositional variation of clinopyroxene from (Di-Hd)$^{cpx}$ to (Ts-Jd-En-Fs)$^{cpx}$ during crystal growth kinetics (Mollo et al., 2013a), the attainment of (near)equilibrium conditions between the selected clinopyroxene phenocrysts and erupted Etnean magmas is verified through the Eqn. DiHd, yielding appropriate values of ΔDiHd < 0.1 (Table 5S). However, as a further test for equilibrium, $D_{Ti}$ for natural clinopyroxene-melt pairs are compared with those predicted by the $P$-$T$-$H_2O$-lattice strain model (Fig. 11). The regression statistics ($R^2$ = 0.70 and SSE = ±0.11) are good with an average deviation of ±0.09, translating to a relative error from 6% to 18%. The correspondence between near-equilibrium ΔDiHd values and the relative high precision of the lattice strain model confirms the relationship between the cation exchange [$^T$Si, $^{M2}$(Mg, $Fe^{2+}$)] ↔ [$^T$Al, $^{M1}$(Al, $Fe^{3+}$)] and the accommodation of Ti in the M1 site of clinopyroxene. During magma dynamics at Mt. Etna volcano, any deviation from equilibrium crystallization is due primarily to rapid crystal growth at shallow crustal levels where magmas are more differentiated and undergo strong degassing and undercooling during ascent (Métrich et al., 2004; Spilliaert et al., 2006; Collins et al., 2009). Degassing-driven crystallization is therefore the key mechanism controlling final crystal growth when most of the $H_2O$ dissolved in the melt is rapidly exsolved (Mollo et al., 2015b). Under such circumstances, near-equilibrium crystal growth in the magmatic reservoir shifts to a rapid crystal growth condition in the conduit (Applegarth et al., 2013). The disequilibrium crystallization of clinopyroxene is frequently accompanied by aluminium enrichments (Mollo et al., 2012; Hammer et al., 2016; Mollo and Hammer, 2017) that facilitates incorporation of highly charged cations, such as HFSE (Wood and Trigila, 2001; Marks et al., 2004). Eqn. DiHd allows to correctly identify clinopyroxene compositions that reflect equilibrium crystallization of the system rather than rapid growth of Al-rich crystals. As a consequence, the partitioning of Ti measured for natural compositions closely agrees with the thermodynamically-derived value of $D_{Ti}$ through the energetics of the different charge balanced/imbalanced configurations (Fig. 11). Considering the electrostatic effect produced by insertion of a heterovalent trace cation into the crystal lattice (e.g., $Ti^{4+}$ for $Mg^{2+}$), the proportion of M1 sites charge-balanced by a tetravalent trace element decreases with increasing $^{M1}Mg$. Owing to charge balancing, $D_{Ti}$ is positively correlated with $^{T}Al$ because Ti much more readily enters the M1 site of $CaAl_2SiO_6$ (3+ charge on M1) than $CaMgSi_2O_6$ (2+ charge on M1). Thus, under kinetically-controlled crystallization conditions, $D_{Ti}$ may increase towards disequilibrium values during substitution of diopside with Tschermak molecules and replacement of divalent Mg with trivalent Fe and Al (Wood and Trigila, 2001; Mollo et al., 2013b).

An important outcome from this study is that, for each single eruptive product at Mt. Etna, the $P$-$T$-$H_2O$-lattice strain model provides a spectrum of $D_{REE+Y}/D_{HFSE}$ values as a function of the crystallization conditions of magma, as well as the clinopyroxene and melt compositions (Table 5S). Evidently the partitioning of trace elements between clinopyroxene and melt is controlled by a number of parameters that may or may not change simultaneously along the differentiation path of the magma. Under such circumstances, a set of nearly constant partition coefficients can be derived only when the magnitudes of parameters exerting two or more opposing effects on $D_{REE+Y}/D_{HFSE}$ are reciprocally compensated (Sun and Liang, 2012). To understand this phenomenon quantitatively, multiple regression analyses have been conducted assuming $D_0^{+3}$ and $D_0^{+4}$ as dependent variables, whereas $P$, $T$, $H_2O$, $^{T}Al$, $[Ca^{2+}/(M^{+}+M^{2+})]^{melt}$, and the crystal electrostatic effects on trace element partitioning are assigned as independent variables. We then determine statistically the influence of each independent variable on the strain-free partition coefficients. Specifically, the standardized regression coefficient equals the original unstandardized regression coefficient of the independent variable, multiplied by the standard deviation of the independent variable and divided by the standard deviation of the dependent variable (Table 9S). For the studied eruptions, results from calculations are plotted in Fig. 12 and Fig. 13 for $D_0^{+3}$ and $D_0^{+4}$, respectively, showing that any single independent variable may differently influence the magnitude of the strain-free partition coefficient. For example, $^{T}Al$ mostly controls the value of $D_0^{+3}$ for eruptions in January 1992 (35%), October 2006 (33%), and 2011-2013 lava fountains (54%) (Fig. 12). Conversely, crystal electrostatic effects are the most influential variable for eruptions occurred on March 1992 (42%), July 2001 - Lower vents (47%), October-November 2002 - Northern fissures (58%), October 2002-January 2003 - Southern fissures (42%), July 20, 2006 (53%), May 2007 (45%), and September 2007 - South-East crater (42%) (Fig. 12). Although $^T$Al and crystal electrostatic effects primarily govern the value of $D_0^{+3}$, there are some exceptions: September-October 2004 - Phase 2 is 88% controlled by $[Ca^{2+}/(M^++M^{2+})]^{melt}$,), July 15, 2006 (72% $P$), and September 2006 (90%, $T$) (Fig. 12). Moreover, $D_0^{+3}$ and $D_0^{+4}$ are differently influenced by the same type of variable. $^T$Al controls the value of $D_0^{+4}$ with a magnitude (2-34%) lower than that (13-54%) measured for $D_0^{+3}$ (Fig. 13). From a crystallochemical point of view, the substitution of REE+Y onto M2 is facilitated for the molecule (REE+Y)MgAlSiO$_6$ where the charge on M1 does not increase with increasing $^T$Al. Indeed, M2 is almost exclusively occupied by divalent Ca, Mg, and Fe. Thus, charge imbalances associated with the entry of trivalent REE+Y cations in M2 cause D$_{REE+Y}$ to increase with $^T$Al in concert with the increasing probability of achieving charge-neutral local configurations as the number of surrounding tetrahedral aluminium atoms increases (Blundy et al., 1998; Hill et al., 2000). In contrast, HFSE enter the smaller M1 octahedral site, and the correlation between D$_{HFSE}$ and aluminium reflects the increasing charge on this site with increasing the number of CaAl$_2$SiO$_6$ and CaFeAlSiO$_6$ substitutions over CaMgSi$_2$O$_6$ (Wood and Trigila, 2001; Marks et al., 2004). Thus, entry of +4 and +5 ions into M1 is enhanced by the substitution of Al$^{+3}$ for Si$^{+4}$ in the tetrahedral site (Hill et al., 2000; Mollo et al., 2017). Regression analysis of the data indicates also that $D_0^{+4}$ is moderately influenced by $P$ (22-41%) and greatly influenced by $T$ (31-97%) for eleven eruptions over a total of eighteen studied (Fig. 13). The melt variable $[Ca^{2+}/(M^++M^{2+})]^{melt}$ is more effective for September 2004 - Phase 1 (49%), September-October 2004 - Phase 2 (77%), and July 15, 2006 (97%) eruptions. The magnitude of the electrostatic effects is weakly influential for eruptions on December 1991 (15%), July 2001 - Lower vents (15%), July 20, 2006 (20%), and May 2007 (19%) (Fig. 13). On the other hand, the concentration of water in equilibrium with clinopyroxene has minor control (1-10%) on both $D_0^{+3}$ (Fig. 12) and $D_0^{+4}$ (Fig. 13), suggesting that a restricted H$_2$O variation of ~1-4 wt.% (Fig. 8) does not change significantly the activity of chemical components in the magma (Wood and Blundy, 2002). In general, the regression analyses confirm that D$_{REE+Y}$/D$_{HFSE}$ are positively correlated with $^T$Al in clinopyroxene (Hill et al., 2000; Wood and Blundy, 2001; Wood and Trigila, 2001) and negatively correlated with $T$ and H$_2$O dissolved in the melt (Wood and Blundy, 2002; Sun and Liang, 2012). However, isolating the individual contributions of these variables is not straightforward, since they are interrelated in natural, multicomponent systems. This is particularly true for the case of pressure whose structural effect is to reduce the volume of M1 and M2 sites in clinopyroxene, making them stiffer (Levien and Prewitt, 1981). Variations in the compressibility between different cations affects the result of a pressure change on the crystal lattice, but incompressibility effects limit the influence of $P$ on $D_{REE+Y}/D_{HFSE}$. Although partition coefficients increase with increasing $P$ due to the positive volume of fusion of silicate minerals, changes in $D_{REE+Y}/D_{HFSE}$ with $P$ are mostly attributable to the covarying $T$ and crystal/melt compositions (Colson and Gust, 1989). At crustal pressures, a small effect of $P\Delta V$ (where $\Delta V$ is the change in volume of the reaction) is expected since volume changes of metal-metal exchange equilibria in silicate systems are typically small (Navrotsky, 1978). If the liquidus temperature of the system does not change with increasing $P$, the amounts of Na and $^{M1}Al$ in clinopyroxene likely increase in concert with a decrease in Ca and $^{T}Al$ contents (Adam and Green, 1994). The higher proportion of $^{M1}Al$ would reduce the volumes of both M1 and M2 sites (Oberti and Caporuscio, 1991), leading to the exclusion of some relatively large and/or highly-charged cations (i.e., Ti, Sr, and REE+Y) and inclusion of low charge and/or radius cations (i.e., Mg and Na) (Green and Pearson, 1985; Adam and Green, 1994). However, opposite compositional changes occur for the temperature-dependent Jd-DiHd and CaTs-DiHd exchange equilibria, due to the positive correlation between $T$ and $^{T}Al$ (Putirka et al., 1996). Considering that in natural magmas liquidus and solidus temperatures increase with pressure, Di content decreases, while Jd, CaTs, and CaTiTs increase, favouring REE and HFSE incorporation into clinopyroxene (Forsythe et al., 1994; Bennett et al., 2004; Hill et al., 2011). Specifically, the increase in proportions of Na and $^{T}Al$ cations serves to increase $D_{REE+Y}/D_{HFSE}$ by increasing the charge-balanced configurations available to accommodate the REE and HFSE in the M2 and M1 sites respectively (Lundstrom et al., 1994; Schosnig and Hoffer, 1998; Hill et al., 2000). Since crystal composition and liquidus temperature change systematically as a function of pressure, global regression fits of experimentally-derived data show that the effect of $P$ may be unimportant for trace element partitioning between clinopyroxene and mafic melts (Sun and Liang, 2012).

To quantitatively assess the role played by clinopyroxene fractionation on the trace element pattern of mafic alkaline magmas, the range of $D_{REE+Y}/D_{HFSE}$ values calculated for each eruption (Table 5S) has been used as input data for the Rayleigh fractional crystallization equation (FC):

$$C_l = C_0 F^{(D_{REE+Y/HFSE}-1)} \tag{39}$$

$C_l$ is the trace element concentration in the residual melt, $C_0$ is the concentration in the starting composition, and $F$ is the fraction of melt remaining. $D_{REE+Y}/D_{HFSE}$ change in concert with the variable physicochemical conditions of the system (i.e., $P$, $T$, $H_2O$, and crystal/melt compositions;

Table 5S). The extent of clinopyroxene fractionation (7-24%) were varied so that the resulting values of $C_l$ reproduced the bulk trace element concentrations of erupted products. Stepwise calculations were performed changing the clinopyroxene-melt pairs at each step of fractionation, in order to derive a suite of different partition coefficients rather than one single value. Errors on model results were calculated by normal error propagation, considering the calibration uncertainties of pressure (1.5 kbar), temperature (28 °C), and melt-H$_2$O content (0.45 wt.%) from Eqns. A$_{MAM}$, 33$_{MAM}$ and H$_{MAM}$ respectively. The effect of minor liquidus olivine (e.g. <8% in post-1971 trachybasalts; Mollo et al., 2015b) was not considered due to its negligible incorporation of REE+Y/HFSE. For all calculations, the Monte Maletto composition was assumed as the parent magma, providing $C_0$ values of La = 50.7, Yb = 1.96, Zr = 191, and Hf = 4.3. It is worth noting that, in terms of fluid-immobile trace elements, the Monte Maletto composition is very similar to that (La = 46.6, Yb = 2.03, Zr = 182, and Hf = 4.1; i.e., sample 041102A from Corsaro et al., 2009) alternatively suggested as the progenitor of the post-1971 eruptions (R. A. Corsaro, personal communication). Replicate calculations performed using sample 041102A did not provide appreciable variations for the FC trajectories. Results from fractional crystallization modeling (Fig. 14) are plotted in La vs. Yb (i.e., REE+Y) and Zr vs. Hf (i.e., HFSE) diagrams, reporting also the natural compositions (i.e., bulk rock analyses) of eruptions. Modeled FC vectors match very well the evolutionary paths of magmas, confirming that the fractionation of clinopyroxene at depth controls the concentrations of REE+Y/HFSE in the eruptive products (Fig. 14). The total content (24%) of crystals fractionated is consistent with that estimated by previous thermodynamic and experimental studies investigating the origin of more differentiated trachybasalts from parental Monte Maletto magma (Armienti et al., 2004; Mollo et al., 2015b). Notably, the 2002-2003 flank eruptions were accompanied by the occurrence of low-potassium oligophyric (i.e., LKO) magmas with remarkably low K, Rb, and Nb contents that points to a petrochemical character similar to historic hawaiitic to mugearitic lavas of the pre-1971 activity (Tanguy et al., 1997; Schiano et al., 2001; Armienti et al., 2004; Ferlito et al., 2008). Consequently, the assumption of Monte Maletto as a progenitor of LKO products leads to FC trajectories that do not capture the natural bulk rock analyses. This discrepancy is resolved when FC calculations are reiterated using as starting composition one of the most primitive hawaiites (MgO ≈ 7 wt.%) of the 1669 AD eruption (La = 63.1, Yb = 2.03, Zr = 216, and Hf = 4.5; sample 669D from Corsaro and Cristofolini, 1996). The petrographic features of some lavas emplaced before the voluminous 1669 AD event are also dominated by abundant (>30%) and large (up to centimeter-sized) plagioclase crystals (Armienti et al., 1997, 2004). These historic eruptions relate to the Recent Mongibello activity when the geochemical evolution of magma was occasionally controlled by degassing-driven plagioclase

crystallization in shallow crustal reservoirs (Tanguy et al., 1997; Corsaro and Cristofolini, 1996; Armienti et al., 2004; Lanzafame et al., 2013; Vetere et al., 2015). Under such conditions, the formation of clinopyroxene is reduced by the enlargement of plagioclase stability field via $H_2O$ exsolution (Metrich and Rutherford, 1998; Metrich et al., 2004). This is also testified by the negative Eu anomaly of LKO products (Eu/Eu* = 0.82-0.85), showing values comparable to those (Eu/Eu* = 0.85-0.90) of some pre-1971 eruptions controlled by plagioclase segregation, but markedly different from those of post-1971 lavas (Eu/Eu* = 0.95-1.05). For the shallow level, more degassed magmas the evolutionary behavior of La-Yb and Zr-Hf cannot be modeled by clinopyroxene fractionation alone, resulting in FC trajectories that deviate from the natural trends (Fig. 15a), requiring for an unlikely degree of clinopyroxene fraction (53%). The LKO eruptions are well reproduced by a two-step FC mechanism (Fig. 15b), including clinopyroxene crystallization (13% for the FC1 vector) at the early stage of magma evolution, and subsequently plagioclase (22% for the FC2 vector). Despite the physicochemical conditions of the system changing during fractionation, REE+Y/HFSE remain highly incompatible into the crystal lattice of plagioclase (Dohmen and Blundy, 2014) and, consequently, the final effect is a steep FC2 trajectory ($D_{La}$ = 0.1, $D_{Yb}$ = 0.02, $D_{Zr}$ = $D_{Hf}$ = 0.01; see D'Orazio et al., 1998) that closely reproduces the compositions of natural magmas (Fig. 15b).

**Conclusions**

This review presents a $P$-$T$-$H_2O$-lattice strain model specific to mafic alkaline magmas that has been derived by combining a set of refined clinopyroxene-based barometric, thermometric and hygrometric equations with different REE+Y/HFSE thermodynamically-derived expressions for the quantification of the lattice strain parameters. Through this approach, the following conclusions can be drawn:

1) the clinopyroxene-based $T$-dependent barometer, $P$-$H_2O$-independent thermometer, and $P$-$T$-dependent hygrometer can be used as an effective, integrated tool to monitor the crystallization path of magma;
2) according to the lattice strain theory, the variation of $D_{REE+Y}/D_{HREE}$ as a function of the different physicochemical conditions of the system can be quantified by the strain-free partition coefficient ($D_0$), the site radius ($r_0$), and the effective elastic modulus ($E$);
3) the partitioning behaviour of REE+Y/HFSE into M2/M1 octahedral sites of clinopyroxene is described by a series of covarying parameters that may differently influence the values of $D_{REE+Y}/D_{HREE}$, such as temperature, pressure, melt-water content, tetrahedrally-coordinated aluminum, crystal electrostatic effects, and melt structure;

4) when the *P-T*-H$_2$O-lattice strain model is used to track the geochemical evolution of mafic alkaline magmas, it is found that the fractional crystallization of clinopyroxene closely controls the concentrations of REE+Y/HFSE in the residual melts.

**Acknowledgements**

Thanks go to Manuela Nazzari and Marcel Guillong for their support with, respectively, microprobe and laser ablation analyses. F.V. acknowledges support from the Marie Curie Fellowship IEF_SolVoM #297880 and to European Research Council for the Consolidator grant ERC-2013-CoG proposal no. 612776—CHRONOS to Diego Perugini.

**Figure Captions**

Fig. 1. The crystallization of clinopyroxene from MAM strictly controls the concentration of REE and HFSE in the melt phase when hypothetical weight fractions of olivine ($^{ol}X = 0.05$-$0.15$), clinopyroxene ($^{cpx}X = 0.05$-$0.25$), plagioclase ($^{pl}X = 0.05$-$0.25$), and magnetite ($^{mt}X = 0.05$) are multiplied by the corresponding partition coefficients measured for La (i.e., $^{ol}D_{La} = 0.01$, $^{cpx}D_{La} = 0.2$, $^{pl}D_{La} = 0.01$, and $^{mt}D_{La} = 0.1$) and Zr (i.e., $^{ol}D_{Zr} = 0.04$, $^{cpx}D_{Zr} = 0.4$, $^{pl}D_{Zr} = 0.03$, and $^{mt}D_{Zr} = 0.3$) to derive the bulk partition coefficient. Partitioning data from D'Orazio et al. (1998).

Fig. 2. Schematic map showing the location of Mt. Etna volcano and the topographic location of the summit craters, i.e., Bocca Nuova (BN, including two pit craters BN1 and BN2), Chasm or Voragine (VOR), NE Crater (NEC), SE Crater (SEC), and new SE Crater (NSEC) (redrawn from Neri et al., 2017).

Fig. 3. TAS (total alkali vs. silica) diagram (a) and MgO vs. CaO diagram (b) showing the major element differentiation path of BA, TB, and BT melts from Run#1, Run#2, Run#3, and Run#4 experiments, in comparison with the bulk rock compositions of recent and historic eruptions at Mt. Etna volcano.

Fig. 4. La vs. Yb diagram (a) and Zr vs. Hf diagram (b) showing the trace element (i.e., REE+Y and HFSE) differentiation path of BA, TB, and BT melts from Run#1, Run#2, Run#3, and Run#4 experiments, in comparison with the bulk rock compositions of recent and historic eruptions at Mt. Etna volcano.

Fig. 5. The most common barometric equations from literature have been tested for MAM compositions, in order to derive the $T$-dependent Eqn. A$_{MAM}$ ($R^2 = 0.66$ and SEE = ±1.5 kbar) through the regression fit of the experimental dataset.

Fig. 6. The most common thermometric equations from literature have been tested for MAM compositions, in order to derive the $P$-$H_2O$-independent Eqn. 33$_{MAM}$ ($R^2$ = 0.55 and SEE = ±28 °C) and $P$-$H_2O$-dependent Eqn. 34$_{MAM}$ ($R^2$ = 0.78 and SEE = ±20 °C) through the regression fit of the experimental dataset.

Fig. 7. $H_2O$ vs. $D_{HFSE}$ diagram showing the effect of melt-water content on $D_{Ti}$ ($R^2_{Ti}$ = 0.09), $D_{Hf}$ ($R^2_{Hf}$ = 0.72), and $D_{Zr}$ ($R^2_{Zr}$ = 0.60). The experimental partition coefficients have been obtained for basaltic melts (MgO = 7.74-12.1) coexisting with clinopyroxenes (MgO = 16.86-18.86) equilibrated at variable $P$ (1-1.7 GPa), $T$ (1245-1502), $H_2O$ (0-4.84 wt.%), and $fO_2$ (QFM-2 - NNO+1) (see Table 6S; data from Skulsi et al., 1994; Hauri et al., 1994; Salters and Longhi, 1999; McDade et al., 2003a, 2003b).

Fig. 8. Onuma parabolas obtained by MELTS (Ghiorso and Sack, 1995) thermodynamic simulations conducted on TB magma at 400 MPa, 970-1,120 °C, 0-3 wt.% $H_2O$, and NNO buffer. At constant $T$ = 1,120 °C, the height of the partitioning parabola for REE+Y decreases when the system shifts from anhydrous to hydrous conditions, but the change of $^{T}Al$ is restricted to only ~2% (a). Under anhydrous conditions, the value of $D_{REE+Y}$ decreases with increasing $T$ and decreasing $^{T}Al$ (b). For the case of HFSE, at constant $T$ = 1,090 °C, $^{T}Al$ decreases by ~28% causing that the partitioning parabola moves downward (c). The value of $D_{HFSE}$ decreases with increasing $T$ and decreasing $^{T}Al$ (d).

Fig. 9. The accuracy of the REE+Y/HFSE lattice strain model has been tested by comparing $D_{REE+Y}$/$D_{HFSE}$ measured for BA, TB, and BT from Run#1 and the values predicted by the model.

Fig. 10. The natural compositions of clinopyroxene phenocrysts and melts from eruptions at Mt. Etna volcano have been used as input data (Table 5S) for Eqn. DiHd (error ±0.06), $T$-dependent Eqn. A$_{MAM}$ (error ±1.5 kbar), $P$-$H_2O$-independent Eqn. 33$_{MAM}$ (error ±28 °C), and $P$-$T$-dependent Eqn. H$_{MAM}$ (error ±0.45 wt.%). Results from calculations are plotted in $P$ vs. $T$ and $P$ vs. $H_2O$ diagrams showing as the geochemical evolution of clinopyroxene reconstructs the crystallization path of each single eruption.

Fig. 11. The equilibrium crystallization of natural clinopyroxene phenocrysts has been tested by comparing $D_{Ti}$ measured for clinopyroxene-melt pairs from Etnean eruptions and values predicted by the *P-T-*$H_2O$*-lattice strain model*.

Fig. 12. The influence (%) of *P*, *T*, $H_2O$, $^TAl$, *[$Ca^{2+}$/($M^+$+$M^{2+}$)]$^{melt}$*, and the crystal electrostatic effects on $D_0^{+3}$ is derived by the statistical formula for the standardized regression coefficient.

Fig. 13. The influence (%) of *P*, *T*, $H_2O$, $^TAl$, *[$Ca^{2+}$/($M^+$+$M^{2+}$)]$^{melt}$*, and the crystal electrostatic effects on $D_0^{+4}$ is derived by the statistical formula for the standardized regression coefficient.

Fig. 14. Clinopyroxene fractional crystallization modeling. $D_{REE+Y}/D_{HFSE}$ change continuously for the trace element of interest, in concert with the variable physicochemical conditions of the system (i.e., *P*, *T*, $H_2O$, and crystal/melt compositions; Table 5S). This is possible by stepwise calculations performed by using different clinopyroxene-melt pairs at each step of fractionation. For all the calculations, the Monte Maletto composition was assumed as the primitive magma (La = 50.7, Yb = 1.96, Zr = 191, and Hf = 4.3).

Fig. 15. Clinopyroxene fractional crystallization modeling. $D_{REE+Y}/D_{HFSE}$ change continuously for the trace element of interest, in concert with the variable physicochemical conditions of the system (i.e., *P*, *T*, $H_2O$, and crystal/melt compositions; Table 5S). This is possible by stepwise calculations performed by using different clinopyroxene-melt pairs at each step of fractionation. For all the calculations, the most primitive hawaiitic rocks belonging to the 1669 AD eruption was assumed as the primitive magma (La = 63.1, Yb = 2.03, Zr = 216, and Hf = 4.5; sample 669D from Corsaro and Cristofolini, 1996). The LKO eruptions are well reproduced by a two-step fractional crystallization mechanism, including clinopyroxene crystallization (13% for the FC1 vector) at the early stage of magma evolution, and subsequent plagioclase formation (22% for the FC2 vector).

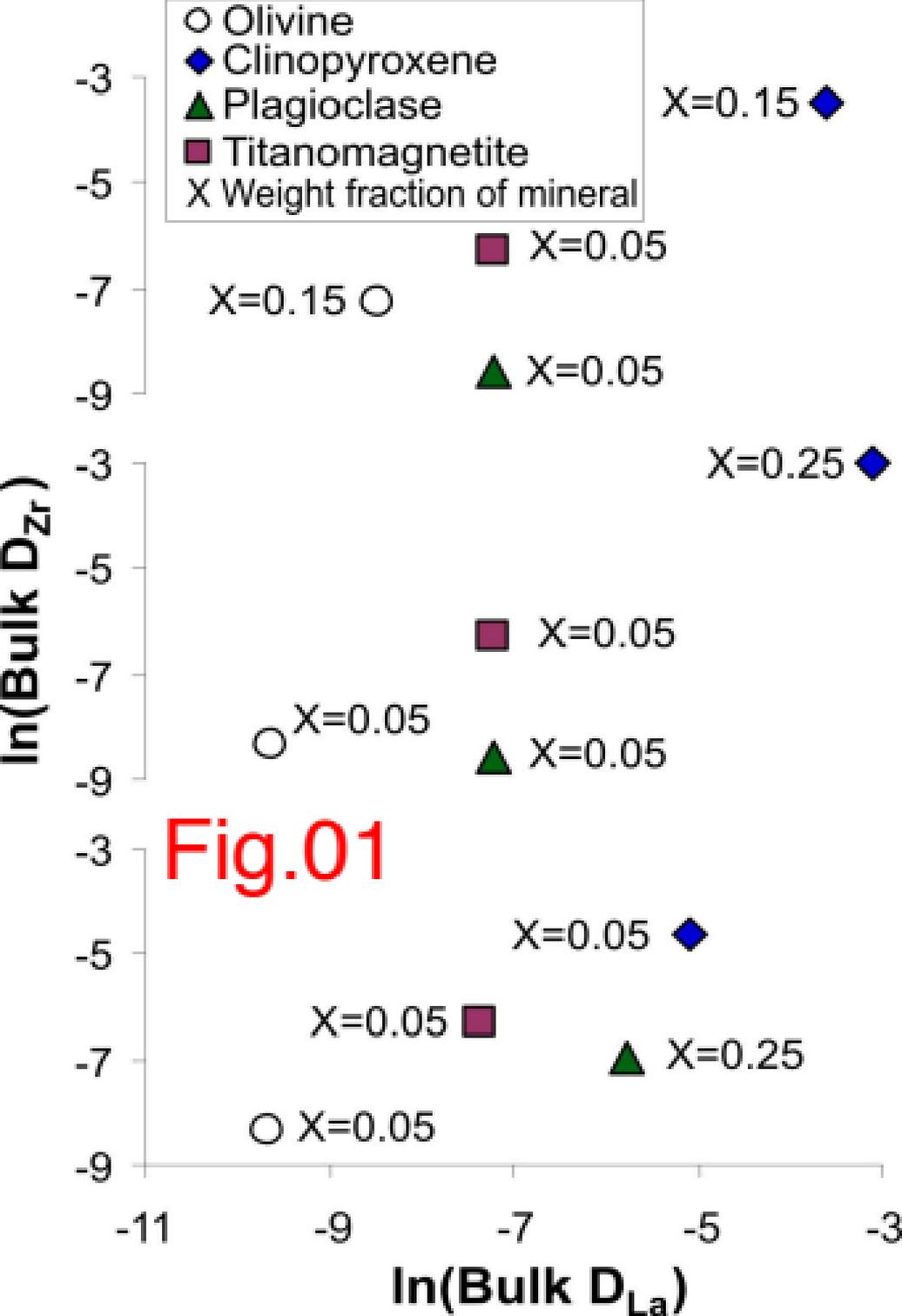

Fig.01

a

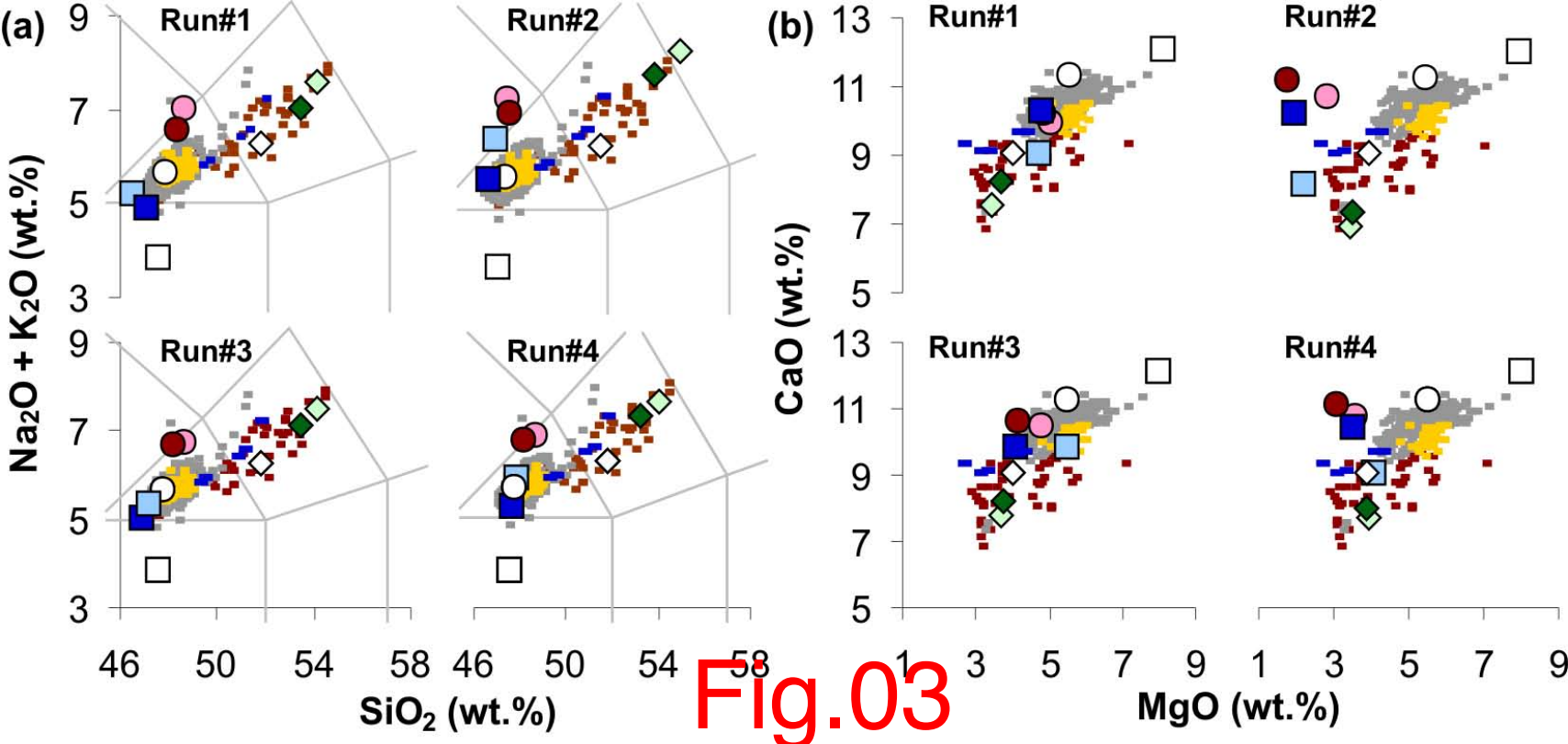

Fig.03

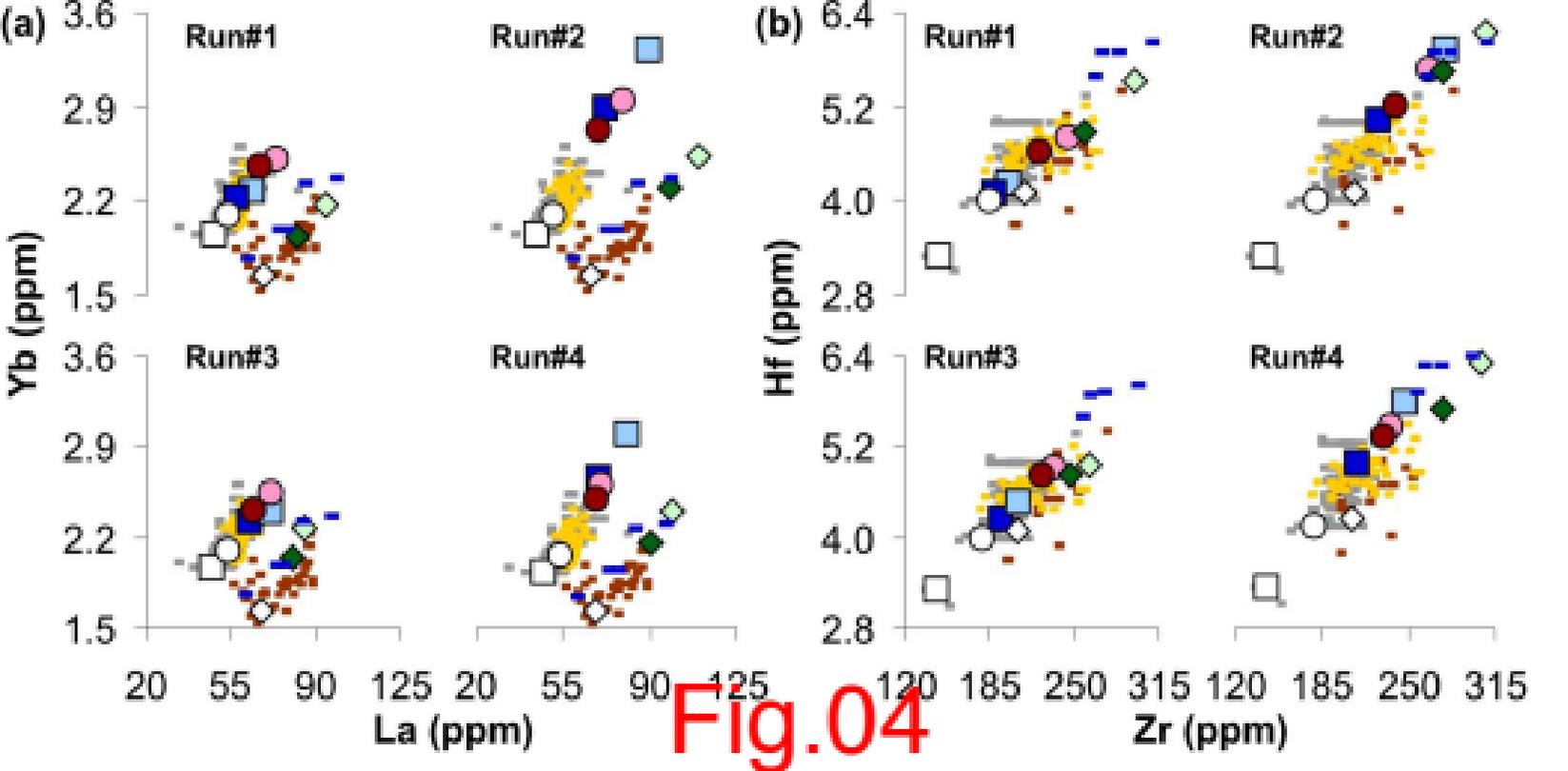

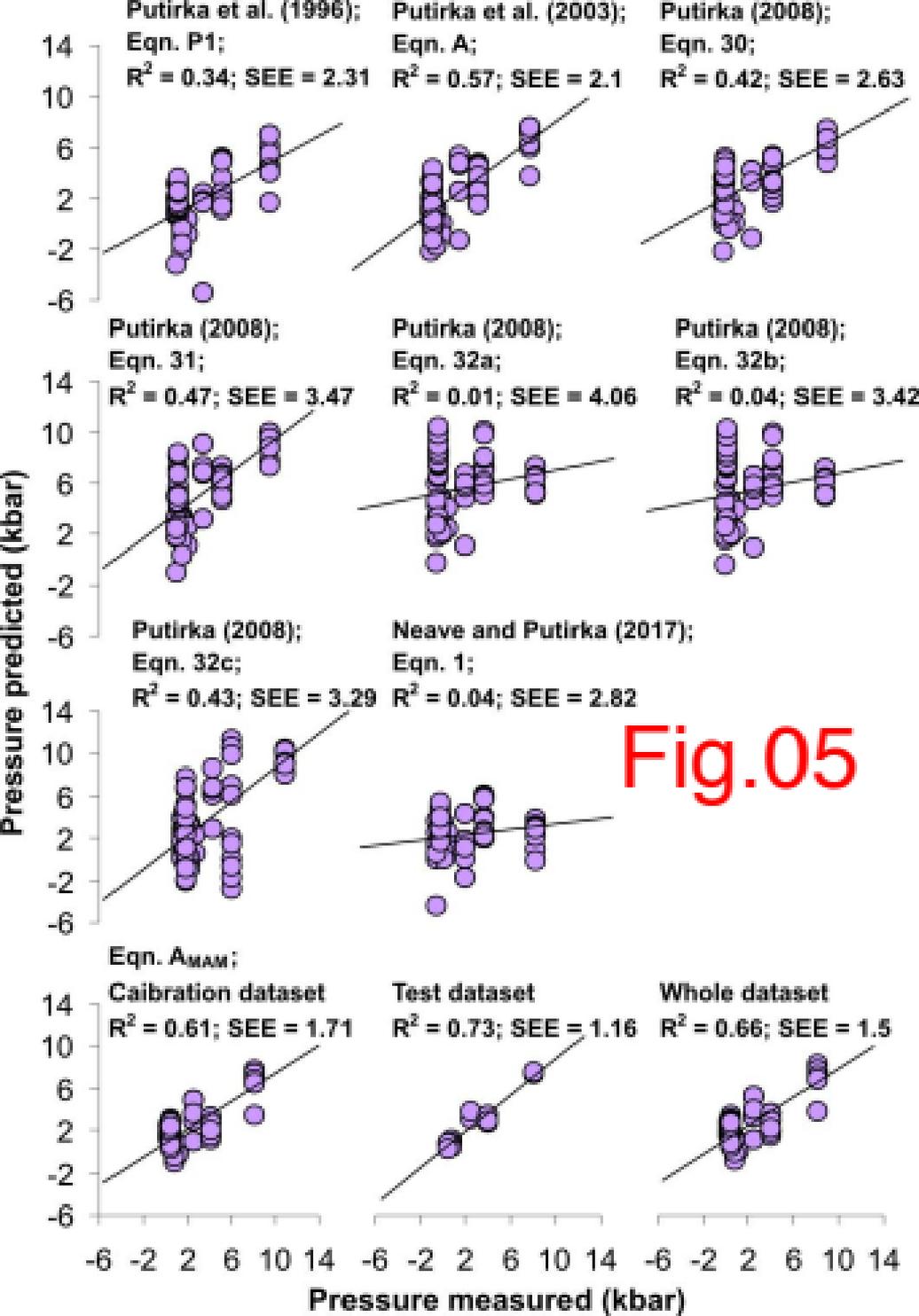

Fig.05

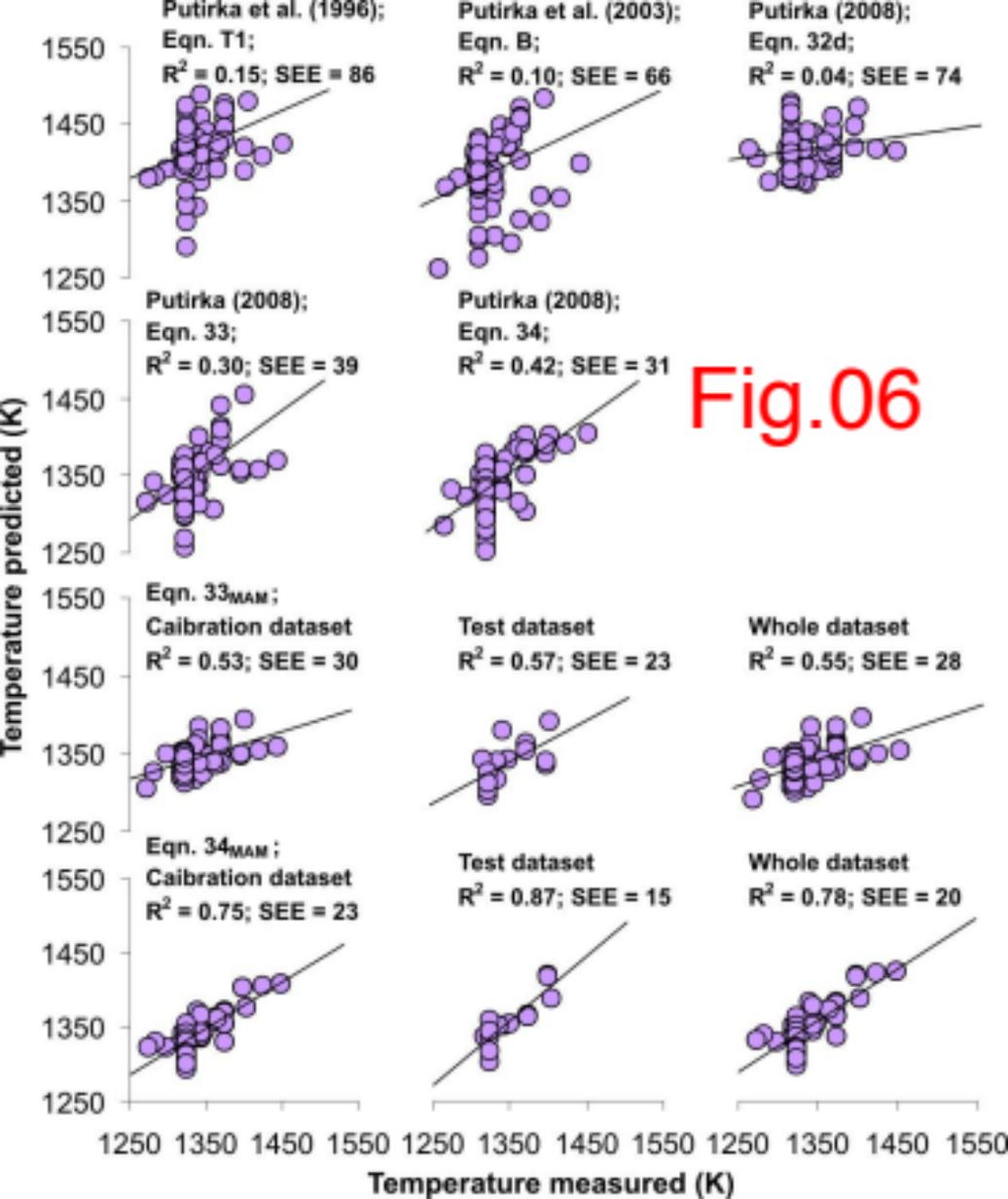

Fig.06

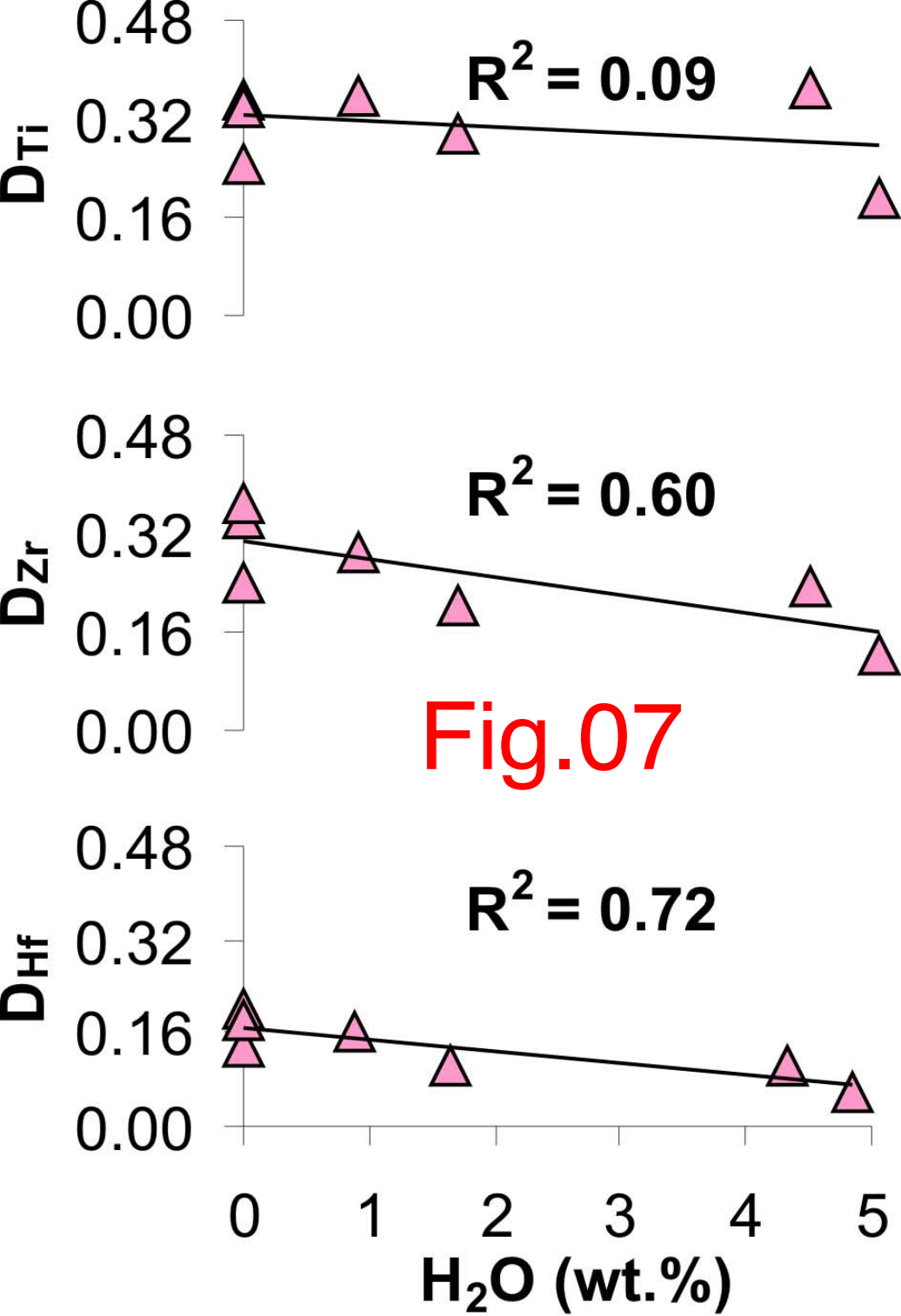

Fig.07

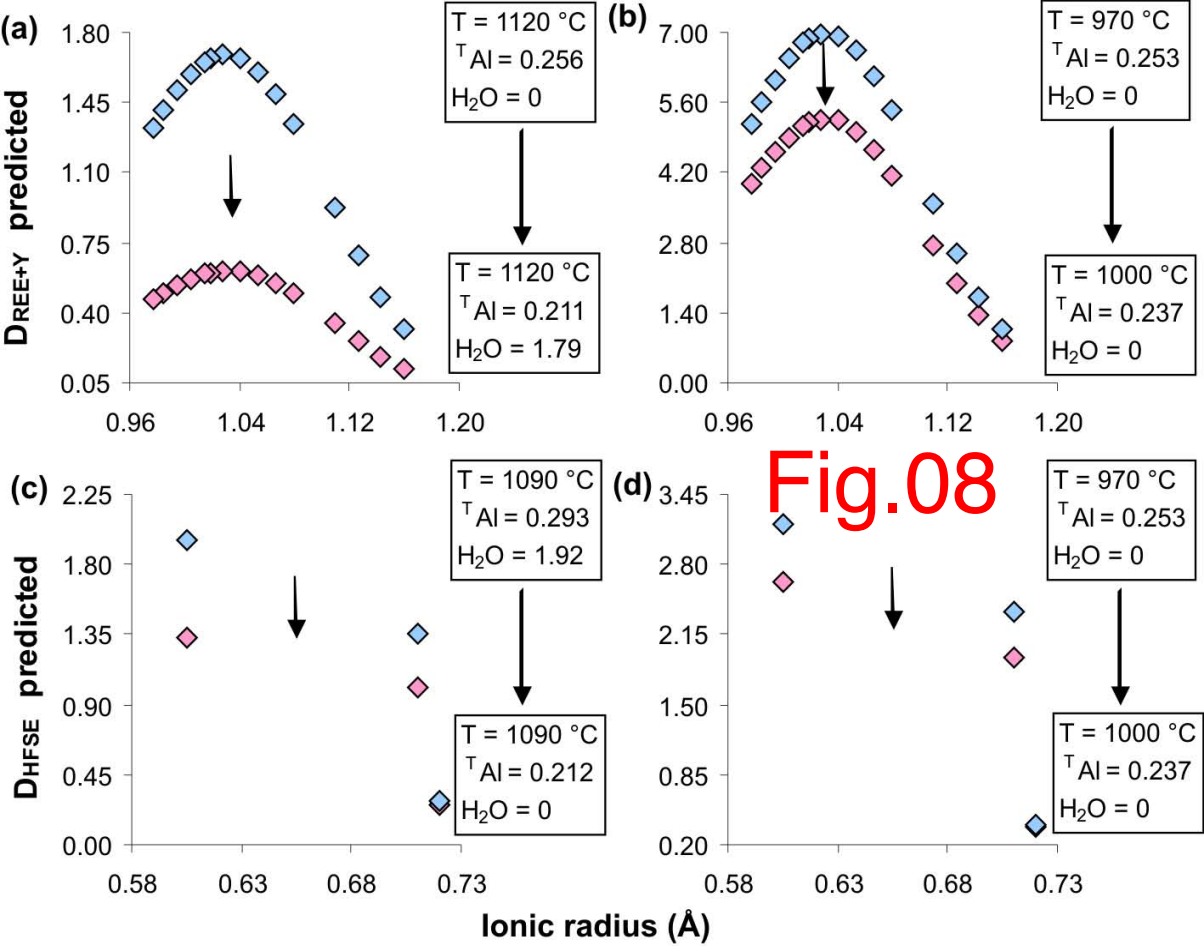

Fig. 08

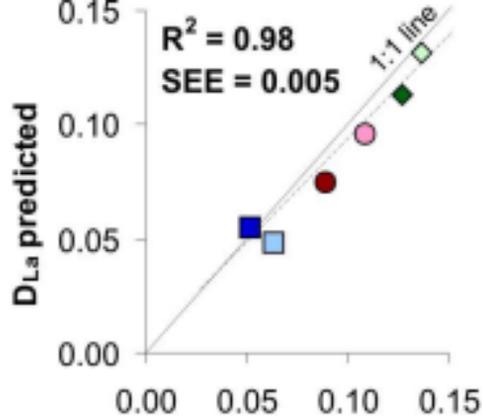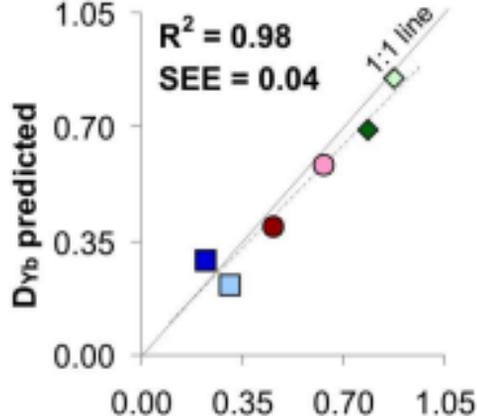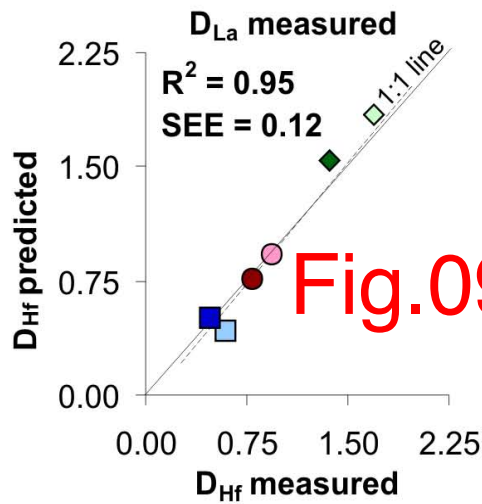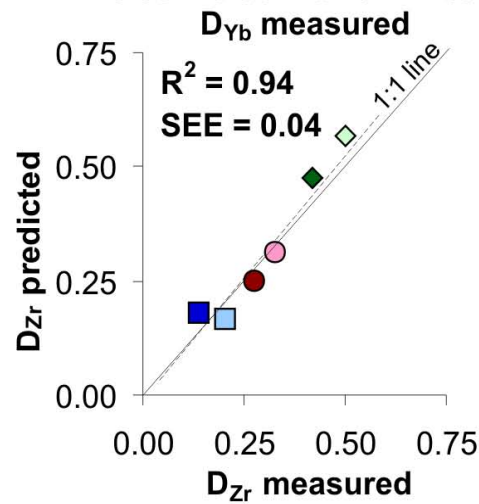

Fig.09

| BA +2 wt.% H$_2$O added | TB +2 wt.% H$_2$O added | BT +2 wt.% H$_2$O added |
| BA +3 wt.% H$_2$O added | TB +3 wt.% H$_2$O added | BT +3 wt.% H$_2$O added |

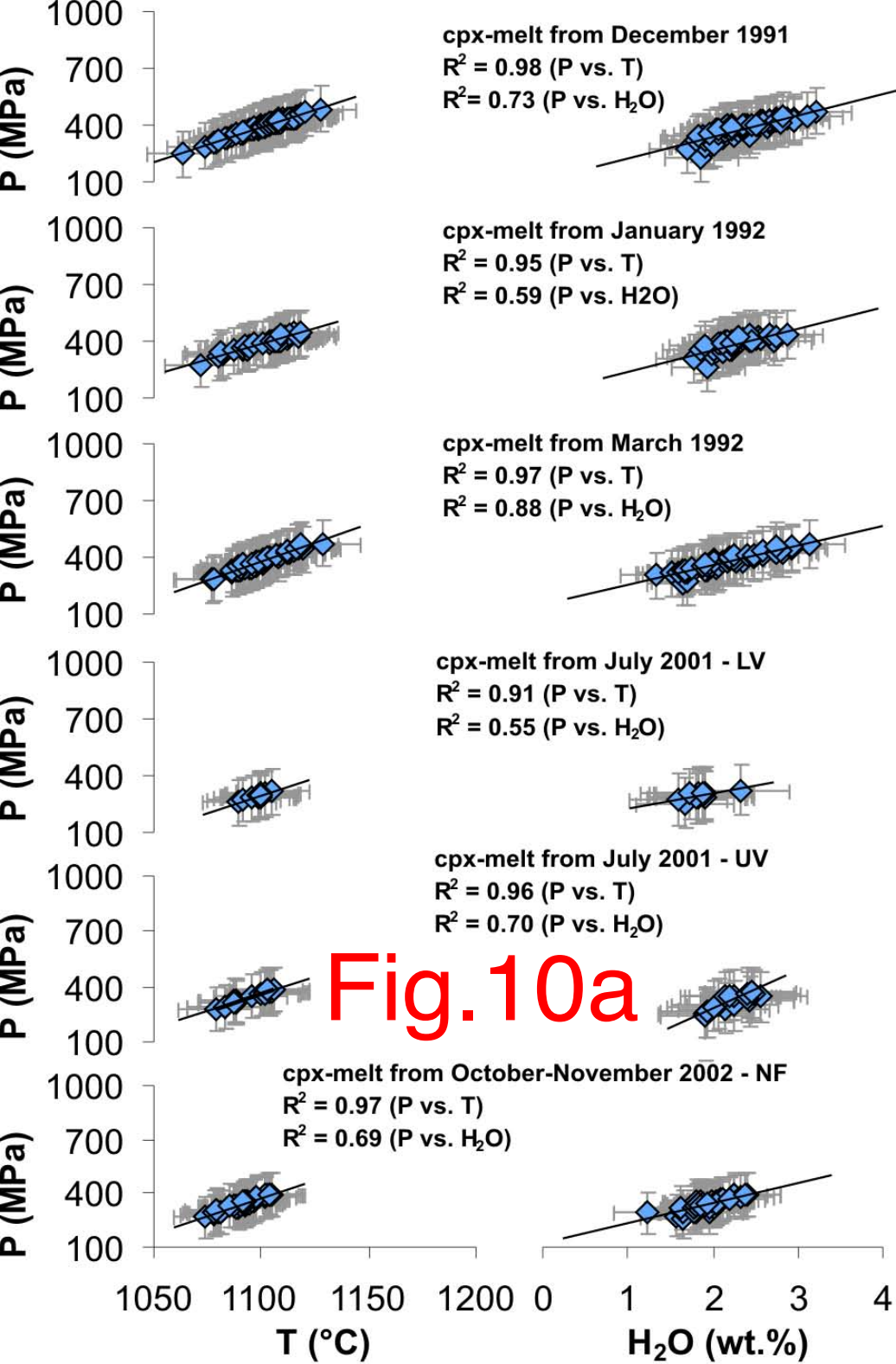

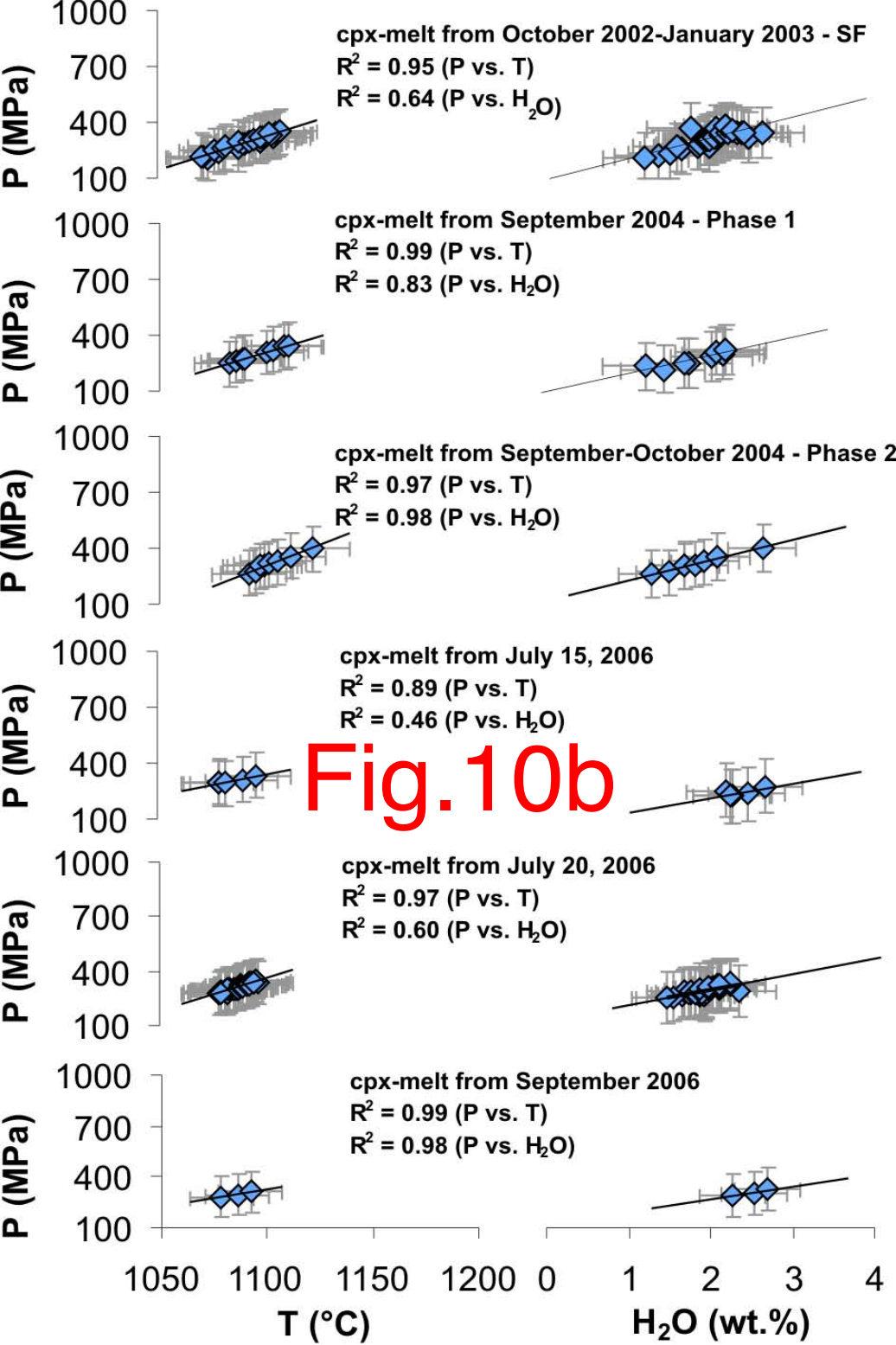

Fig. 10b

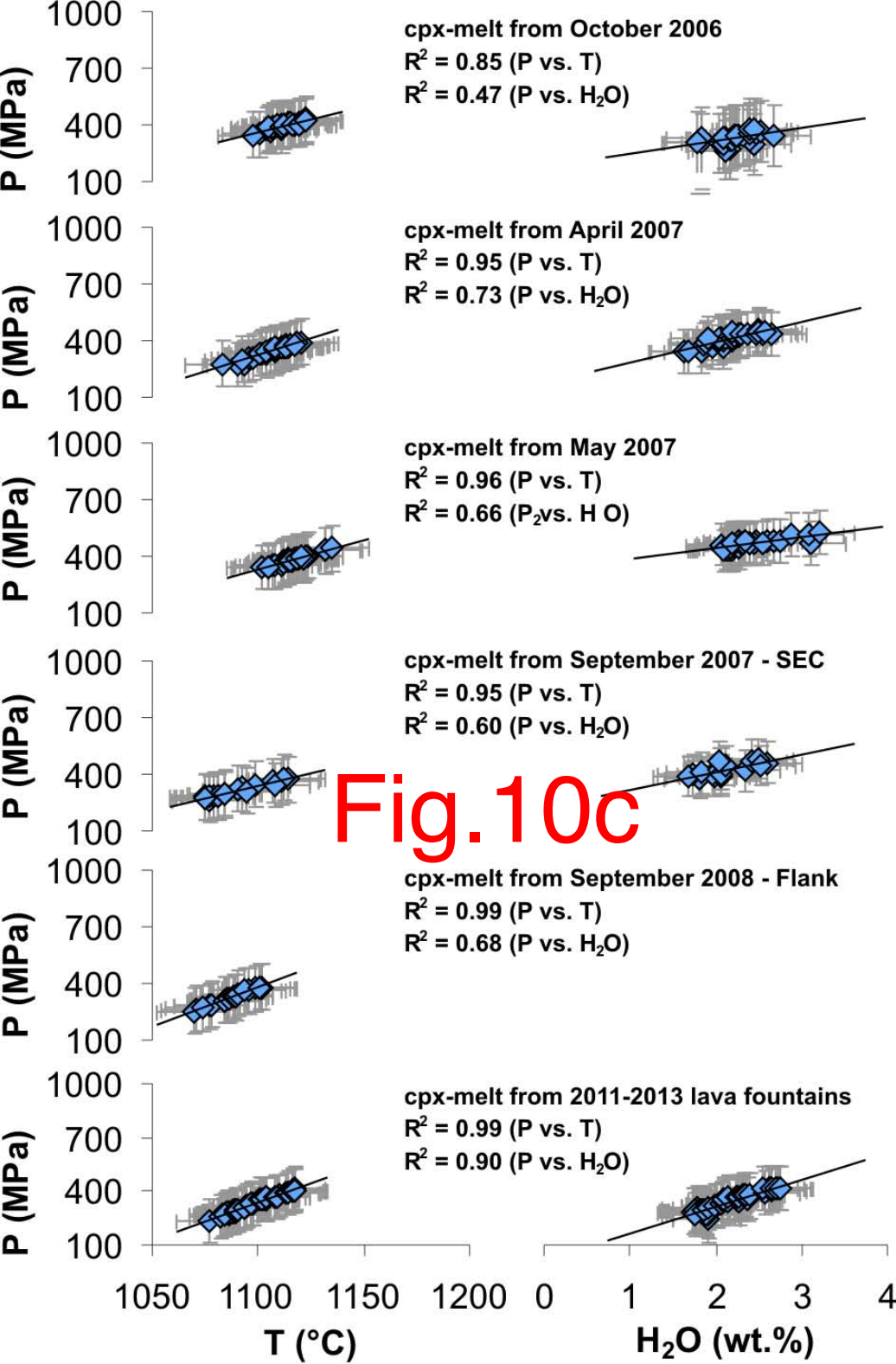

Fig.10c

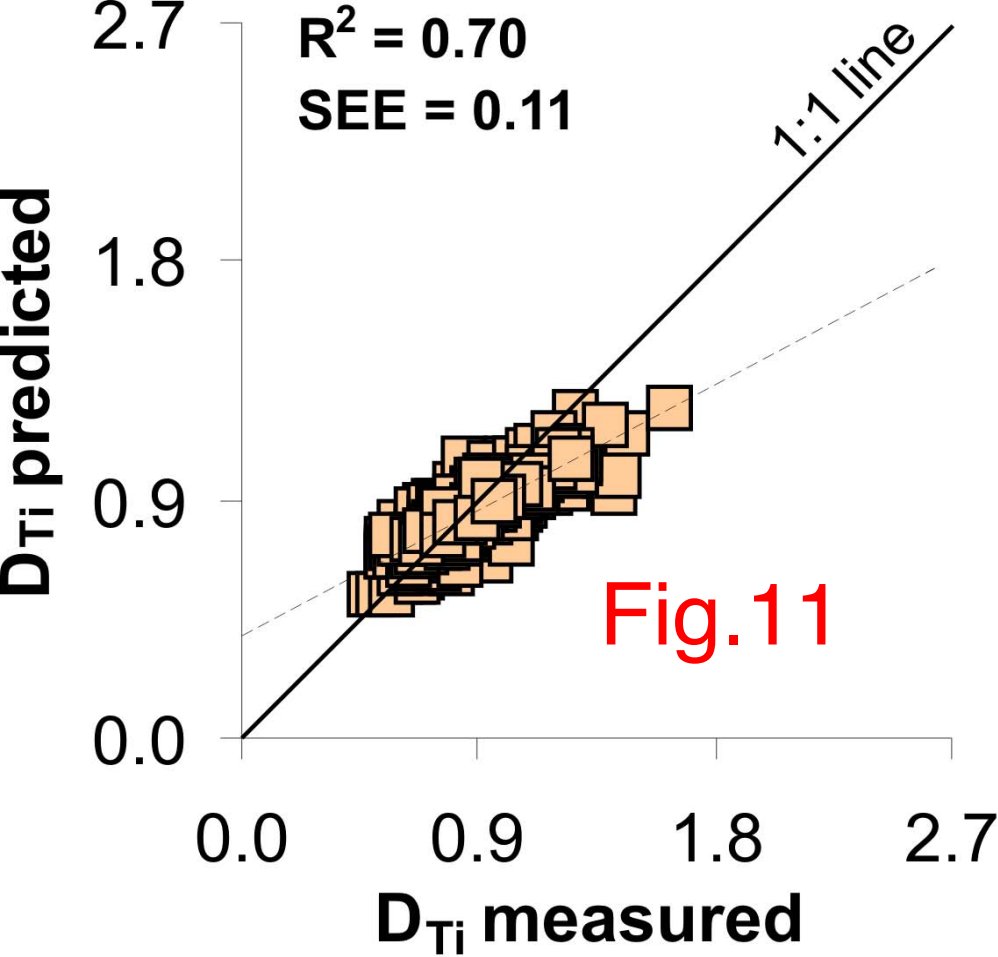

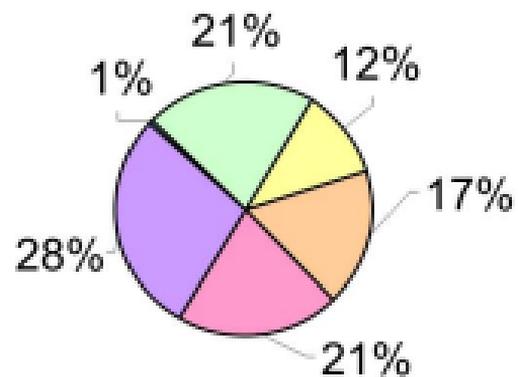
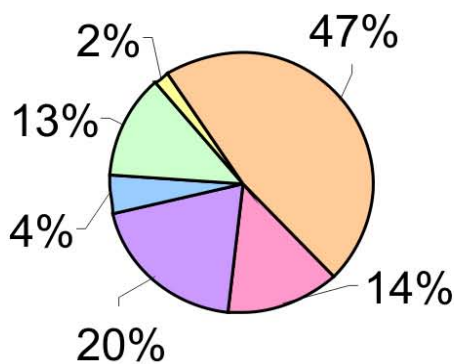
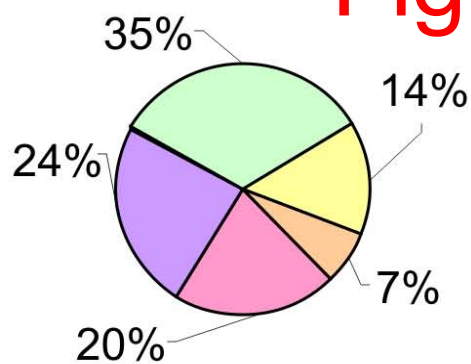
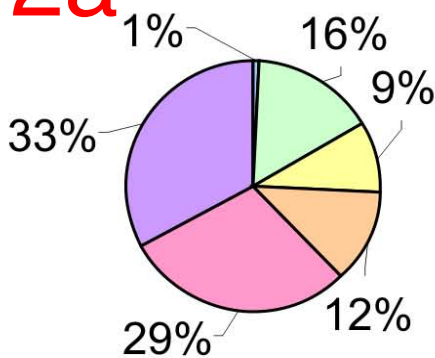
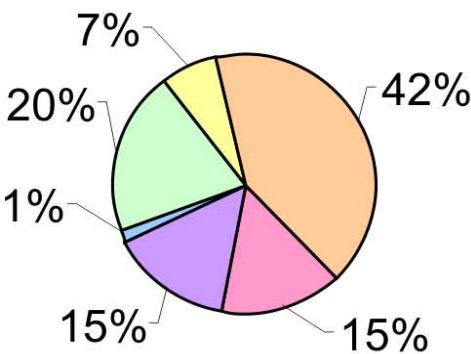
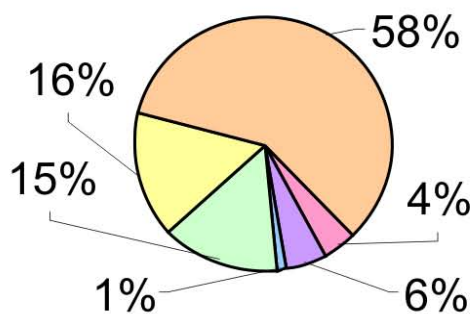

Fig.12a

| | |
|---|---|
| $P$ (GPa) | $T$ (°C) |
| $H_2O$ (wt.%) | $^TAl$ |
| $[Ca^{2+}/(M^+ + M^{2+})]$ | Electrostatic effects |

October 2002-January 2003 - SF 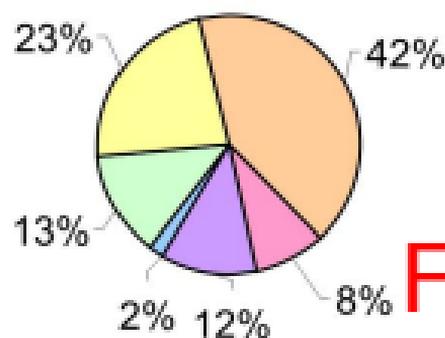 July 15, 2006 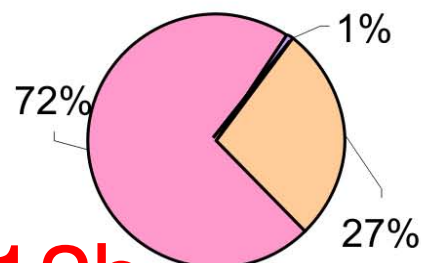

September 2004 - Phase 1 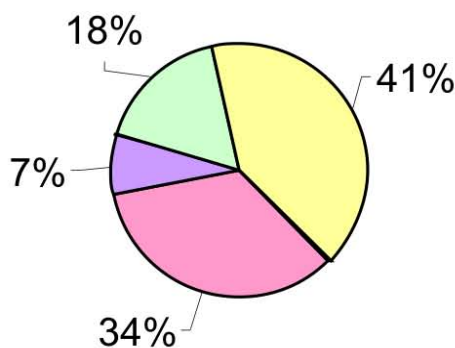 July 20, 2006 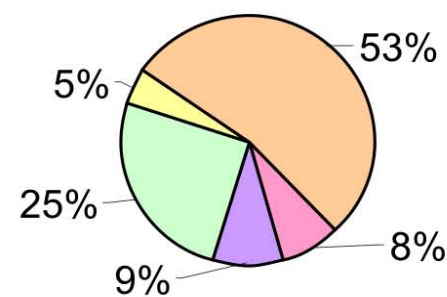

September-October 2004 - Phase 2 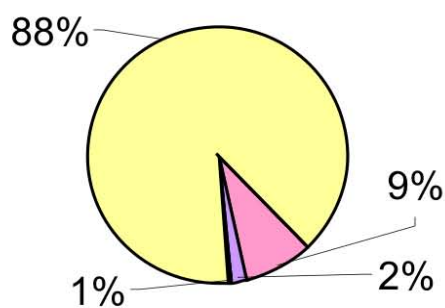 September 2006 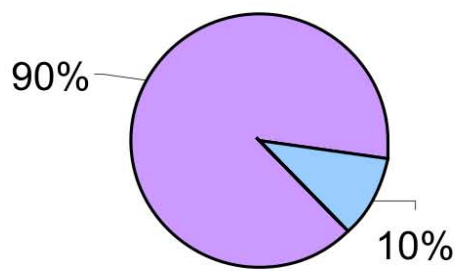

Fig.12b

| | |
|---|---|
| $P$ (GPa) | $T$ (°C) |
| $H_2O$ (wt.%) | $^T Al$ |
| $[Ca^{2+}]/(M^+ + M^{2+})$ | Electrostatic effects |

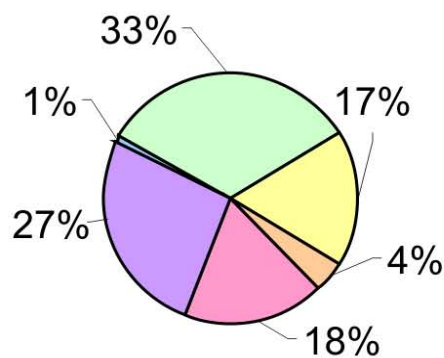 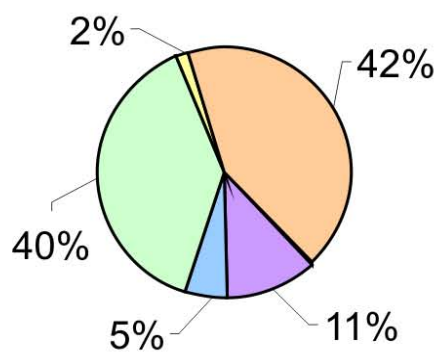
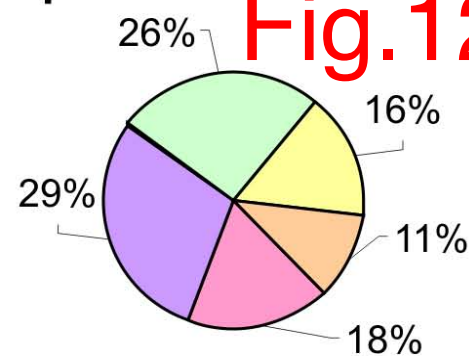 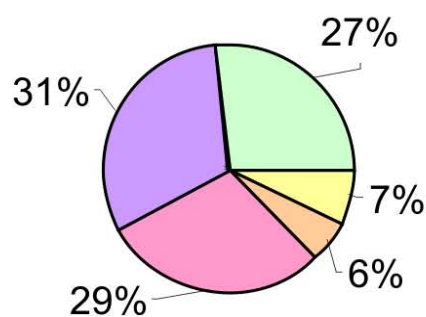
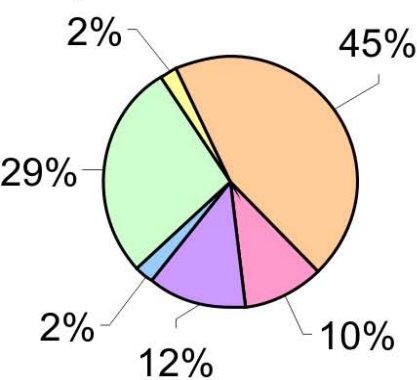 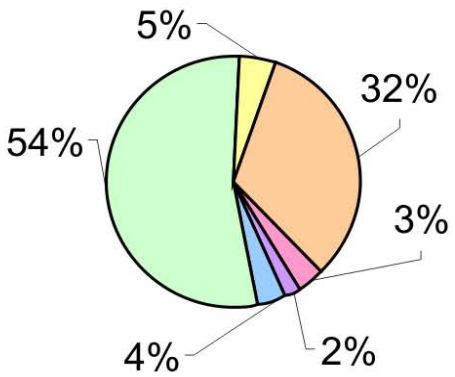

Fig.12c

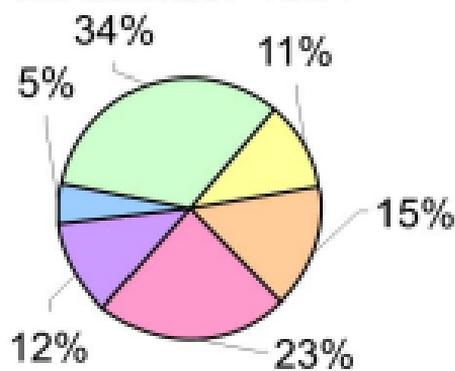
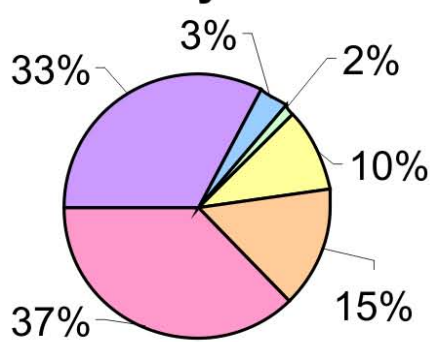

Fig.13a

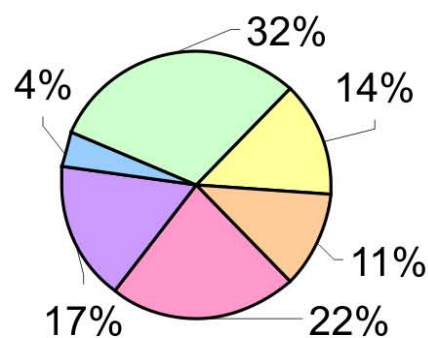
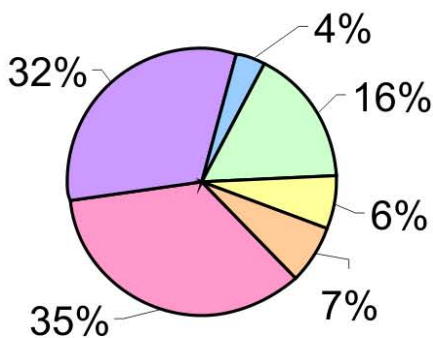
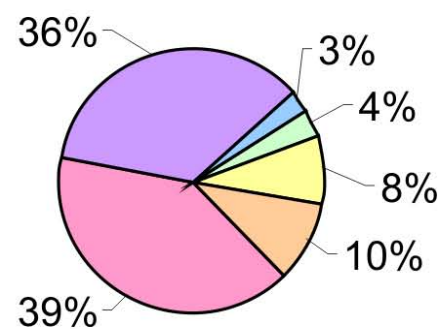
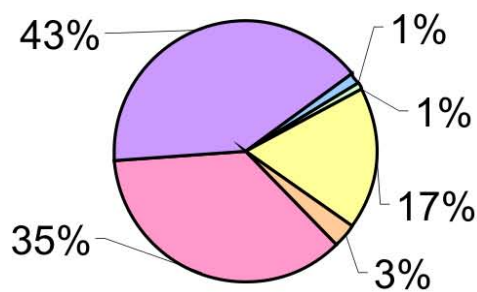

- $P$ (GPa)
- $T$ (°C)
- $H_2O$ (wt.%)
- $^TAl$
- $[Ca^{2+}/(M^+ + M^{2+})]$
- Electrostatic effects

## October 2002-January 2003 - SF

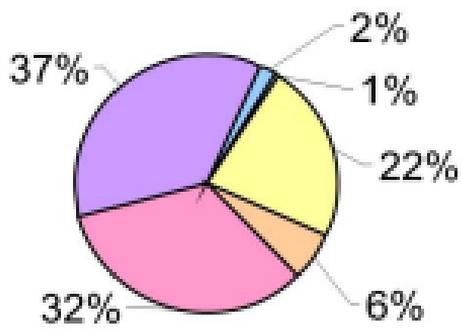

37%, 2%, 1%, 22%, 6%, 32%

## July 15, 2006

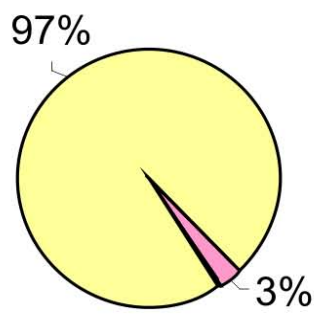

97%, 3%

## September 2004 - Phase 1

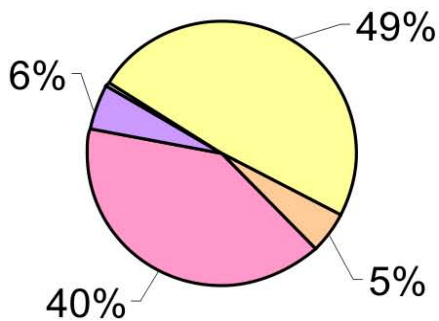

6%, 49%, 5%, 40%

## July 20, 2006

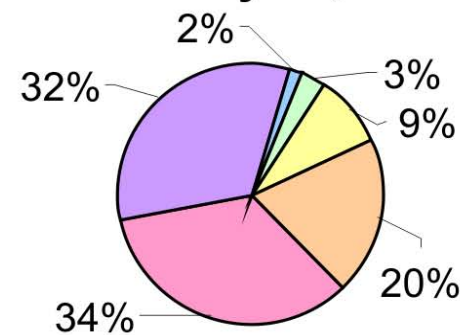

2%, 3%, 9%, 20%, 34%, 32%

## September-October 2004 - Phase 2

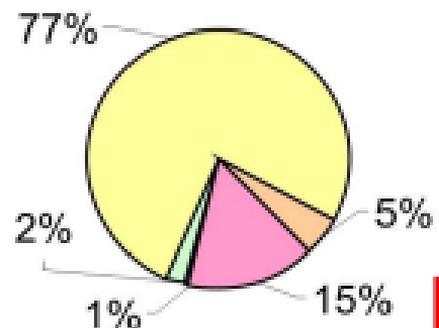

77%, 5%, 15%, 1%, 2%

## September 2006

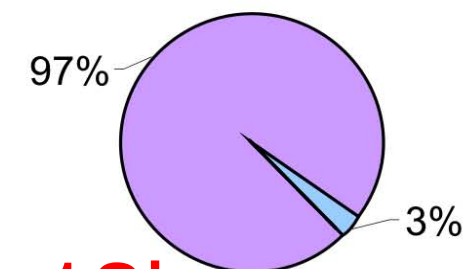

97%, 3%

Fig.13b

- $P$ (GPa)
- $T$ (°C)
- $H_2O$ (wt.%)
- $^TAl$
- $[Ca^{2+}/(M^+ + M^{2+})]$
- Electrostatic effects

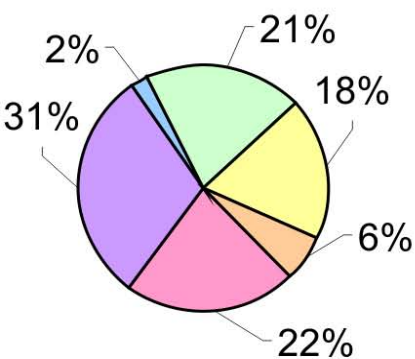
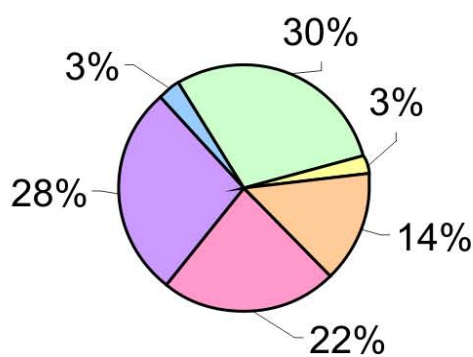
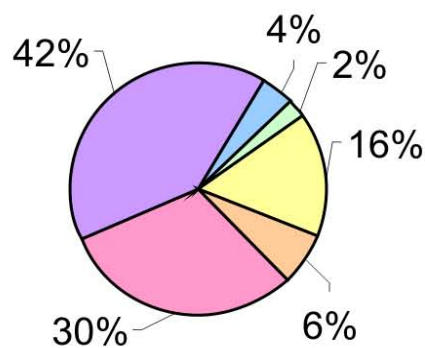
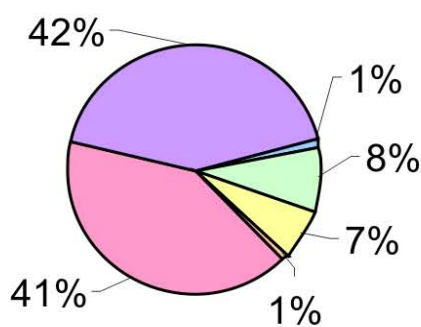
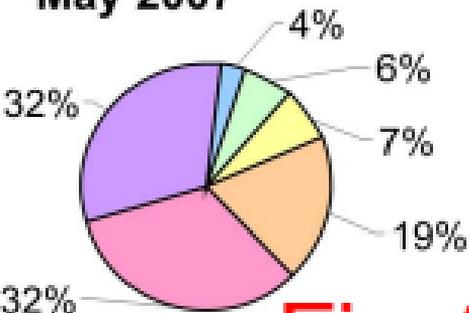
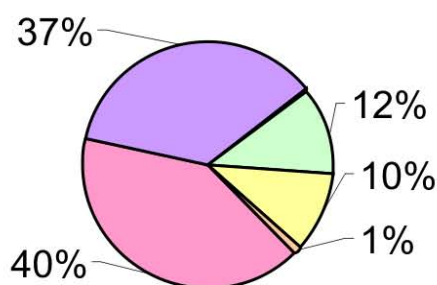

Fig.13c

- $P$ (GPa)
- $T$ (°C)
- $H_2O$ (wt.%)
- $^TAl$
- $[Ca^{2+}/(M^+ + M^{2+})]$
- Electrostatic effects

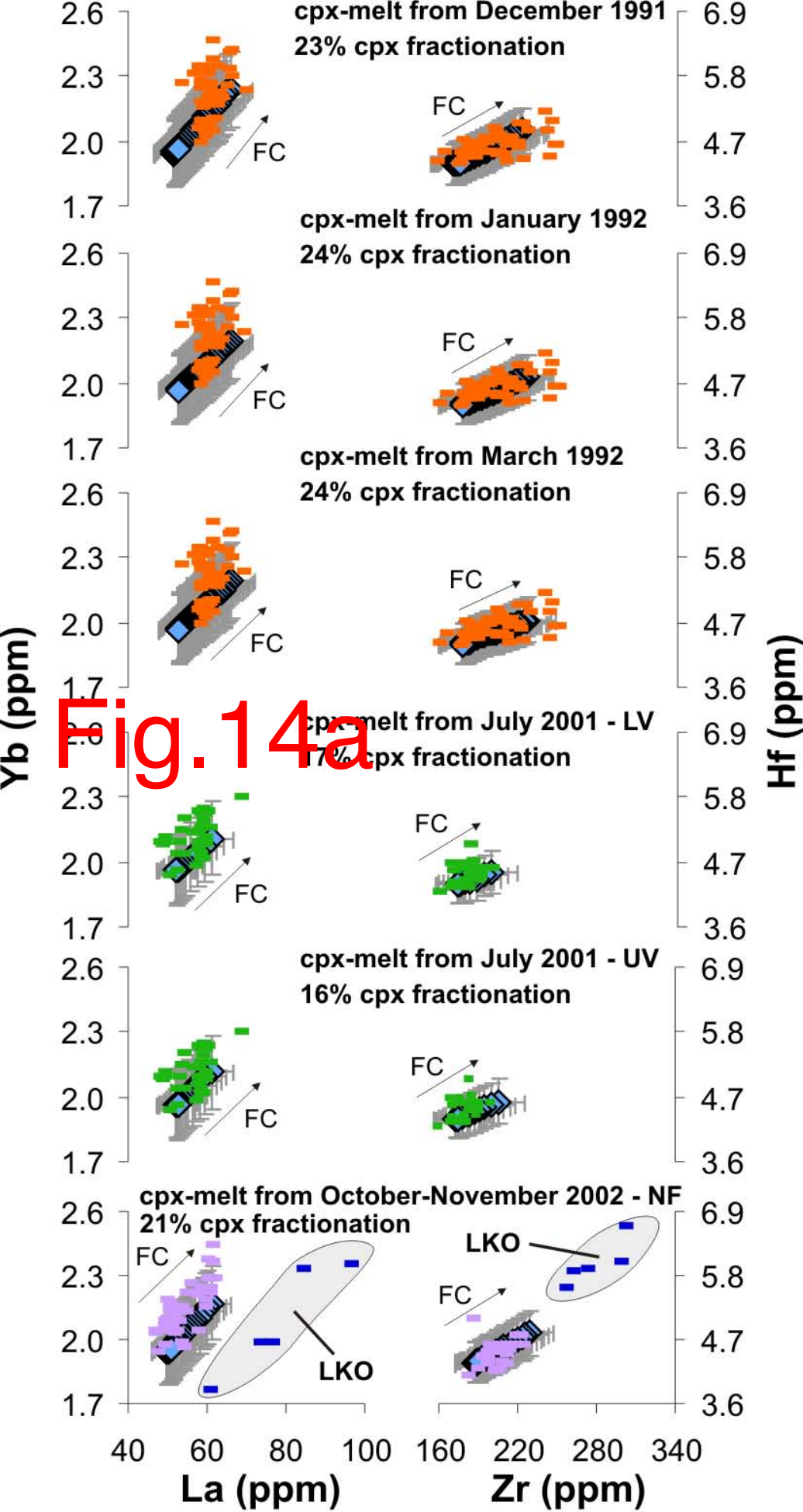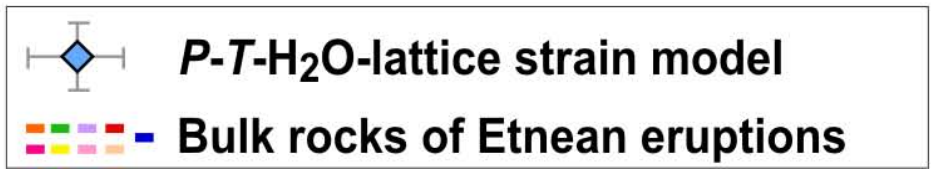

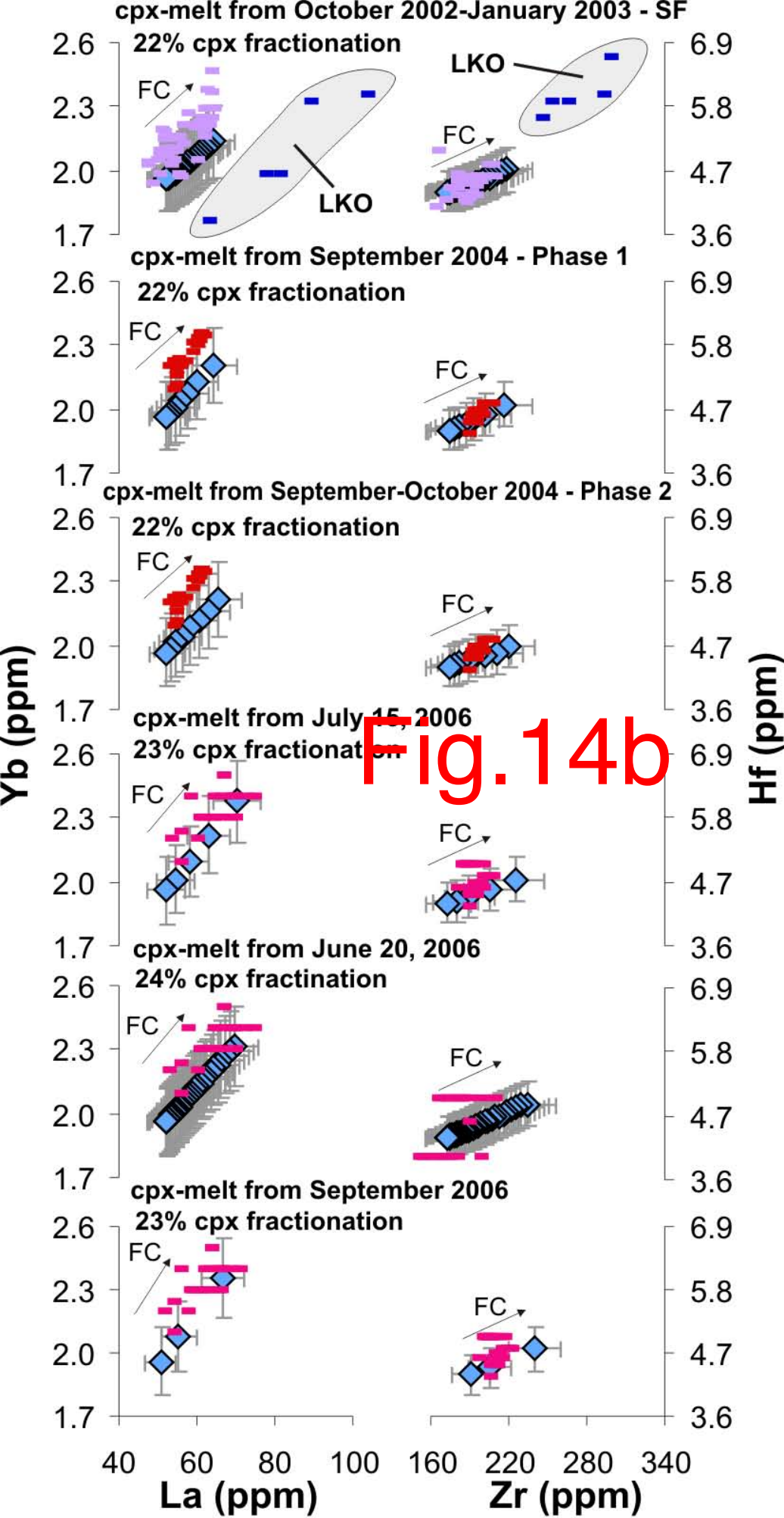

Fig.14b

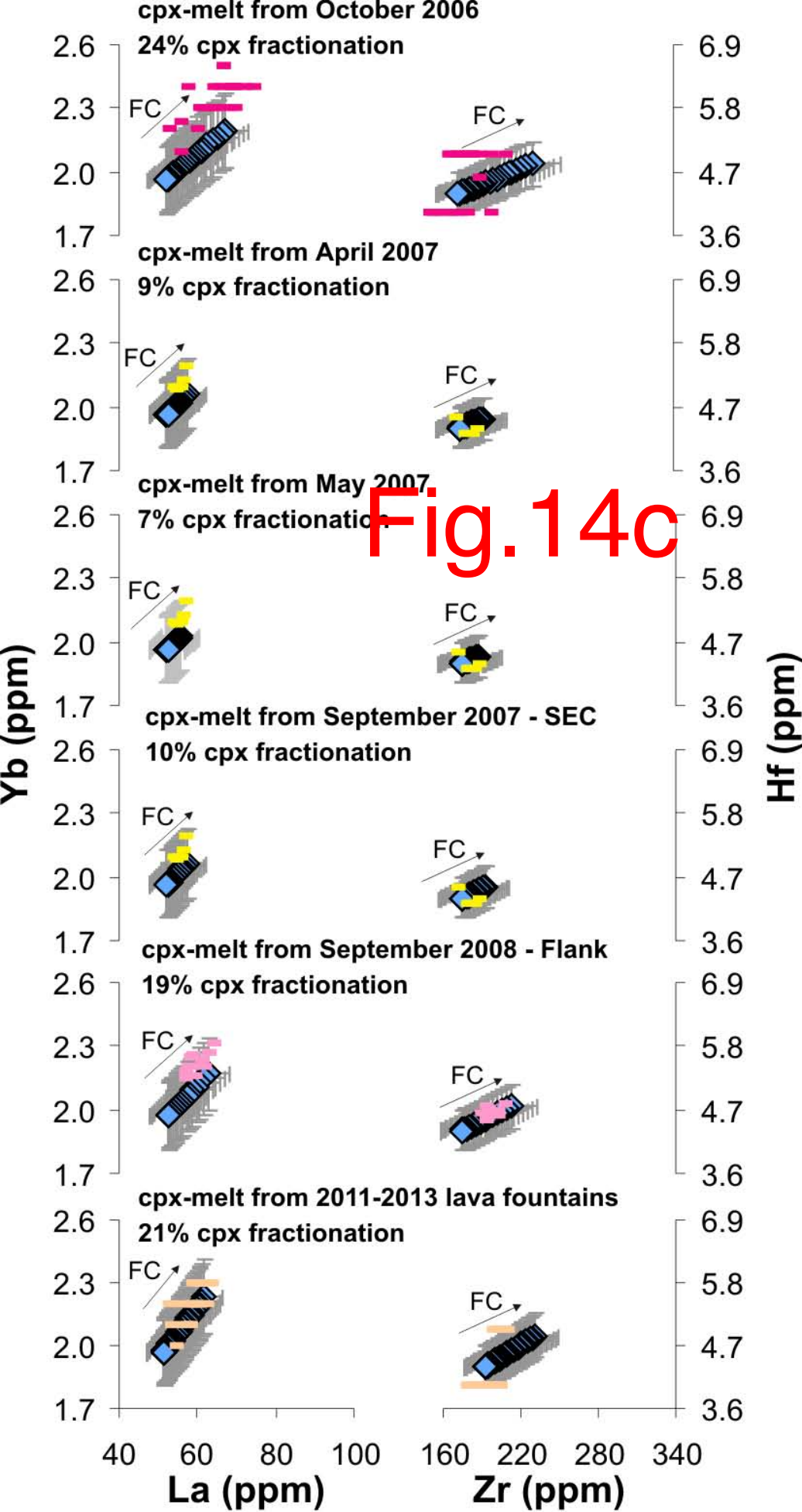
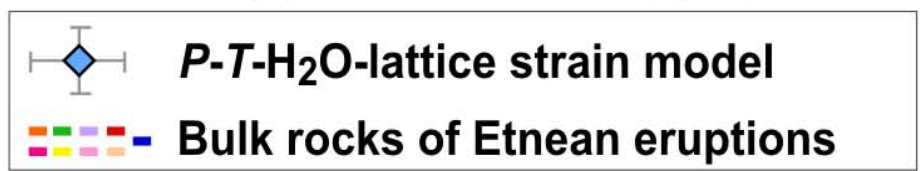

Fig.14c

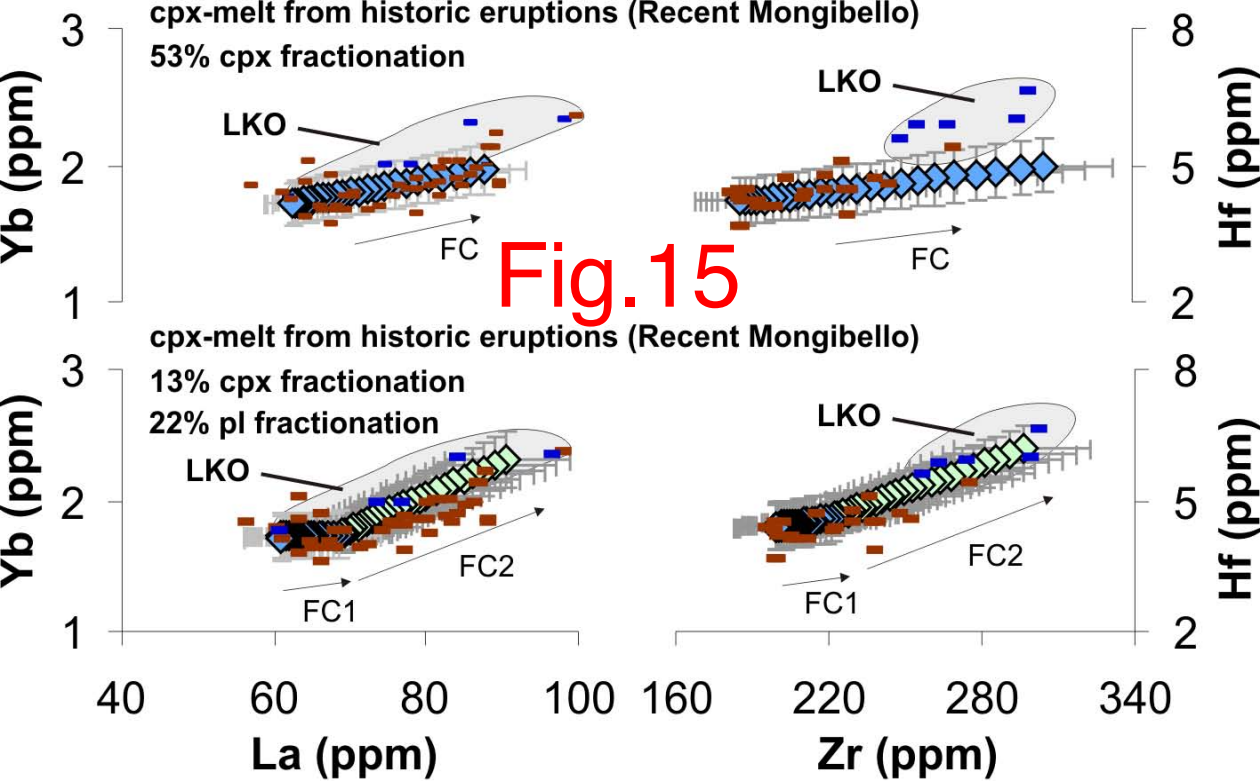

Fig.15

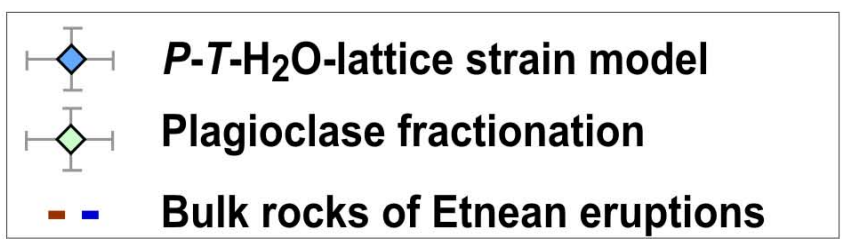